\documentclass[acmsmall]{acmart}

\usepackage{subcaption}
\usepackage{listings}
\usepackage{xspace}
\usepackage{tikz}
\usetikzlibrary{positioning, calc, fit, backgrounds, automata, tikzmark, arrows,shapes}
\usepackage{pgfmath}
\usepackage{algorithm}
\usepackage{algpseudocode}
\usepackage{enumitem}
\usepackage[hang,flushmargin,bottom]{footmisc}  
\usepackage{amsmath}
\usepackage{amsfonts}

\usepackage{amssymb}
\usepackage[capitalize]{cleveref}
\usepackage{multirow}
\usepackage{stmaryrd}
\usepackage{wrapfig}
\usepackage{syntax}
\usepackage{mdframed}
\usepackage{booktabs}
\usepackage{cancel}
\usepackage{soul}

\usepackage[dvipsnames]{xcolor} % colors

\newcommand{\sys}{\textsc{Scion}\xspace}
\newcommand{\bonsai}{\textsc{Bonsai}\xspace}

\newcommand{\kw}[1]{\textbf{#1}}

\usepackage{caption}
\lstdefinestyle{cpp-style}{
  language=C++,
  basicstyle=\tiny\ttfamily,
  backgroundcolor=\color{gray!8},
  frame=single,
  commentstyle=\color{teal!70!black}\ttfamily,
  keywordstyle=\color{blue!70!black}\ttfamily,
  stringstyle=\color{orange!70!black}\ttfamily,
  showstringspaces=false,
  breaklines=true,
  tabsize=2,
  numbers=left,
  numbersep=4pt,
  xleftmargin=4pt,
  xrightmargin=4pt,
  numberstyle=\tiny\color{black},
  backgroundcolor=\color{gray!8},
  columns=fullflexible,
  aboveskip=3pt, 
  belowskip=3pt,
  escapechar=§,
  keepspaces=true,
  breakatwhitespace=true,
}

\definecolor{FireBrickRed}{RGB}{178,34,34}
\definecolor{DarkCyan}{RGB}{30,144,255}
\definecolor{LessDarkCyan}{RGB}{20,110,255}
\lstdefinelanguage{layout}{
  keywords={layout, field, derive, group, split, ptr, direct, indirect, from, by, parent, build, this, root},
  keywordstyle=\color{DarkOrchid}\textbf,
  identifierstyle=\color{black},
  sensitive=true,
  comment=[l]{//},
  commentstyle=\color{Gray},
  morekeywords = [2]{
    f32x3, f32x4, f16x3, q16x3, u8x3, i8x3,
    u64x4, u64x8, i32x4, u32x3, u16x3,
    u1, u2, u4, u5, u8, u16, u28, u30, u32, u64, u128,
    i1, i2, i4, i8, i16, i32, i64, i128,
    f16, f32, f64, f128, u32x8, 
    f32x3x4, f32x3x8, f32x4x3, u8x3x4, i8x4x3,
    void, bool, boolx3, qbox3x8, qbox3
  },
  keywordstyle=[2]\color{DarkCyan},
  morekeywords=[3]{
    Interior, Leaf, Node,
    BVH, BVHm, BVHt, BVHd, LinearBVH,
    Triangle, Sphere, Point, BezierSegment, Primitive, 
    Ray, FInterval, TriangleIntersection,
    AABBNode, OBBNode, AABB, OBB, DOP14
  },
  keywordstyle=[3]\color{LessDarkCyan}\textbf,
  morekeywords=[4]{if, else, as, to, type, func, mut, match, foreach, let, return, set, option},
  keywordstyle=[4]\color{FireBrickRed}\textbf,
}

\lstdefinestyle{layout-style-base}{
  language=layout,
  columns=fullflexible,
  aboveskip=3pt, 
  belowskip=3pt,
  keepspaces=true,
  breakatwhitespace=true,
  numbers=left,
  numbersep=4pt,
  numberstyle=\tiny\color{black},
  backgroundcolor=\color{gray!8},
  frame=single,
  framesep=3pt,
  xleftmargin=4pt,
  xrightmargin=4pt,
  escapechar=§,
  literate=
    {_}{{\_}}1
    {<}{{<}}1
    {>}{{>}}1
}

\lstdefinestyle{layout-style}{
  style=layout-style-base,
  breaklines=false,
  basicstyle=\ttfamily\tiny,
}
\lstdefinestyle{layout-style-full-size}{
  style=layout-style-base,
  breaklines=true,
  basicstyle=\ttfamily\small,
}

\NewDocumentCommand{\code}{m}{%
  \lstinline[style=layout-style-full-size]|#1|%
}

\AtBeginDocument{%
  }

\setcopyright{cc}
\setcctype{by}
\acmDOI{10.1145/3808253}
\acmYear{2026}
\acmJournal{PACMPL}
\acmVolume{10}
\acmNumber{PLDI}
\acmArticle{175}
\acmMonth{6}
\acmSubmissionID{pldi26main-p19-p}
\received{2025-11-07}
\received[accepted]{2026-04-03}

\begin{document}

\title{Decoupling Data Layouts from Bounding Volume Hierarchies}

\author{Christophe Gyurgyik}
\orcid{0000-0001-8493-1133}
\affiliation{%
  \institution{Stanford University}
  \city{Stanford}
  \country{USA}
}
\email{cpg@cs.stanford.edu}

\author{Alexander J Root}
\orcid{0000-0001-6221-1389}
\affiliation{%
  \institution{Stanford University}
  \city{Stanford}
  \country{USA}
}
\email{ajroot@cs.stanford.edu}

\author{Fredrik Kjolstad}
\orcid{0000-0002-2267-903X}
\affiliation{%
  \institution{Stanford University}
  \city{Stanford}
  \country{USA}
}
\email{kjolstad@cs.stanford.edu}

% \renewcommand{\shortauthors}{Gyurgyik et al.}

\begin{abstract}
Bounding volume hierarchies are ubiquitous acceleration structures in graphics, scientific computing, and data analytics. Their performance depends critically on data layout choices that affect cache utilization, memory bandwidth, and vectorization---increasingly dominant factors in modern computing. Yet, in most programming systems, these layout choices are hopelessly entangled with the traversal logic. This entanglement prevents developers from independently optimizing data layouts and algorithms across different contexts, perpetuating a false dichotomy between performance and portability. We introduce \sys{}, a domain-specific language and compiler for specifying the data layouts of bounding volume hierarchies independent of tree traversal algorithms. We show that \sys{} can express a broad spectrum of layout optimizations used in high-performance computing while remaining architecture-agnostic. We demonstrate empirically that Pareto-optimal layouts (along performance and memory footprint axes) vary across algorithms, architectures, and workload characteristics. Through systematic design exploration, we also identify a novel ray tracing layout that combines optimization techniques from prior work, achieving Pareto-optimality across diverse architectures and scenes.
\end{abstract}



%   \label{figure:teaser}


\maketitle

\section{Introduction}
\label{sec:introduction}

In high-performance graphics, scientific computing, and data analytics, bounding volume hierarchies (BVHs) are essential acceleration structures used for spatial queries such as ray tracing, closest point queries, and collision detection. Extensive research has focused on optimizing the layout of these structures due to their substantial impact on performance~\cite{christensen2025renderman, eisemann2012implicit, howard2019quantized, ericson2004rtcd, kavcerik2024sah, kay1986ray, laine2013megakernels, meister2021survey, pharr2023pbrt, mahovsky2006memory, segovia2010memory, vaidyanathan2016watertight, wald2008simdrt, weghorst1984improvedrt, woop2014hair, ylitie2017incoherentrt, wodniok2013analysis, benthin2018compressed, kim2010racbvh, wachter2006instant, eisemann2008ray, havran2006fastconstruction, lin2020hardware, bauszat2010minimal, mora2011naive, cline2006lightweight, lauterbach2008reducem, wald2007deformable, dammertz2008shallow, ernst2008multi, smits2005efficiency, wachter2007terminating, mahovsky2005ray, stuerzlinger1994optimization, wald2014embree, benthin2017improved, benthin2021ray, chitalu2020binary, liktor2016bandwidth, huang2025aqb8, howard2016efficient, larsson2009bounding}.

However, no single layout is universally optimal. The Pareto frontier of layouts---balancing runtime performance and memory utilization---depends critically on three factors: the algorithm, hardware architecture, and characteristics of the input data. Layouts optimized to maximize performance on CPUs may increase the BVH memory footprint to reduce instruction count or improve cache utilization on latency-bound workloads~\cite{wald2014embree, benthin2018compressed, woop2014hair}. On the other hand, memory-efficient layouts are essential for bandwidth-bound systems such as GPUs~\cite{laine2013megakernels, liu2023lumibench, saed2022vulkan, huang2025aqb8} and memory-constrained platforms such as mobile devices ~\cite{lee2013sgrt, nah2014raycore, kosarevsky2025raytracing, seo2017efficient}, where compact representations directly reduce memory traffic and enable larger scenes to fit in limited memory. Scene properties further complicate the picture, e.g., extremely sparse scenes will benefit from different representations than dense scenes, and the spatial distribution of primitives affects the efficacy of quantization and compression schemes. This produces a substantial exploration space: even a conservative enumeration of $k$ data layouts $\times$ $m$ machines $\times$ $n$ algorithms $\times$ $p$ scenes yields a multifarious set of evaluation contexts, each admitting a potentially different Pareto-optimal solution.

Exploring this design space is complex because general-purpose languages tightly couple data layouts with application logic. Optimized layouts employ diverse techniques: global layout transformations, e.g., struct-of-arrays and hybrid variants~\cite{laine2013megakernels, wodniok2013analysis, woop2014hair, cline2006lightweight, dammertz2008shallow}; bit field exploitation, e.g., union types~\cite{benthin2018compressed, meister2021survey, eisemann2008ray, pharr2023pbrt, cline2006lightweight}; increased branching factors (arity of the tree)~\cite{goldsmith1987automatic, woop2014hair, ylitie2017incoherentrt, dammertz2008shallow, ernst2008multi}; specialized bounding volume representations~\cite{woop2014hair, kavcerik2024sah, kavcerik2025sobb, larsson2009bounding}; implicit indices~\cite{smits2005efficiency, cline2006lightweight, bauszat2010minimal, eisemann2012implicit}; and compression~\cite{howard2019quantized, segovia2010memory, cline2006lightweight, mahovsky2005ray, benthin2018compressed}. Each of these techniques requires pervasive changes throughout the application code, in how fields are accessed, nodes are referenced, and the tree is traversed.

Orthogonally, programming language research has developed ways to decouple \emph{logical representations} from \emph{physical layouts}. This prior work has focused on fine-grain, per-node layout optimization~\cite{baudon2023bitstealing, chen2023dargent} or coarse-grain composition of data structures hidden behind iterator interfaces~\cite{kjolstad2017taco, liu2024unisparse, hu2019taichi, pharr2012ispc}. However, state-of-the-art BVH layout optimization requires coordination across both levels of granularity: how individual nodes are represented and how collections of nodes are organized in memory.

We propose a decoupling with control over both. Our key insight is that manual BVH layout optimizations explore different physical realizations of the same underlying logical structure. This enables a clean separation between a tree's logical specification and its physical representation. This separation preserves the portability and maintainability of traversal algorithms while simultaneously enabling performance engineers to independently tune physical layouts for specific evaluation contexts. Akin to the separation of algorithm and schedule~\cite{jrk2013halide, ikarashi2022exo}, we introduce domain-specific languages to separate algorithm and physical layout. Our primary contributions are the following:

\begin{itemize}[leftmargin=2.5em]
    \item Two domain-specific languages that decouple a tree's logical data structure from its physical representation: a \emph{layout} language for specifying physical-to-logical mappings, and a \emph{build} language for expressing the inverse logical-to-physical transformation.
    \item A compilation strategy for \emph{destructor specialization} that lowers pattern matching onto different physical layouts in accordance with the layout specification.
    \item A complementary compilation strategy for \emph{constructor specialization} that generates layout-specific constructors from canonical algebraic data types.
\end{itemize}

We implement these ideas in \sys{}\footnote{The name draws from botanical terminology: \emph{scion} is the grafted portion of a plant chosen to yield particular traits. Analogously, our approach grafts \emph{data layouts} onto acceleration trees to realize targeted performance properties.}, a system that decouples the specification and traversal logic of tree data structures from their physical representations. This separation enables expressive and portable traversal implementations while supporting systematic exploration of the physical representation design space. Our evaluation demonstrates that layout performance varies significantly across algorithms, architectures, and data characteristics, confirming that design space exploration is essential for Pareto-optimal performance. Through this exploration, we discover a novel layout that composes optimizations from prior work in a previously unexplored way, achieving Pareto optimality across 35 of 42 evaluation contexts. Additionally, we provide evidence that \sys{}-generated code is competitive with state-of-the-art libraries when using equivalent traversal algorithms, establishing that our abstraction imposes no performance penalty.

Finally, we clarify the scope of this work by identifying what we intentionally exclude. First, we treat tree topology as fixed (\emph{logical tree} in \Cref{sec:build-language}) and focus solely on layout optimization. While tree quality plays a significant performance role, it is an orthogonal problem with extensive prior work~\cite{benthin2024hploc, meister2017parallel, apetrei2014fast, karras2012maximizing, gu2013efficient, pantaleoni2010hlbvh, lauterbach2009fast, walter2008fast}. \sys{} explicitly separates logical and physical tree representations to facilitate future integration of topology optimization. Second, we do not explore execution strategies such as vectorization~\cite{woop2014hair, wald2014embree}, ray reordering~\cite{meister2020reordering}, and packet tracing~\cite{boulos2007packet, aila2009understanding}. \sys{} deliberately isolates data layout decisions from scheduling, treating them as separate dimensions of the optimization space. We employ techniques like parallelized outer loops and software-defined stacks for evaluation, but leave a dedicated scheduling language as important future work. While \sys{} does not perform instruction selection, specifying layouts is a critical first step as automatic vectorization relies on regular layouts~\cite{vanhattum2021vectorization, chen2021vegen, ahmad2022vector}. Third, we target read-only acceleration structures, the norm in high-performance systems~\cite{meister2021survey, benthin2018compressed, laine2013megakernels, kavcerik2024sah, howard2016efficient}. Mutability is deferred for future work.

\section{Background: Bounding Volume Hierarchies}
\label{sec:bvh}

\begin{figure*}[t]
% \vspace{-0.25em}
\centering
\begin{subfigure}[t]{0.59\textwidth}
    \includegraphics[width=\linewidth]{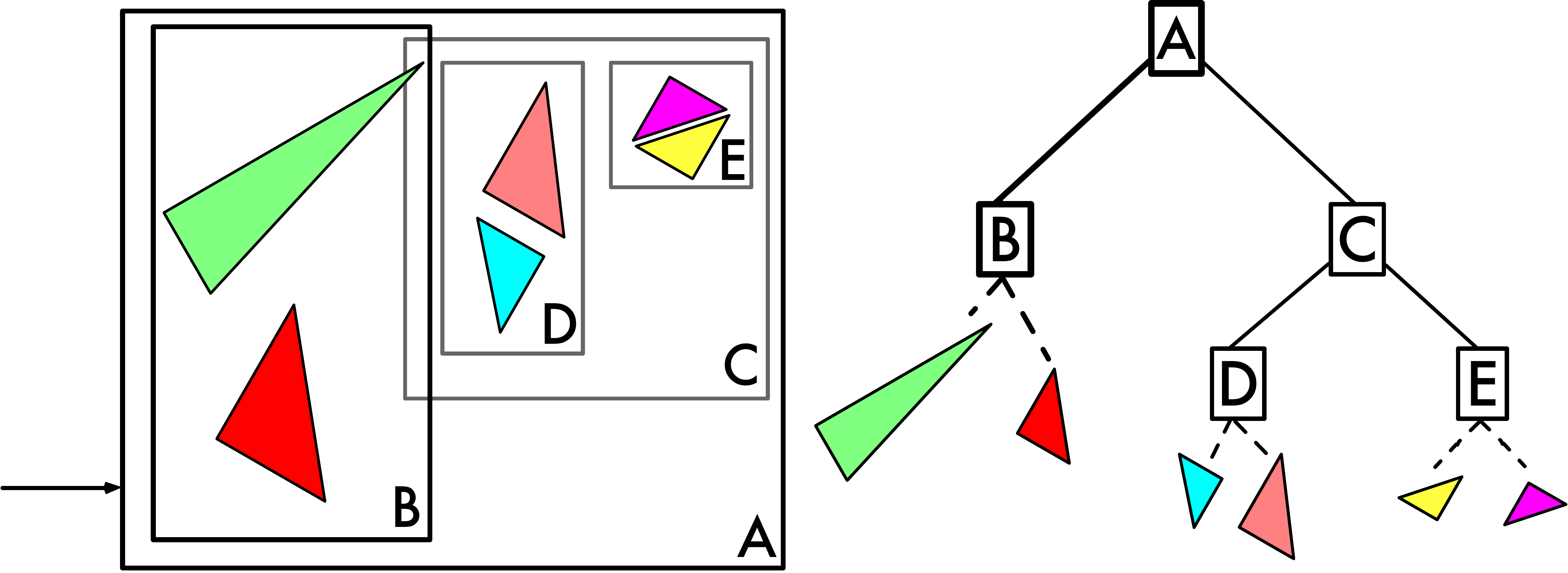}
    \caption[BVH traversal example]{A ray traversing the BVH, visiting bounding volumes \textbf{A} and \textbf{B} before performing intersection tests with contained primitives. Nodes \textbf{C}, \textbf{D}, and \textbf{E} are skipped since the ray misses their boxes.}
    \label{figure:bvh-traversal}
\end{subfigure}
\hfill
\begin{subfigure}[t]{0.37\textwidth}
    \includegraphics[width=\linewidth]{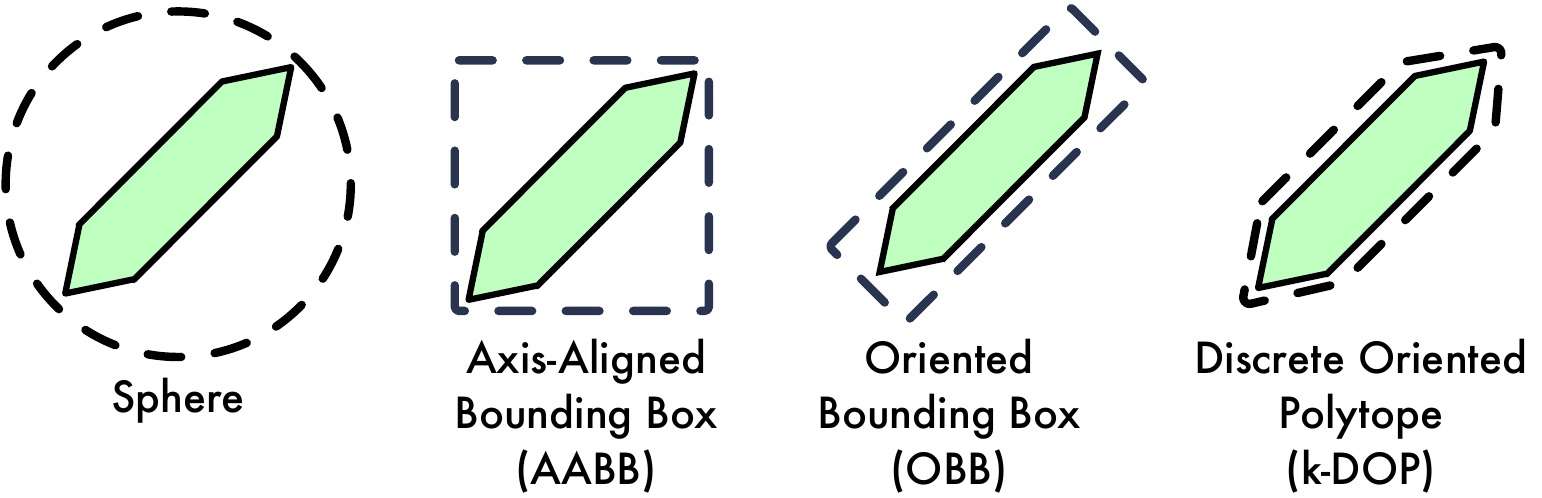}
    \caption[Bounding volume representations]{Different bounding volume representations. Tighter fits generally incur higher memory and computational cost.}
    \label{figure:bv-example}
\end{subfigure}
\label{figure:bvh}
\caption{Visualization of bounding volume hierarchies.}
\vspace{-1.25em}
\end{figure*}

Bounding volume hierarchies (BVHs) are tree-structured indexes that are widely used in data analytics, computer graphics, and scientific computing to accelerate spatial queries. Each internal node stores a bounding volume that encloses the geometry of its descendants, while each leaf stores one or more geometric primitives. As illustrated in \Cref{figure:bvh-traversal}, the root spans the entire scene, forming a spatial decomposition that enables logarithmic-time pruning of the query space~\cite{rubin1980complexscene}.

Despite their conceptual simplicity, BVHs exhibit a wide range of performance tradeoffs stemming from their \emph{data representation}. Implementations vary in how bounding volumes, nodes, and primitives are stored, aligned, and traversed. Layout choices affect traversal cost and memory footprint in complex machine-dependent and data-dependent ways, as demonstrated in prior work~\cite{woop2014hair, kavcerik2024sah, howard2019quantized}. This sensitivity makes BVHs an ideal case study for our system: even modest structural changes, e.g., fusing internal and leaf node arrays, can shift the performance frontier. For a detailed introduction to BVHs and surveys of existing optimization techniques, we refer readers to ~\citet[Ch.~7.3]{pharr2023pbrt}, ~\citet[Ch.~6]{ericson2004rtcd}, and ~\citet{meister2021survey}.

\section{Overview and Programming Model}
\label{sec:overview}

\begin{figure}[b]
\centering
\begin{tikzpicture}
    \node[anchor=south west,inner sep=0] (image) at (0,0) {\includegraphics[width=1.0\textwidth]{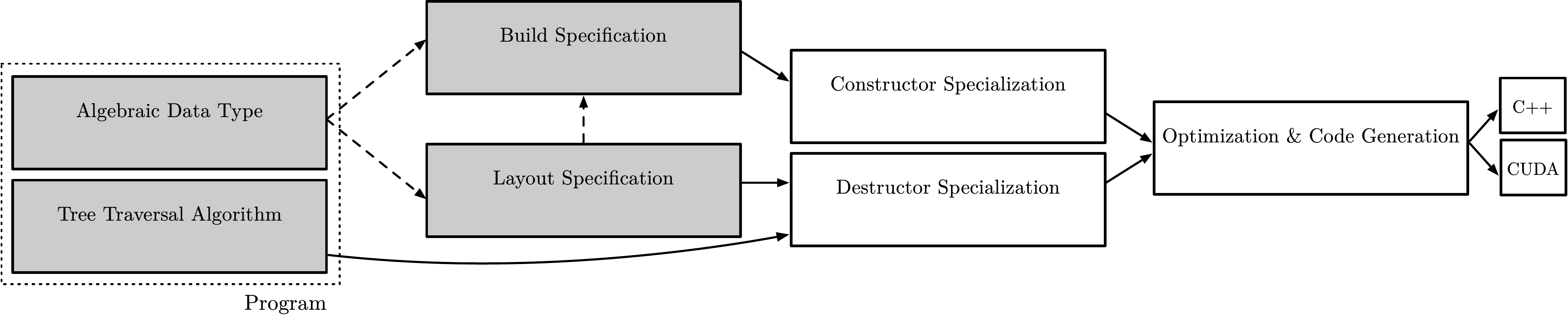}};
    \begin{scope}[x={(image.south east)},y={(image.north west)}, every node/.style={font=\tiny}]
        % Coordinates must be between 0 and 1.
        \node at (0.11,0.55) {\Cref{sec:overview}};
        \node at (0.11,0.225) {\Cref{sec:overview}};
        \node at (0.375,0.78) {\Cref{sec:build-language}};
        \node at (0.375,0.335) {\Cref{sec:layout-language}};
        \node at (0.6,0.625) {\Cref{sec:compilation-construction}};
        \node at (0.6,0.315) {\Cref{sec:compilation-destruction}};
        \node at (0.835,0.475) {\Cref{sec:compilation-backend}};
    \end{scope}
\end{tikzpicture}
\caption[\sys{} overview]{The \sys{} system overview. Gray boxes are inputs, solid arrows denote lowering dependencies, and dashed arrows represent use dependencies, e.g., the layout specification is written with respect to the ADT.}
\label{figure:overview}
\end{figure}

\begin{wrapfigure}{r}{0.38\textwidth}
\vspace{1em}
\tiny
\begin{minipage}{\linewidth}
\begin{tabular}{@{}l@{}}
\hline
\\[-0.8em]
$F$ : Function ::= \kw{func} $x$\texttt{(}$p^*$\texttt{)} \texttt{->} $T$ \texttt{=} $s$ \\[0.3em]
$p$ : Parameter ::= $x$\texttt{:} \kw{mut}$^?$ $T$ $(\texttt{=}\ e)^?$ \\[0.3em]
$s$ : Stmt ::= $e$ \\
\phantom{$s$ : } | $x$ \texttt{=} $e$ \hfill store \\
\phantom{$s$ : } | $s$ \texttt{;} $s$ \hfill sequence\\
\phantom{$s$ : } | \kw{return} $e^?$ \hfill return \\
\phantom{$s$ : } | \kw{match} $e$ \texttt{\{} $A^+$ \texttt{\}} \hfill pattern match \\
\phantom{$s$ : } | \kw{let} $x$\texttt{:} $T$ \texttt{=} $e$ \hfill local binding \\
\phantom{$s$ : } | \kw{if} $e_1$ \texttt{\{} $s_1$ \texttt{\}} $(\kw{elif}\ e_2\ \texttt{\{}\ s_2\ \texttt{\}})^*$ $(\kw{else}\ \texttt{\{}\ s_3\ \texttt{\}})^?$  \hfill conditional \\
\phantom{$s$ : } | \kw{foreach} $x$ \kw{in} $e$ \texttt{\{} $s$ \texttt{\}} \hfill loop \\ [0.3em]
$e$ : Expr ::= $x$ \hfill variables \\
\phantom{$e$ : } | $n$ \hfill literals \\
\phantom{$e$ : } | $x(e^*)$ \hfill function call \\
\phantom{$e$ : } | $T(e^*)$ \hfill constructor \\
\phantom{$e$ : } | $e_1$\texttt{[}$e_2$\texttt{]} \hfill index \\
\phantom{$e$ : } | $e_1$\texttt{[}$e_2$\texttt{:}$e_3$\texttt{]} \hfill slice \\
\phantom{$e$ : } | $e$ \kw{as} $t$ \hfill type cast \\
\phantom{$e$ : } | $e$ \kw{to} $t$ \hfill bit cast \\
\phantom{$e$ : } | $e_1 + e_2 \mid e_1 - e_2 \mid \dotsc$ \hfill operators \\[0.3em]
$A$ : Arm ::= \texttt{|} $q$ \texttt{->} $s$ \\[0.3em]
$q$ : Pattern ::= $T_v$\texttt{(}$(x : T)^*$\texttt{)} \hfill variant \\
\phantom{$q$ : } | \texttt{\_} \hfill wildcard \\[0.3em]
$T$ : Type ::= $t$ \hfill primitives\\
\phantom{$T$ : } | $T$\texttt{[}$n$\texttt{]} \hfill statically-sized fixed array \\
\phantom{$T$ : } | $T$\texttt{[}$x$\texttt{]} \hfill dynamically-sized fixed array \\
\phantom{$T$ : } | \kw{set}\texttt{[}$T$\texttt{]} \hfill set \\
\phantom{$T$ : } | \kw{option}\texttt{[}$T$\texttt{]} \hfill optional \\
\phantom{$T$ : } | ($T_1$, $\dotsc$, $T_i$) \hfill tuple \\
\phantom{$T$ : } | $T \times n$ \hfill vector \\[0.3em]
\hline
\end{tabular}
\end{minipage}
\vspace{-0.5em}
\caption{Syntax of the traversal language.}
\label{figure:tree-traversal-syntax}
\vspace{-2em}
\end{wrapfigure}

\Cref{figure:overview} illustrates the \sys{} system overview. Application developers express tree traversal algorithms against algebraic data type (ADT) specifications that define the logical structure of the tree, while performance engineers separately specify how those ADTs are physically realized using the \emph{layout} language (\Cref{sec:layout-language}) and \emph{build} language (\Cref{sec:build-language}). The layout specification enables \emph{destructor\footnote{We use \emph{destructor} in the categorical sense: operations that eliminate ADT values, dual to \emph{constructors} that introduce them.} specialization}: the compiler generates code that extracts an ADT term's logical values from physical storage (\Cref{sec:compilation-destruction}). Conversely, the build specification enables \emph{constructor specialization}: the compiler generates layout-specific constructors that populate physical memory in accordance with the layout mapping (\Cref{sec:compilation-construction}). Lastly, the compiler performs domain-specific optimizations and emits backend code for CPUs or GPUs (\Cref{sec:compilation-backend}).

We note that, while automatic layout inversion is appealing, it is impractical for many important optimizations. Critical optimizations, e.g., quantization for watertight traversal~\cite{benthin2018compressed, howard2019quantized, huang2025aqb8, meister2021survey}, involve non-injective operations that preclude straightforward synthesis (evident in more complex examples provided in Appendix~\ref{app:layout-implementations}). Therefore, the build specification must be explicitly stated.

\begin{figure}[b]
\vspace{-0.75em}
\begin{lstlisting}[style=layout-style, mathescape=true]
type Ray(origin: f32x3, direction: f32x3, tmax: f32 = $\infty$);
type AABB(low: f32x3, high: f32x3); type Triangle(p0: f32x3, p1: f32x3, p2: f32x3);
type BVH(bounds : AABB) = Interior(left: BVH, right: BVH) | Leaf(nprims: u16, data: Triangle[nprims]); §\label{line:closest-hit-adt}§
func closest_hit(ray: Ray, bvh: BVH, best: mut (f32, Triangle)) = §\label{line:closest-hit-traversal-begin}§
  match bvh {
  | Interior(bounds, left, right) ->
    if intersects(ray, bounds) && (distmin(ray, bounds) < best[0]) { 
      closest_hit(ray, left,  best);  closest_hit(ray, right, best);
    }
  | Leaf(bounds, nprims, data) ->
    if intersects(ray, bounds) {
      foreach t in data { if intersects(ray, t) && distmin(ray, t) < best[0] {  best = (distmin(ray, t), t); } } 
    } 
  } §\label{line:closest-hit-traversal-end}§
\end{lstlisting}
\vspace{-0.25em}
\caption{A closest-hit ray tracing query algorithm written with respect to the logical BVH specification.}
\label{figure:closest-hit}
\vspace{-0.25em}
\end{figure}

~\Cref{figure:closest-hit} illustrates a closest-hit ray tracing query over a standard binary BVH written in \sys{}'s tree traversal language (syntax provided in ~\Cref{figure:tree-traversal-syntax}). The \code{BVH} ADT is declared with two variants: \code{Interior} nodes, which contain a bounding box and child references, and \code{Leaf} nodes, which contain a bounding box and triangle primitives. For brevity, the syntax on Line~\ref{line:closest-hit-adt} states that \emph{each} variant stores a bounding box (\code{bounds}). The traversal algorithm on Lines~\ref{line:closest-hit-traversal-begin}--\ref{line:closest-hit-traversal-end} matches on these variants, and only recurses on \code{Interior} nodes or checks for ray-triangle intersection for \code{Leaf} nodes after verifying the ray intersects the bounding box. We acknowledge that writing efficient tree traversal algorithms is equally critical to overall performance. We leverage \bonsai{}~\cite{root2026bonsai} to automatically generate all traversal algorithms presented in this paper.

Crucially, this algorithm is written with respect to the logical field names, e.g., \code{left}, and makes no assumptions about how these fields are represented in memory. This separation realizes a form of \textit{data independence}~\cite{codd1970relational} for performance-critical data structures. Logical definitions state \textit{what} the data structure contains and the subsequent operations performed, while separate physical specifications state \textit{how} that data is represented, stored, and accessed.

\section{Physical Layout Language}
\label{sec:layout-language}

Separating layout from algorithm requires addressing how representation choices affect both storage and traversal. An ADT term might be referenced via a stored 32-bit array index or by implicitly computing its location from the parent's position. Each choice changes not only the physical footprint but also the operations needed to traverse the structure, and thus requires specifying how to locate ADT terms in memory and how to interpret their physical representation once located. To address this, the layout language expresses two complementary pieces of information: (1) the concrete type that encodes an ADT \emph{reference}---the representation used to uniquely identify a term in memory---and (2) the interpretation of that representation, i.e., how to determine which variant it denotes and how to derive its fields. We begin by examining a layout from a standard rendering textbook~\cite{pharr2023pbrt}, showing how typical optimizations alter both representation and interpretation in intertwined ways, and then define layout primitives that generalize these transformations.

\subsection{PBRTv4 Layout Example}

\begin{figure}[t]
% \vspace{-1.1em}
\centering
\begin{subfigure}[t]{0.48\textwidth}
\begin{lstlisting}[style=cpp-style]
struct LinearBVH {
  uint32_t P;
  uint32_t N;
  Triangle* primitives; 
  LinearBVHNode* nodes;
};

struct alignas(32) LinearBVHNode {
  AABB bounds;
  union {
    uint32_t p_o;   // Leaf     (primitive offset)
    uint32_t c_o;   // Interior (2nd child offset)
  };
  uint16_t nprims;  // 0 -> Interior, > 0 -> Leaf
}; 
\end{lstlisting}
\caption{Original PBRTv4 layout in C++.}
\label{figure:pbrt-cpp}
\end{subfigure}
\hfill
\begin{subfigure}[t]{0.45\textwidth}
\begin{lstlisting}[style=layout-style]
layout LinearBVH(I: u32) {  §\label{line:pbrt-signature}§
  P: u32; N: u32; primitives: Triangle[P]; §\label{line:pbrt-globals}§
  group nodes[size=N, align=32] by I { §\label{line:pbrt-node-begin}§
    bounds: AABB;  §\label{line:pbrt-field1}§
    split nprims {  §\label{line:pbrt-split-leaf-begin}§ §\label{line:pbrt-split-begin}§
      > 0 -> Leaf {
        p_o : u32; data = primitives[p_o : p_o + nprims];
      }; §\label{line:pbrt-split-leaf-end}§
      0 -> Interior {
        c_o : u32; left = I + 1; right = I + c_o; §\label{line:pbrt-left-right}§
      }; 
    }; §\label{line:pbrt-split-end}§
    nprims: u16; §\label{line:pbrt-field2}§ 
  }; §\label{line:pbrt-node-end}§
}; 
\end{lstlisting}
\caption{PBRTv4 layout in \sys{}.}
\label{figure:pbrt-layout}
\end{subfigure}
\vspace{-0.25em}
\caption{Comparison of PBRTv4 BVH layouts for the logical \code{BVH} on Line \ref{line:closest-hit-adt} of \Cref{figure:closest-hit}.}
\label{figure:pbrt-comparison}
\vspace{-1.5em}
\end{figure}

The PBRTv4 rendering system~\cite{pharr2023pbrt} performs ray tracing as illustrated in \Cref{figure:closest-hit}, but with the C++ data structures shown in ~\Cref{figure:pbrt-cpp} for its BVH data structure. Its design incorporates several optimizations that deviate from a standard pointer-based encoding of the BVH ADT:
\begin{enumerate}[leftmargin=2.5em]
    \item Nodes reside in a contiguous array; children (\code{left}, \code{right}) are 32-bit unsigned indices into it.
    \item The \code{left} child is stored directly after the parent, eliminating an explicit reference.
    \item \code{Interior} data and \code{Leaf} data are unified, discriminated by the \code{nprims} field.
    \item Triangles reside in a separate array; \code{Leaf} nodes store an offset into it and a primitive count.

\end{enumerate}

To achieve the goal of combining clarity of ADTs with specialized layout performance, \sys{} provides a separate declarative layout language. \Cref{figure:pbrt-layout} presents the PBRTv4 layout written in \sys{}. Line~\ref{line:pbrt-signature} specifies that an ADT term can be referenced by \code{I} of type \code{u32}. Because multiple arrays may be indexed by the same reference, e.g., separate arrays for different variants that share a common index space, the reference type is declared in the layout signature rather than within individual \code{group} declarations. This enables \code{I} to serve as a reference across one or more groups.
Line~\ref{line:pbrt-globals} declares global fields: the number of primitives, the number of nodes, and an array of triangles. Lines~\ref{line:pbrt-node-begin}--\ref{line:pbrt-node-end} declare an array of nodes. The \code{group} declaration establishes the size of the array named \code{nodes} and the alignment of each element, as well as how the array is indexed (via \code{I}). Lines~\ref{line:pbrt-field1} and \ref{line:pbrt-field2} specify fields that each node stores: the bounding box and number of primitives respectively. Lines~\ref{line:pbrt-split-begin}--\ref{line:pbrt-split-end} define a safe tagged union discriminated by the number of primitives. This is justified by the BVH invariant that only leaves contain primitives.

\subsection{Layout Primitives}

\sys{}'s layout language provides a small set of composable primitives for describing the exact layout of a tree in memory (abstract syntax provided in ~\Cref{figure:layout-syntax}). A layout specification defines:
\begin{enumerate}[leftmargin=1.5em]
    \item A \textit{reference} type: the concrete type that represents a logical reference to an ADT term. This enables coarse-grain optimization, such as storing ADT terms in struct-of-arrays format. 
    \item How that reference type should be interpreted: which ADT variant it represents and how the fields of that variant should be loaded. This enables fine-grain optimization, such as bit-stealing (using bits that are otherwise reserved for padding or pointer alignment).
\end{enumerate}

\begin{wrapfigure}{r}{0.35\textwidth}
\vspace{-1.75em}
\tiny
\begin{minipage}{\linewidth}
\begin{tabular}{@{}l@{}}
\\ \hline
\\[-0.8em]
$\ell$ : Layout ::= \kw{layout} $x(p^*)$ \texttt{\{} $M$ \texttt{\}} \\[0.3em]
$p$ : Parameter ::= $x$\texttt{:} $T$ $(\texttt{=}\ e)^?$ \\[0.3em]
$M$ : Member ::= \\
\phantom{$M$ : } | $x\texttt{:}\ T$ \hfill field \\
\phantom{$M$ : } | $n$ \hfill padding \\
\phantom{$M$ : } | $x\ \texttt{=}\ e$ \hfill derive\\
\phantom{$M$ : } | \kw{group} $id$ [$K^*$] (\kw{by} $x$)$^?$ \texttt{\{} $M$ \texttt{\}} \hfill group \\
\phantom{$M$ : } | $\kw{indirect}$ \kw{group} $id$ \texttt{\{} $M$ \texttt{\}} \hfill \ \ \ \ \ indirect group \\
\phantom{$M$ : } | \kw{split} $e$ \texttt{\{} $A^+$ \texttt{\}} \hfill split \\
\phantom{$M$ : } | \kw{let} $x$\texttt{:} $T$ \texttt{=} $e$ \hfill local binding \\
\phantom{$M$ : } | $M;M$ \hfill sequence \\[0.3em]
$A$ : Arm ::= $p \to M$ \\
\phantom{$A$ : } | $p \to T$ \kw{from} $x$\texttt{[}$e$\texttt{]} \hfill foreign key \\[0.3em]
$K$ : Attribute ::= $n$ \hfill literals \\
\phantom{$K$ : } | $size=n$  \hfill cardinality\\
\phantom{$K$ : } | $align=n$ \hfill alignment\\[0.3em]
$p$ : Pattern ::= $n$ \hfill literals \\
\phantom{$p$ : } | $(> | < | \geq | \leq | \sim)$ $n$ \hfill compare \\[0.3em]
\phantom{$p$ : } | \texttt{\_} \hfill wildcard \\[0.3em]
$e$ : Expr ::= \kw{parent}.$x$ \hfill tree dependency \\
\phantom{$e$ : } $\cup \big\{ e \in \texttt{Tree Traversal Language} \big\}$ \hfill
\\[0.3em] \hline
\end{tabular}
\end{minipage}
\caption{Syntax of the layout language.}
\label{figure:layout-syntax}
\vspace{-1.25em}
\end{wrapfigure}

A layout in \sys{} begins by declaring the reference type: Line~\ref{line:pbrt-signature} of \Cref{figure:pbrt-layout} does this via \code{layout LinearBVH(I: u32)}, which specifies that each term of the ADT is uniquely identified by a value of type \code{u32}. The parameter \code{I} serves as a bound identifier for the reference in the rest of the layout specification. Reference types can be raw pointers, indices into arrays, composite structures with multiple indices, or richer encodings described in~\Cref{sec:relational-constructs}. The primitives specifying the interpretation of a reference type operate at three conceptual levels: 
\begin{enumerate}[leftmargin=2.5em]
    \item \textit{Local}: how bits correspond to fields, 
    \item \textit{Structural}: how values \emph{compose} into collections, and
    \item \textit{Relational}: how dependencies \emph{span} collections.
\end{enumerate}

\subsubsection{Local}
\label{sec:local-constructs}
\sys{} supports two notions of \textit{fields}, a \textit{stored} field and a \textit{derived} field. \textit{Stored fields} assign an identifier to a region of bits in memory. For example, \code{x:f16} defines a field named \code{x} that occupies 16 bits and has type \code{f16}. Conceptually, a field behaves like an l-value ~\cite{strachey2000fundamental}: it refers to a location in memory that can be read or written. Accessing a field performs a load from memory. In PBRT's BVH in \Cref{figure:pbrt-layout}, \code{bounds} on Line~\ref{line:pbrt-field1} is an example of a stored field. The offsets \code{p_o} and \code{c_o} are also stored fields, but they do not directly correspond to \textit{logical} fields of the ADT. They are instead used when deriving other logical fields of the ADT.

\textit{Derived fields} bind a name to an expression, i.e. an r-value\footnote{Also referred to as a \emph{method call} on a class in object-oriented programming or a \emph{computed column} in databases.}, as in \code{x = 42}. Some fields might be pure functions of other fields, i.e., chosen not to be physically stored in the layout. Derived fields are evaluated lazily, and at most once per destruction of the associated field; this may be elided by compiler optimizations. \code{left} and \code{right} on Line~\ref{line:pbrt-left-right} of \Cref{figure:pbrt-layout} are examples of derived fields: they are computed from the reference and a stored field, \code{c_o}. Together, stored and derived fields span the entire spectrum of compute–memory tradeoffs, from purely stored to purely derived representations. The remaining primitives serve only to provide useful abstractions that express the layout of \textit{collections} of terms.

\subsubsection{Structural}
\label{sec:structural-constructs}
Structural primitives specify how stored and derived fields are organized into aggregates and variants. They provide the operators that enable the composition of fields into variants (via \code{split}) and control the layout of \emph{collections} of fields or variants (via \code{group}).

\begin{figure*}[t]
% \vspace{-0.25em}
\centering
\begin{subfigure}[t]{0.64\textwidth}
    \includegraphics[width=\linewidth]{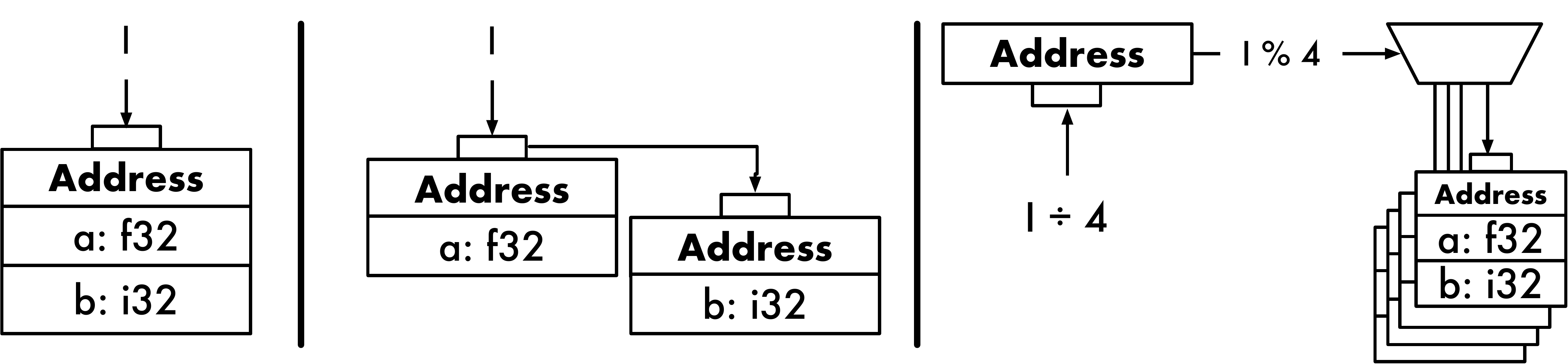}
    \caption[Data flow in groups]{Global transformations of the same fields (left to right):
      \\ \textbf{\small{AoS}}: \code{group G by I \{a: f32; b: i32;\};}
      \\ \textbf{\small{SoA}}: \code{group G1 by I \{a: f32;\}; group G2 by I \{b: i32;\};}
      \\ \textbf{\small{AoSoA}}: \code{group Go by I \{ group Gi[4] \{a: f32; b: i32;\}; \};}
    }
    \label{figure:data-flow-group}
\end{subfigure}
\hfill
\begin{subfigure}[t]{0.34\textwidth}
    \includegraphics[width=\linewidth]{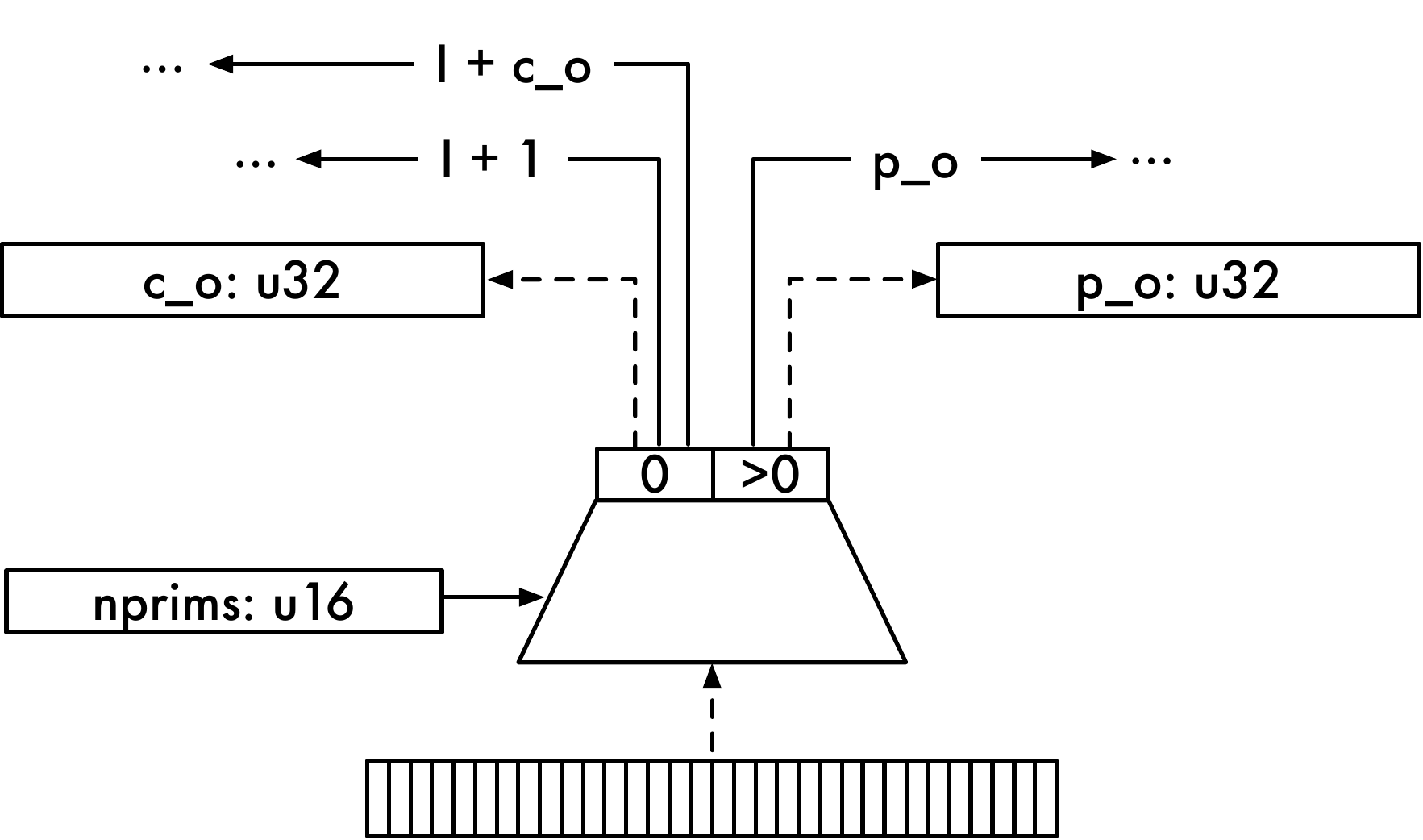}
    \caption[Data flow in splits]{The \code{split} construct in \Cref{figure:pbrt-layout} branches on field \code{nprims}: the left path (\code{= 0}) interprets the bit field as \code{c_o}, and the right path (\code{> 0}) as \code{p_o}.}
    \label{figure:data-flow-split}
\end{subfigure}
\label{figure:data-flow-example}
\caption{Example sub-layout diagrams. Solid arrows indicate the flow of data (either a derivation or loaded field), and dashed arrows represent interpretation of previously unnamed bits by the \code{split} construct.}
\vspace{-1.25em}
\end{figure*}

\textit{Groups} define an ordered collection of sub-layouts and provide the mechanism for expressing global layout transformations such as tiling data along the array-of-structs (AoS) to struct-of-arrays (SoA) spectrum, including hybrid variants used in vectorized code~\cite{woop2014hair, benthin2018compressed}. Every group is indexed by a value that determines how its elements are addressed in memory. Typically, the value is an integral-type index, indicating a contiguous array, but could alternatively be a pointer type, stipulating a logical grouping of structures that are stored arbitrarily in memory.

Groups come in two forms: direct and indirect. Direct groups are indexed by a component of the reference type and correspond to primary storage. In the PBRTv4 layout, the group starting on Line~\ref{line:pbrt-node-begin} of \Cref{figure:pbrt-layout} is a direct group: it is indexed by the reference value \code{I}. Indirect groups, labeled via \code{indirect}, represent auxiliary storage (akin to \emph{foreign tables} in relational databases) and are accessed from direct groups using \code{from} (a relational primitive, discussed in ~\Cref{sec:relational-constructs}). As demonstrated in ~\Cref{sec:case-study-hair}, indirect groups are particularly effective for heterogeneous data, where different ADT variants have disparate storage requirements.

Composing groups facilitates specifying complex data layouts. An AoS layout uses a single group with multiple fields, while a SoA layout uses adjacent single-field groups that share a common index; these are illustrated in ~\Cref{figure:data-flow-group}. Hybrid layouts, e.g., AoSoA, are specified by nesting groups, where the inner group defines a tile size. We also provide a primitive for declaring SoA layouts within a single group when each group shares the same index. The separator \code{---} splits a group into multiple adjoining groups while preserving lexical scope so derivations can refer to one another.

\textit{Splits} provide a mechanism for safe tagged unions, as in the \textsc{Ribbit} compiler~\cite{baudon2023bitstealing}. The \code{split e} construct expresses conditional branching and bit interpretation based on the value of the expression \code{e}. We implement a subset of the split construct provided in \textsc{Ribbit}, tailored for the tagged unions found in BVH layout optimization. Each arm of the split defines a unary condition, where simple constants like \code{0} are implicitly treated as equality tests, and other comparisons must be stated explicitly, e.g., \code{< 42} or $\geq$ \code{0}. The wildcard symbol \code{\_} serves as a catch-all for any unmatched cases. The right-hand side of a split arm specifies the corresponding sub-layout and its variant type. For example, Lines~\ref{line:pbrt-split-leaf-begin}--\ref{line:pbrt-split-leaf-end} of \Cref{figure:pbrt-layout} specify that when \code{nprims} is greater than \code{0}, the sub-layout storing the \code{p\_o} field and deriving the \code{data} array should be interpreted as a \code{Leaf} variant.

The primitives so far describe how individual ADT terms and the entire collection are organized in memory. We further need to describe how groups are connected, i.e., how local primitives of an indirect group are accessed, and how recursive datatypes are instantiated.

\subsubsection{Relational}
\label{sec:relational-constructs}

\sys{}’s relational primitives define how data in one part of the layout refers to or depends on data in another. These operators generalize pointer dereferences and \emph{foreign-key} lookups, enabling layouts to specify both explicit references between disjoint groups and implicit dependencies along a tree’s hierarchy. \sys{} supports two core relational primitives: a \code{from} operator that acts as a load from an \code{indirect} group via a foreign key and the construction of complex primary keys, i.e., the reference type, with \emph{tree-carried dependencies}.

An \code{indirect} group is not directly indexed by a component of the reference 
type. Consequently, fields in indirect groups have different scoping rules than 
those in direct groups, as we discuss in~\Cref{sec:well-formedness}. To access the fields represented in an \code{indirect} group, the \code{from} operator is used in conjunction with a key or range of keys. A concrete example of this is provided in ~\Cref{sec:case-study-hair}.

While \code{from} handles relationships between disjoint groups, recursive references capture relationships \emph{within} the hierarchy. These references must respect the layout’s declared reference type, which determines how a node identifies and communicates with its children. Similarly, any field that is originally a recursive datatype in the ADT language must be loaded or derived with a type that matches the specified reference type of the tree. While the reference type is typically an index, a set of indices, e.g., nested direct groups, or a pointer type, \sys{} additionally supports tree-carried dependencies, also referred to as hierarchical encoding~\cite{mahovsky2006memory, segovia2010memory}.

Tree-carried dependencies enable terms to share information with recursive ADT fields, reducing per-term storage by passing shared data into the reference when destructing. Since a reference becomes incomplete without parent context, the parent information is incorporated into the reference itself and thus the \code{layout} signature. The \code{parent} prefix is used to specify these dependencies. For example, \code{parent.x} is the parent's value of \code{x}. As demonstrated in ~\Cref{sec:case-study-tcd}, the layout signature must indicate that such dependencies are components of the reference type.

\subsection{Layout Diagrams}
\label{sec:data-flow-diagram}

\begin{wrapfigure}{r}{0.40\textwidth}
\vspace{-1em}
\centering
\includegraphics[width=0.40\textwidth]{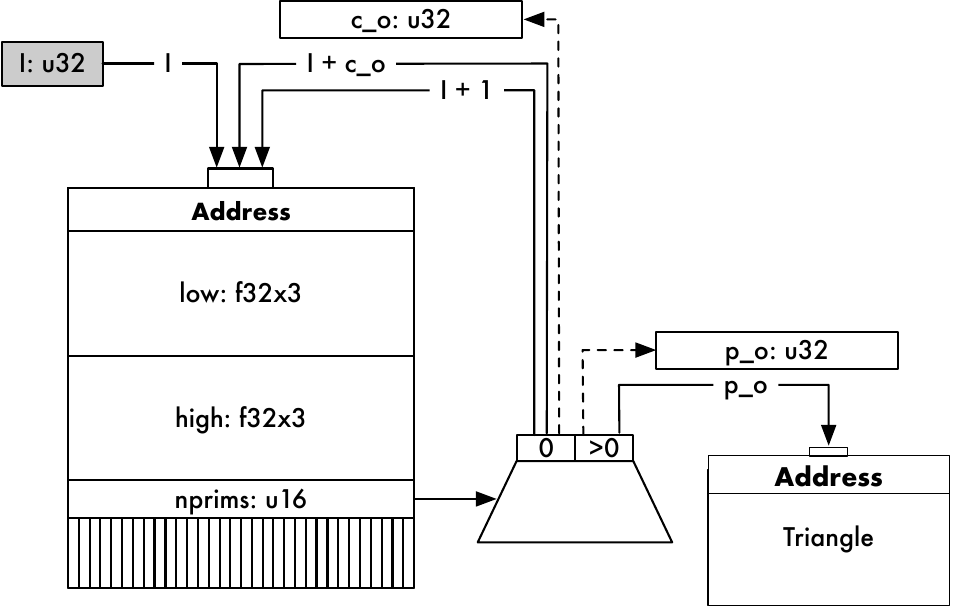}
\caption{Layout diagram for PBRTv4.}
\label{figure:pbrt-diagram}
\vspace{-0.55em}
\end{wrapfigure}

To aid understanding of layout specifications, we introduce \emph{layout diagrams} that visualize how field accesses propagate through layout language constructs. These diagrams trace the path from the initial reference (illustrated as a shaded box) through \code{group} constructs, \code{split} primitives, and \code{indirect} lookups, terminating at either stored fields (loads from memory) or derived fields (computed values). Solid arrows represent data flow with associated semantics: an unlabeled arrow with a field origin denotes a direct memory load, while a labeled arrow with an expression, e.g., \code{I + c_o} in~\Cref{figure:pbrt-diagram}, denotes a derived field computed from previously accessed values. Dashed arrows represent type reinterpretation of unnamed bit fields (visualized as individual bits) performed by the \code{split} construct. At any point in the diagram, all previously encountered named fields are accessible and bound to their corresponding memory addresses or computed values. The diagram thus captures both the data flow through structural and relational primitives, and scoping of local primitives during traversal, further explained in ~\Cref{sec:well-formedness}. For brevity, we elide derivation expressions for fields whose values are directly copied from memory.

Importantly, layout transformations correspond directly to structural manipulations of these diagrams: global layout transformations like AoS-to-SoA duplicate group structure (\Cref{figure:data-flow-group}), \code{split} constructs introduce conditional branching or reinterpretation of bits (\Cref{figure:data-flow-split}), and relational lookups, i.e., \code{from}, create inter-group edges (\Cref{sec:case-study-hair}). We now describe how these primitives compose \textit{correctly}. We next discuss how the build language specifies the \emph{constructor} (logical-to-physical mapping).

\section{Build Language} \label{sec:build-language} A goal of \sys{} is to provide a declarative specification for transforming logical trees into their physical representations while maintaining clarity of the algebraic structure. Recall that the \emph{logical} tree is a straightforward realization of the algebraic data type, where each constructor maps to a variant and fields are stored with their declared types. The \emph{physical} tree is the optimized representation obtained by transforming the logical tree according to the build\footnote{In this work, \emph{build} and \emph{construction} have distinct definitions. Build refers to the transformation from \emph{logical} tree to \emph{physical} tree (in scope), and construction refers to the assembly of the logical tree (out of scope).} specification. The layout language specifies the memory layout, while the build language specifies the transformation from logical tree to physical tree.

Together, layout and build languages express a bidirectional mapping between logical and physical representations. The \emph{layout} defines the destructor: how to extract logical fields from a physical tree (physical $\rightarrow$ logical). The \emph{build} defines the constructor: how to populate physical storage from a logical term (logical $\rightarrow$ physical). 

These specifications must be consistent with each other. A field that appears in a layout's \code{split} branch of \code{group} must be populated by a corresponding \code{build} statement. Conversely, every \code{build} statement must target a field that exists in the layout. However, the two specifications serve complementary roles and need not be symmetric, e.g., the layout may define additional derivations, while the build only populates stored fields.

\subsection{Two-Phase Construction}
Prior work has used two strategies for construction of acceleration structures. \emph{Single-phase} construction builds the specialized physical layout directly during tree construction, avoiding intermediate representations. This was the approach taken by~\citet{burtscher2011efficient} for a class of space-partitioning trees, and is attractive when construction latency is critical, e.g., dynamic scenes. \emph{Two-phase} construction first builds a canonical tree (with an unoptimized layout) and then packs it into the target physical layout~\cite{pharr2023pbrt, barbier2025fused}. This approach incurs additional latency and memory overhead from intermediate na\"ive layout construction, but cleanly separates tree topology optimization from physical layout optimization, thus simplifying end-to-end construction. 

We adopt the two-phase approach: \sys{} assumes the tree topology is fixed before layout specialization begins. This assumption is natural for static scenes, where construction costs are amortized over many traversal queries, and it additionally enables the build language to treat the logical tree as immutable input.

\subsection{PBRTv4 Build Example}

\begin{wrapfigure}{r}{0.51\textwidth}
\vspace{-4.5em}
\begin{lstlisting}[style=layout-style]
build LinearBVH[order=pre] { §\label{line:pbrt-build-attr}§
 build Interior(bounds : AABB, left: BVH, right: BVH) { §\label{line:pbrt-build-interior}§
  build bounds; build nprims = 0; §\label{line:pbrt-build-interior-bounds}§
  build left; let R: u32 = build right; §\label{line:pbrt-build-interior-traversal}§
  build c_o = R - this; return this; §\label{line:pbrt-build-interior-return}§
 }; 
 build Leaf(bounds : AABB, nprims: u16, data: Triangle[nprims]) {
  build bounds; build nprims; §\label{line:pbrt-build-leaf-bounds}§
  build p_o = append(data, nprims); §\label{line:pbrt-build-leaf-append}§
  return this; §\label{line:pbrt-build-leaf-return}§
 }; 
};
\end{lstlisting}
\vspace{-0.5em}
\caption{Build specification for the PBRTv4 layout.}
\label{figure:pbrt-build-example}
\vspace{-0.75em}
\end{wrapfigure}

We illustrate the build language via the PBRTv4 example before defining its primitives. \Cref{figure:pbrt-build-example} presents the build specifications for the LinearBVH layout. Line~\ref{line:pbrt-build-attr} declares that the tree is \emph{laid out} in preorder traversal, i.e., the parent node is built before its children are recursively visited. Line~\ref{line:pbrt-build-interior} declares the \code{Interior} variant signature, containing the original fields from the logical ADT. Line~\ref{line:pbrt-build-interior-bounds} copies a field from the logical tree, i.e., \code{build bounds}, and stores the discriminant value (\code{build nprims = 0}). Line~\ref{line:pbrt-build-interior-traversal} builds the left and right subtrees, saving the reference of the right child. Line~\ref{line:pbrt-build-interior-return} computes the relative offset to the right child (\code{build c_o = R - this}), leveraging the preorder layout to elide storing the left child reference. Lastly, it returns the current node's reference. The \code{Leaf} variant follows a similar pattern: copying stored fields (Line~\ref{line:pbrt-build-leaf-bounds}), appending primitive data to the indirect array (Line~\ref{line:pbrt-build-leaf-append}), and returning the node reference (Line~\ref{line:pbrt-build-leaf-return}). Together, these specifications completely determine the logical-to-physical tree constructor, using the layout mapping to determine \emph{where} these values live in memory. Next, we define the semantics of primitives in the build language.

\subsection{Build Primitives}

The build language provides a small set of primitives augmented by the tree traversal language for describing how to construct physical trees from logical trees. Each build specification defines, for each variant of the algebraic data type, the sequence of operations required to derive that variant in the physical layout. The constructs operate at two conceptual levels: \emph{local} primitives and \emph{structural} primitives. The abstract syntax is provided in ~\Cref{figure:build-syntax}.

\subsubsection{Local}
\label{sec:build-local}

\begin{wrapfigure}{r}{0.3\textwidth}
\vspace{-1.5em}
\tiny
\begin{minipage}{\linewidth}
\begin{tabular}{@{}l@{}}
\\ \hline
\\[-0.8em]
$\beta$ : Build ::= \kw{build} $x[K]$ \texttt{\{} $V^+$ \texttt{\}} \\[0.3em]
$K$ : Attribute ::= \\
\phantom{$K$ : } $order=\{$pre, post$\}$ \hfill ordering \\[0.3em]
$V$ : Variant ::= \kw{build} $T_v$\texttt{(}$(x\texttt{:}\ T)^*$\texttt{)} \texttt{\{} $R^?$ \texttt{;} $s$ \texttt{\}} \\[0.3em]
$R$ : Root ::= \kw{build} \kw{root} \texttt{\{} $s$ \texttt{\}} \\[0.3em]
$s$ : Stmt ::= \kw{build} $x$ $(\texttt{=}\ e)^?$ \hfill build field \\
\phantom{$s$ : } | \kw{return} $e$ \hfill return reference \\
\phantom{$s$ : } | \kw{let} $x$\texttt{:} $T$ \texttt{=} $e$ \hfill local binding \\
\phantom{$s$ : } | $s$ \texttt{;} $s$ \hfill sequence \\[0.3em]
$e$ : Expr ::= \kw{this} \hfill current reference \\
\phantom{$e$ : } | \kw{append}$(x, e)$ \hfill append to array \\
\phantom{$e$ : } $\cup \big\{ e \in \texttt{Tree Traversal Language} \big\}$
\\[0.2em] \hline
\end{tabular}
\end{minipage}
\vspace{-0.5em}
\caption{Build language syntax.}
\label{figure:build-syntax}
\vspace{-1em}
\end{wrapfigure}

Local primitives describe how stored (physical) fields are populated for each variant type. \textit{Build} statements populate physical fields from logical values. The statement \code{build x} copies the field \code{x} directly from the logical representation to the corresponding physical location in the layout specification. For example, in \Cref{figure:pbrt-build-example}, \code{build low} copies the \code{low} field from the logical tree to its layout-specified memory location (handled automatically by the compiler). Build statements can also compute values: in \code{build x = e}, \code{e} is evaluated in the context of the current logical node, and the result is stored in \code{x}'s physical location. For example, the \code{Interior} node declares \code{build nprims = 0} to store 0 in the layout's \code{nprims} field. This primitive enables layout-specific encodings absent from the logical representation, such as discriminant tags or auxiliary metadata. We also provide syntactic sugar for appending to an \code{indirect} group: \code{append(x, n)} appends \code{n} elements of \code{x} and returns the starting position, with the location automatically inferred by the layout mapping.

\subsubsection{Structural}
\label{sec:build-structural}

Structural primitives control the global organization of terms in memory and the construction of references between terms. For \code{build} statements of fields with recursive algebraic data types, the fields are constructed in a recursive, depth-first manner. For example, \code{build left} fully materializes the left child and all its descendants before any subsequent statements execute. The \code{order} attribute controls when the current term is materialized in memory relative to its children: preorder (\code{order=pre}) emits the parent node before recursively visiting children, while postorder (\code{order=post}) emits the parent node after all descendants have been visited. Critically, \code{order} affects only the memory layout. It does not constrain dependency resolution, e.g., the parent node may compute fields that depend on values from its children in preorder layouts, provided those values have already been constructed. This is demonstrated concretely on Line~\ref{line:pbrt-build-interior-return} in ~\Cref{figure:pbrt-build-example}, where the current node's field \code{c_o} depends on the reference to the right child \code{R}.

\textit{This} is a special identifier that refers to the unique reference of the current term, e.g., an index into a contiguous array. Uses of \code{this} enable computing relative offsets, e.g., in \Cref{figure:pbrt-build-example}, we compute the right child's offset by subtracting its index from the currently-visited node's index. We can also modify the reference type to hold additional information. For example, a pointer type may only use a 48 bits within a 64-bit field, leaving 12 bits that can be used to store auxiliary information, e.g., \code{this[48:63] = metadata}.

\textit{Return} statements produce the reference to the constructed term. The expression \code{e} in \code{return e} must evaluate to a value with the reference type declared in the layout specification. When building a field with a recursive ADT, this value may be used by the parent to realize inter-term dependencies.

\textit{Root} specifications handle special initializations required at the root of the tree. Some layouts require computing base-case values for inductively defined tree-carried dependencies or initializing global parameters. The \code{root} block executes before any variant-specific build procedures and can populate these global values. For instance, a quantization scheme might compute the world bounding box and store it as a global parameter that all nodes use during quantization. Such an example is provided in Appendix~\ref{app:pbrt-q16-implementation}.

\section{Well-Formedness}
\label{sec:well-formedness}

A key concern for BVH developers adopting \sys{} is what correctness guarantees the compiler provides. Specifically, can every logical field be unambiguously accessed in the physical layout, and does encoding a tree and decoding it back preserve semantics? We address this through a combination of static analyses that cover the full language, and proof sketches for the correctness of \emph{derivation-free} specifications. This subset includes non-derived \code{field} and structural primitives (\code{group}, \code{split}), but excludes other constructs (derived \code{field} and relational primitives) as these may introduce arbitrary computation. Verifying round-trip soundness of derived specifications would require proving functional equivalence of two programs, which is (in general) undecidable or at least requires separate proof obligations~\cite{rice1953classes}. We thus trust that programmers of derived specifications implement correct code.

\subsection{Static Analysis}
\label{sec:static-analysis}

The tree traversal language uses Hindley-Milner typing~\cite{hindley1969principal}. For destructor specialization (Section~\ref{sec:compilation-destruction}) to be well-defined, field resolution must produce an unambiguous memory access or derivation for each logical field reference in the traversal code. The compiler statically verifies this by checking that every layout admits a unique physical path to every field of every variant; the algorithm is provided in Appendix~\ref{app:path-uniqueness}. We also provide additional static checks to eliminate certain classes of bugs. For example, the \code{split} construct must cover the full domain of its discriminant, the dependency graph built by derived fields must be acyclic, and fields in the physical layout must preserve the typing of the logical specification.

\subsection{Formal Guarantees}
\label{sec:formal-guarantees}
In Appendix~\ref{app:wf}, we state the well-formedness conditions for the derivation-free subset of \sys{}. This necessarily deviates from any guarantees about layouts with arbitrary compute possible in the complete language. Informally, we show that if the layout and build are well-formed, then for every variant $V$ and every logical field $(f : T) \in V$, there exists exactly one physical path to it. Consequently, in the derivation-free subset, encoding a tree (via the \emph{build} language) and decoding the tree (via the \emph{layout} language) will always result in the same tree.

The derivation-free subset captures the structural core of \sys{} but excludes features essential for state-of-the-art BVH optimizations. Derived fields enable bounding-volume quantization and hierarchical compression; tree-carried dependencies permit compression of information shared inductively; and indirect groups support multi-buffer layouts. Extending the round-trip proof sketch to cover them requires substantial proof obligations on semantic correctness, e.g., that quantization followed by dequantization conservatively encloses the original bounding volume. Formalizing these obligations, or intelligently restricting the expressiveness of \sys{} to guarantee semantic preservation by construction, is important future work.

\section{Compilation}
\label{sec:compilation}

The \sys{} compiler transforms programs into machine code through a set of stages. Layout and build validation ensure well-formedness: unambiguous field paths, complete coverage, and type consistency. These constraints are not merely safety checks. They directly simplify the lowering process by guaranteeing that every field access has a unique resolution path and that every ADT term has a corresponding physical representation. Layout specialization translates declarative specifications into explicit memory address computations, relying on unambiguous field paths to perform straightforward syntax-directed expansion. Build lowering emits constructor code by matching layout fields against algebraic data type fields, automatically generating the logical-to-physical tree conversion. We then run compiler optimization passes, and finally, the backend emits either CUDA or C++ for existing compiler toolchains.

\subsection{Destructor Specialization}
\label{sec:compilation-destruction}

\begin{wrapfigure}{r}{0.41\textwidth}
\vspace{-7em}
\centering
\begin{minipage}{0.41\textwidth}
\begin{algorithm}[H]
\tiny
\caption{Field Resolution}
\label{alg:concretization}
\begin{algorithmic}[1]
\State $P$, current memory access path
\State $M$, layout member to traverse  
\State $F$, target field to resolve  
\State $T_{v}$, current variant type for \textsc{Split} disambiguation
\State $I$, map of indirect group identifier to (layout, path) pairs
\State $S$, map for caching derivation paths
\Function{Concretize}{$P, M, F, T_{v}, I, S$}
    \State \textbf{match} $M$ \textbf{with}
    \State \hspace{0.5em} $|$ \textsc{Field}$(id)$ $\Rightarrow$ 
    \State \hspace{1em} \textbf{if} $id = F$ \textbf{then return} $(P, id)$
    \State \hspace{1em} \textbf{else} cache $(id, (P, id))$ in $S$ for \textsc{Derive}
    \State \hspace{0.5em} $|$ \textsc{Group}$(T_{G}, body, index)$ $\Rightarrow$ 
    \State \hspace{1em} \textbf{match} $T_{G}$ \textbf{with}
    \State \hspace{1.5em} $|$ \textsc{Direct} $\Rightarrow$
    \State \hspace{2em} \textbf{if} $index$ \textbf{is ptr then} 
    \State \hspace{2.5em} \textbf{return} \Call{Concretize}{$*P$, $body$, $F$, $T_{v}$, $I$, $S$}
    \State \hspace{2em} \textbf{else} 
    \State \hspace{2.5em} \textbf{return} \Call{Concretize}{$P[index]$, $body$, $F$, $T_{v}$, $I$, $S$}
    \State \hspace{1.5em} $|$ \textsc{Indirect}($name$) $\Rightarrow$
    \State \hspace{2em}  \textbf{insert} ($name$, ($body$, $P$)) into $I$; \textbf{return} $\bot$
    \State \hspace{0.5em} $|$ \textsc{From}$(name, key)$ $\Rightarrow$
    \State \hspace{1em} \textbf{let} $(body, P_{i}) := I[name]$ \textbf{in} \Comment{Fail if $name \notin I$}
    \State \hspace{1em} \textbf{return} \Call{Concretize}{$P_{i}[key]$, $body$, $F$, $T_{v}$, $I$, $S$}
    \State \hspace{0.6em} $|$ \textsc{Split}$(discriminant, arms)$ $\Rightarrow$
    \State \hspace{1em} \textbf{for each} $Arm(C, T_{a}, body) \in arms$ \textbf{do}
    \State \hspace{1.5em} \textbf{if} $T_{a} = T_{v}$ \textbf{then}
    \State \hspace{2em} \textbf{return} \Call{Concretize}{$P$, $body$, $F$, $T_{v}$, $I$, $S$}
    \State \hspace{0.5em} $|$ \textsc{Derive}$(id, E)$ $\Rightarrow$
    \State \hspace{1em} $V \gets$ \Call{Evaluate}{$E$, $S$}
    \State \hspace{1em} cache $(id, V)$ in $S$
    \State \hspace{1em} \textbf{if} $id = F$ \textbf{then return} $V$
    \State \hspace{0.5em} $|$ \textsc{Sequence}$(layouts)$ $\Rightarrow$
    \State \hspace{1em} \textbf{for each} $L \in layouts$ \textbf{do}
    \State \hspace{1.5em} \textbf{let} $r := $ \Call{Concretize}{$P$, $L$, $F$, $T_{v}$, $I$, $S$} \textbf{in}
    \State \hspace{1.5em} \textbf{if} $r \neq \bot$ \textbf{then return} $r$
    \State \textbf{return} $\bot$ \Comment{field not found}
\EndFunction
\end{algorithmic}
\end{algorithm}
\end{minipage}
\vspace{-1em}
\end{wrapfigure}

Destructor specialization is the process of mapping logical field access in the ADT (destruction) to physical memory, as specified by the layout (specialization). This is conducted through a field resolution process: given a destructed field from the ADT, the compiler determines the corresponding memory access or derivation in the layout specification. \Cref{alg:concretization} presents a simplified version of this process. The algorithm operates as a syntax-directed macro expansion: it traverses the layout specification to find the declaration of the requested field, then recursively expands that declaration into concrete memory operations. This approach exploits the compositionality of layout transformations---a field access through multiple primitives expands by recursively applying each transformation's interpretation rules, yielding a \emph{path}. For example, accessing the \code{bounds} field in~\Cref{figure:pbrt-layout} yields the path ($\mathcal{B}$, \code{nodes[I]}, \code{bounds}): start at the base structure $\mathcal{B}$ (a compiler-generated name), index into the direct group \code{nodes} via reference \code{I}, and load field \code{bounds}. ~\Cref{figure:pbrt-lower-example} provides the complete destructor specialization of closest-hit ray tracing.

The full implementation of ~\Cref{alg:concretization} handles additional complexities, e.g., scope management for nested contexts. Despite these simplifications, the algorithm embodies the essential mechanism: systematic traversal of the layout to resolve concretized field access patterns that respect all layout transformations. Field resolution is linear in the number of layout members, resulting in quadratic complexity for concretizing an entire tree traversal. This remains practical as layouts are generally compact (tens of members), and the compiler amortizes this cost by caching resolved fields.

\begin{wrapfigure}{r}{0.525\textwidth}
\vspace{-1em}
\centering 
\begin{lstlisting}[style=layout-style, mathescape=true] 
func closest_hit(
  ray: Ray, I: u32, PT: LinearBVH, best: mut (f32, Triangle)) = 
  if !(PT.nodes[I].nprims > 0) {
    if intersects(ray, PT.nodes[i].bounds) &&
       distmin(ray, PT.nodes[i].bounds) < best[0] {
      closest_hit(ray, I + 1, best);
      closest_hit(ray, I + PT.nodes[I].c_o, best);
    }
  } else {
    if intersects(ray, PT.nodes[i].bounds) {
      foreach t in PT.primitives[
          PT.nodes[I].p_o : PT.nodes[I].p_o + PT.nodes[I].nprims] {
        if intersects(ray, t) && distmin(ray, t) < best[0] {
          best = (distmin(ray, t), t); 
        } 
      } 
    } 
  }
\end{lstlisting} 
\vspace{-0.35em}
\caption{Destructor code generation for PBRTv4 layout.}
\label{figure:pbrt-lower-example}
\vspace{-0.5em}
\end{wrapfigure}

Critically, the well-formedness constraints discussed in \Cref{sec:well-formedness} ensure this traversal is always well-defined. Field coverage (and uniqueness) guarantees every queried field has an unambiguous representation, acyclic dependencies ensure derivations can be evaluated in order, and switch exhaustiveness ensures variant disambiguation always succeeds. Consequently, this enables layout-polymorphic queries: the same field access in the traversal algorithm automatically compiles to different concretized data representations depending on the layout specification.

\subsection{Constructor Specialization}
\label{sec:compilation-construction}

Constructor specialization generates code that transforms logical ADT terms into 
their physical representation according to the build specification. Lowering proceeds in linear time, through two phases. First, the compiler analyzes layout specifications to compute the size of each contiguous buffer, ensuring that each buffer requires only a single allocation. Second, the compiler generates code to recursively traverse the logical tree, applying variant-specific build operations at each node: \code{build} statements handle field values and process children, and \code{return} statements produce specialized references. Buffer offsets are tracked automatically to ensure correct placement. Each logical node thus produces exactly one specialized node with all physical transformations applied. The complete C++ lowering for Closest Hit Ray Tracing with PBRTv4 layout appears in Appendix~\ref{app:generated-code-pbrt}.

\subsection{Optimization \& Backend Code Generation}
\label{sec:compilation-backend}

A key advantage of DSL-level optimization is the ability to exploit semantic guarantees unavailable to general-purpose language compilers. Consider the field resolution algorithm, which na\"ively stamps in fields during tree traversal even when a path was already accessed, resulting in redundant memory operations. Eliminating these redundancies requires sophisticated data flow and memory analysis, which is prohibitively difficult in general-purpose languages. Our tree traversal DSL simplifies this analysis by guaranteeing BVH structures remain immutable during traversal. This immutability guarantee enables aggressive elimination of redundant control flow, and, in conjunction with a simpler intermediate representation, enables straightforward reuse of common subexpressions across control flow boundaries. These optimizations alone yield approximately $1.2$--$1.4\times$ speedups compared to \sys{} when DSL-specific compiler optimizations are disabled.

The compiler targets both C++ and CUDA, performing several lowering passes to bridge the gap between the high-level representation and backend languages. The lowering process is relatively straightforward since our tree traversal language is already similar to C. For the CUDA backend, we also perform necessary memory transfers between host and device. Where possible, the compiler maps arithmetic operations to architecture-specific intrinsics, e.g., on CUDA, this includes leveraging built-in math intrinsics for bounding volume quantization following~\citet{howard2016efficient}.

\section{Case Studies}
\label{sec:case-studies}

We demonstrate \sys{}'s expressiveness through three case studies that span the design space of data layout optimizations for BVHs. The first case study, Discrete Oriented Polytopes~\cite{kavcerik2024sah}, shows how memory access pattern transformations improve cache utilization by rearranging data. The second, Strand-Based Geometry~\cite{woop2014hair}, illustrates heterogeneous representation by combining multiple bounding volume types within the same ADT. The third, Tree-Carried Dependencies~\cite{eisemann2008ray, bauszat2010minimal}, demonstrates hierarchical compression, where child nodes inherit and refine parent state rather than storing redundant information. Each example highlights how \sys{}'s compositional primitives enable concise specification of real data representation optimizations.

\subsection{Improving Locality for Discrete Oriented Polytopes}
\label{sec:case-study-dop}

\begin{figure*}[t]
\begin{subfigure}[t]{0.46\textwidth}
\begin{lstlisting}[style=layout-style]
type BVHd(lo1: f32x3, hi1: f32x3, lo2: f32x4, hi2: f32x4)
  = Interior(left: BVHd, right: BVHd) 
  | Leaf(nprims: u4, data: Triangle[nprims]);
\end{lstlisting}
\includegraphics[width=\linewidth]{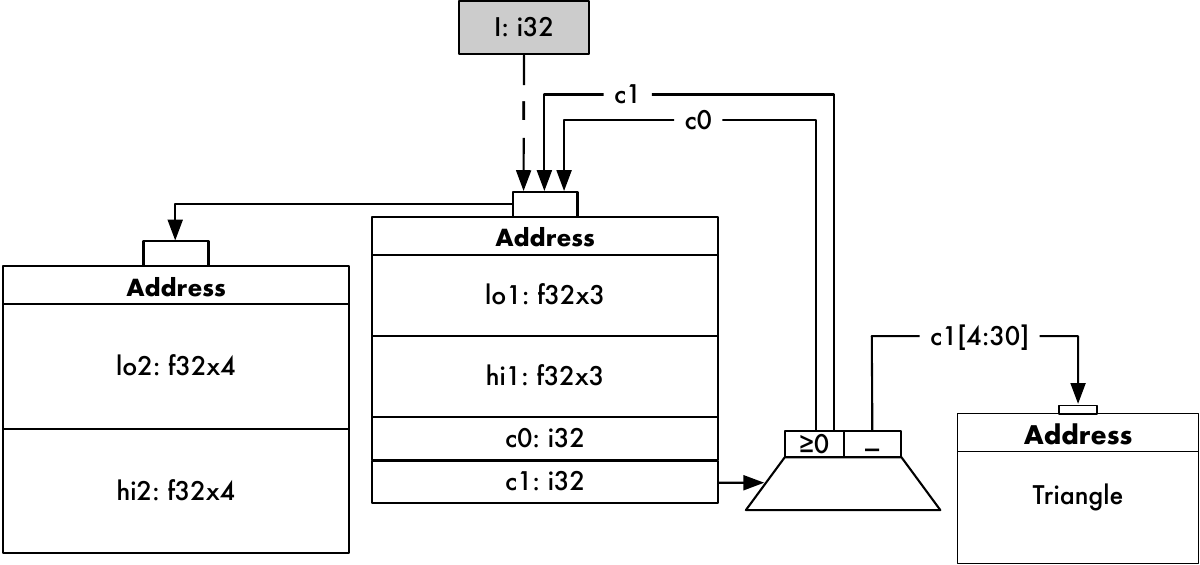}
\label{figure:dop14-type-flow}
\end{subfigure}
\hfill
\begin{subfigure}[t]{0.49\textwidth}
\begin{lstlisting}[style=layout-style]
layout BVHd(I: i32) {
  P: u32; N: u32; primitives: Triangle[P];
  group nodes[size=N] by I {
    lo1: f32x3; hi1: f32x3; // planes 0-2 §\label{line:dop14-planes02}§
    c0: i32; c1: i32;
    split c1 {
      >= 0 -> Interior { left = c0; right = c1; };
      _ -> Leaf {
        nprims = c1[0:3]; tri_address = c1[4:30];
        data = primitives[tri_address : tri_address + nprims];
      };
    };
    ---  §\label{line:dop14-soa}§
    lo2: f32x4; hi2: f32x4; // planes 3-6 §\label{line:dop14-planes36}§
  }; 
};
\end{lstlisting}
\label{figure:dop14-layout}
\end{subfigure}
\vspace{-0.5em}
\caption{BVH with DOP-14 bounding volumes~\cite{kavcerik2024sah} specified in \sys{}.}
\label{figure:dop14}
\vspace{-1.5em}
\end{figure*}

The choice of bounding volume introduces another axis of physical layout decisions, as seen in prior work for spheres  ~\cite{hubbard2002collision, palmer1995collision, kim1997fast},  axis-aligned bounding boxes (AABBs) ~\cite{howard2019quantized, vaidyanathan2016watertight, cohen1995collide, cline2006lightweight}, oriented bounding boxes (OBBs) ~\cite{woop2014hair, gottschalk1996obbtree, eberly2002dynamic}, k-ary discrete oriented polytopes (k-DOPs) ~\cite{kavcerik2024sah, coming2007velocity, klosowski1998efficient}, and hybrid variants~\cite{larsson2009bounding, kavcerik2025sobb}. ~\citet{kavcerik2024sah} demonstrate that separating 14-plane discrete oriented polytope (DOP-14) bounding volumes into two arrays improves traversal performance by enabling independent memory access patterns; this is illustrated in~\Cref{figure:dop14}. Their work introduces a novel culling strategy where they first check the AABB planes of the bounding volume (Line~\ref{line:dop14-planes02}), and only test planes 3--6 (Line~\ref{line:dop14-planes36}) if warranted~\cite[Sec 4.4]{kavcerik2024sah}. The authors implemented separate traversal algorithms: one for the AoS layout that loads and checks all seven planes at once, and another for the SoA layout that processes separated planes. Our layout language can express the latter transformation by enclosing these fields in a new group (Line~\ref{line:dop14-soa}). Fields \code{lo1}, \code{hi1}, \code{c0}, and \code{c1} correspond to these AABB planes occupying the first array, and are accessed immediately during traversal. Fields \code{lo2} and \code{hi2} reside in a separate array, and are accessed only when the finer-grain intersection test is needed.

This case study demonstrates \sys{}'s ability to express seemingly complex transformations in a concise manner. Additional changes, e.g., laying out planes in AoSoA for vectorization, can also be readily explored without requiring the traversal to be rewritten. \sys{} cleanly separates the logical bounding volume specification (which planes comprise a DOP-14) from the physical organization (which planes are co-located in memory). This separation facilitates exploration of alternative bounding volume representations that yield significant performance implications.

\subsection{Bit-stealing for Strand-Based Geometry Rendering}
\label{sec:case-study-hair}

\begin{figure*}[b]
\vspace{-1em}
\begin{subfigure}[t]{0.51\textwidth}
\begin{lstlisting}[style=layout-style]
type BVHm 
 = AABBNode(C: BVHm[4], low: f32x3x4, high: f32x3x4)
 | OBBNode(C: BVHm[4], low: f32x3x4, high: f32x3x4, or: f32x4x3)
 | Leaf(nprims: u4, data: BezierSegment[nprims]);
\end{lstlisting}
\includegraphics[width=\linewidth]{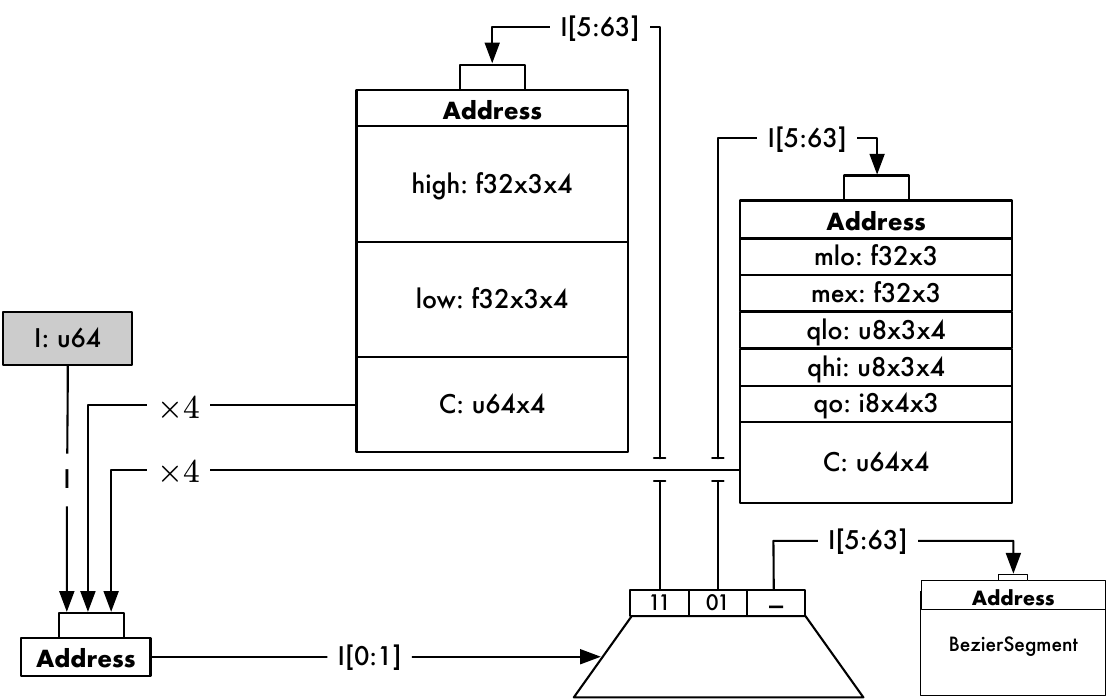}
\label{figure:embree-type-flow}
\end{subfigure}
\hfill
\begin{subfigure}[t]{0.44\textwidth}
\begin{lstlisting}[style=layout-style]
layout BVHm(I: u64 = 0'b11) {
  P: u64; primitives : BezierSegment[P];
  indirect group AABBs { §\label{line:hair-aabb-begin}§
    low: f32x3x4; high: f32x3x4; C: u64x4; 
  }; §\label{line:hair-aabb-end}§
  indirect group OBBs { §\label{line:hair-obb-begin}§
    mlo: f32x3; mex: f32x3;   // merged parent
    qlo: u8x3x4; qhi: u8x3x4; // quantized offset
    qo: i8x4x3;               // quantized orientation
    C: u64x4;                 // children
    obb_low  = dequantize_bounds(mlo, mex, qlo); §\label{line:hair-quantize-begin}§
    obb_high = dequantize_bounds(mlo, mex, qhi);
    or       = dequantize_orientation(qo); §\label{line:hair-quantize-end}§
  }; §\label{line:hair-obb-end}§
  group by I {
    let o: u64 = I[5:63]; §\label{line:hair-bit-steal-begin}§
    split I[0:1] {
      0b'11 -> AABBNode from AABBs[o];
      0b'01 -> OBBNode from OBBs[o];
      _ -> Leaf { 
        nprims = I[1:4] + 1; §\label{line:hair-bit-steal-end}§
        data = primitives[o : o + nprims];
      }; 
    }; 
  }; 
};
\end{lstlisting}
\label{figure:embree-layout}
\end{subfigure}
\vspace{-0.25em}
\caption{BVH for strand-based geometry with heterogeneous volumes and quantized OBBs specified in \sys{}.}
\label{figure:embree}
\vspace{-1em}
\end{figure*}

Strand-based geometry, e.g., hair and fur, presents unique challenges for ray tracing due to high primitive counts and poor spatial coherence. We demonstrate \sys{}'s expressiveness through an implementation of the mixed bounding volume strategy from ~\citet{woop2014hair}, which employs AABBs near the tree root for fast traversal, then transitions to OBBs at deeper levels for tighter volume fits. This heterogeneous representation combines sum types for variant discrimination, a quantized bounding volume encoding, and bit-field exploitation for compact node encoding.

~\Cref{figure:embree} presents Woop's layout specification in \sys{}. Lines~\ref{line:hair-bit-steal-begin}--\ref{line:hair-bit-steal-end} illustrate bit-stealing on index \code{I} to encode the node type in bits 0--1, the primitive count for leaves in bits 1--4 (unused in interior nodes), and the offset in bits 5--63 (primitive offset for leaves or node offset for the two different interior variant types). AABB nodes store uncompressed floating-point bounds (Lines~\ref{line:hair-aabb-begin}--~\ref{line:hair-aabb-end}). OBB nodes (Lines~\ref{line:hair-obb-begin}--\ref{line:hair-obb-end}) apply a quantization scheme where a merged-parent bounding volume provides the quantization frame, and per-child bounds encode 8-bit offsets and a shared quantized orientation matrix amortized across all four children as described in ~\citet[Sec~3.5]{woop2014hair}. For brevity, we do not show the definitions of the OBB dequantization functions used in Lines~\ref{line:hair-quantize-begin}--\ref{line:hair-quantize-end}.

This case study illustrates two aspects of \sys{}'s design philosophy. First, the algebraic data type cleanly separates the logical tree structure (three node variants with different fields) from the physical layout (quantization, bit packing, and indirection). Second, the compositional specification enables straightforward exploration of alternative representations: we can easily evaluate applying the AABB compression from ~\citet[Sec~4.1]{benthin2018compressed}, or reorganize memory by storing bounding volumes in a shared buffer to improve spatial locality during the AABB-to-OBB transition. These variations require modifying only the layout specification and build language, with the low-level traversal code automatically regenerated by the compiler.

\subsection{Reducing Storage Through Tree-Carried Dependencies}
\label{sec:case-study-tcd}

\begin{figure*}[b]
\vspace{-1em}
\begin{subfigure}[t]{0.46\textwidth}
\begin{lstlisting}[style=layout-style, mathescape=true]
type BVHt(low: f32x3, high: f32x3) 
  = Interior(left: BVHt, right: BVHt) 
  | Leaf(nprims: u5, data: Point[nprims]);
\end{lstlisting}
\includegraphics[width=\linewidth]{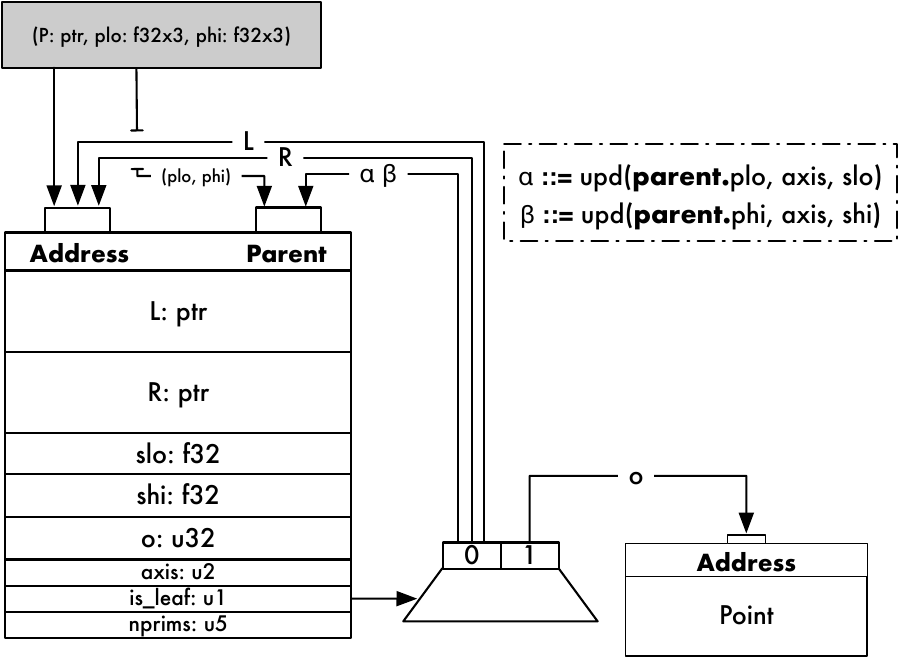}
\label{figure:tcd-type-flow}
\end{subfigure}
\hfill
\begin{subfigure}[t]{0.49\textwidth}
\begin{lstlisting}[style=layout-style, mathescape=true]
func upd(v: f32x3, axis: u2, S: f32) -> f32x3 {
  t : mut f32x3 = v; t[axis] = S;  return t;
}
layout BVHt(P: ptr, plo: f32x3, phi: f32x3) {
  N: u32; primitives: Point[N];
  group node by P {
    L: ptr; R: ptr;
    slo: f32; shi: f32; // low, high split §\label{line:tcd-split-plane}§
    o: u32; // primitive offset
    axis: u2; is_leaf: u1; nprims: u5; §\label{line:tcd-split-axis}§
    low  = parent.plo;
    high = parent.phi;
    split is_leaf {
      0 -> Interior {
        let $\alpha$ : f32x3 = upd(parent.plo, axis, slo); §\label{line:tcd-begin}§
        let $\beta$ : f32x3 = upd(parent.phi, axis, shi);
        left  = (L, $\alpha$, $\beta$); right = (R, $\alpha$, $\beta$); §\label{line:tcd-end}§
      };
      1 -> Leaf { 
        data = primitives[o : o + nprims]; 
      }; 
    }; 
  }; 
};
\end{lstlisting}
\label{figure:tcd-layout}
\end{subfigure}
\vspace{-0.5em}
\caption{BVH for the shared slab optimization with tree-carried dependencies specified in \sys{}.}
\label{figure:tcd}
\end{figure*}

Many BVH compression techniques exploit dependence (or hierarchical) relationships where child node representations depend on parent node state~\cite{ylitie2017incoherentrt,segovia2010memory}. The canonical example is the AABB shared slab optimization~\cite{eisemann2008ray, bauszat2010minimal} expressed in ~\Cref{figure:tcd}. When a node splits along a single axis, the bounding planes orthogonal to that axis remain unchanged for both children. Rather than storing all six bounding planes per child, each interior node stores only the split axis (Line~\ref{line:tcd-split-axis}) and two split plane positions (Line~\ref{line:tcd-split-plane}). Children nodes inherit the four unchanged planes by carrying parent bounds through the traversal (Lines~\ref{line:tcd-begin}--\ref{line:tcd-end}). This optimization reduces redundant storage by having children inherit invariant properties from ancestors rather than storing them explicitly.

This example demonstrates \sys{}'s support for more complex reference types and thus stateful traversal through its parent mechanism. The logical specification, a binary tree of axis-aligned bounding volumes, remains unchanged, while the physical layout exploits parent-child relationships to reduce memory footprint. This is essential for representing field compression in recursive ADTs.

\section{Evaluation}
\label{sec:evaluation}

We provide evidence that data representation is a first-order concern in the optimization of tree traversal algorithms over bounding volume hierarchies. We also show that the productivity advantages of \sys{} do not incur performance overhead when compared to state-of-the-art systems. The evaluation is structured into two parts:

\begin{enumerate}[leftmargin=1.5em]
    \item We conduct an extensive design space exploration demonstrating that Pareto-optimal (performance versus BVH memory footprint) data representations are fundamentally dependent on the characteristics of the input data, target machine, and tree traversal algorithm.
    \item We position \sys{}-generated kernels relative to hand-optimized kernels to establish a performance baseline for three applications: ray tracing, closest point query, and collision detection.
\end{enumerate}

\subsection{Experimental Methodology}
\label{sec:methodology}

To demonstrate portability, we evaluate \sys{} on three hardware platforms. The x86 evaluation platform is an Intel Core i9-14900K processor with 24 CPU cores (8 performance, 16 efficiency) and AVX-512 vector extension support. It has a three-level cache hierarchy: 896 KiB of L1 data cache, 32 MiB of L2 cache, and 36 MiB of L3 cache.  The ARM evaluation platform is an Apple MacBook M2 Pro with 10 CPU cores, Neon (128-bit) vector extensions, and 16 GiB of unified memory. The GPU evaluation platform is an NVIDIA GeForce RTX 4090 with 24 GiB of GDDR6X memory.

CUDA kernels are compiled with CUDA driver 12.6 and \code{-O3}; C++ kernels are compiled using Clang 19.1.7 and \code{-O3 -march=native}. Each benchmark conducts a warm-up run, then executes 9 times, dropping the lowest and highest 2 runs, and reports the weighted average of the remaining 5 runs. Unless stated otherwise, we use a software stack-based traversal (64 entries), consistent with the systems we're comparing to. Implementations for each evaluated algorithm appear in Appendix~\ref{app:tree-traversal}: closest-hit ray tracing \texttt{(CHRT)}, closest point query \texttt{(CPQ)}, and collision detection \texttt{(CD)}.

\textit{Layouts.} \Cref{table:layouts} enumerates the layouts we evaluate using \sys{}, spanning binary and octonary tree topologies. These layouts draw from established layout optimization techniques~\cite{pharr2023pbrt, benthin2018compressed, woop2014hair, howard2019quantized} while introducing novel compositions and application-specific refinements.

\begin{table}[b]
\centering
\tiny
\begin{minipage}[t]{0.48\textwidth}
\centering
\begin{tabular}{@{}lcl@{}}
\toprule
\textbf{2-ary Layout} & \textbf{Node Size (B)} & \textbf{Description} \\
\midrule
\texttt{pbrt}               & 32  & PBRTv4 LinearBVH~\cite{pharr2023pbrt} \\
\texttt{ptr}                & 48  & \texttt{pbrt} + children pointers \\
\texttt{pbrt-align16}       & 32  & \texttt{pbrt} + 16B align \\
\texttt{pbrt-soaos}         & 32  & \texttt{pbrt} + SoAoS \\
\texttt{pbrt-soaos-align16} & 32  & \texttt{pbrt-soaos} + 16B align \\
\texttt{sg-pbrt-q16}        & 16  & novel (\Cref{sec:pbrt-q16}) \\
\texttt{sg-pbrt-q16-soaos}  & 16  & \texttt{sg-pbrt-q16} + SoAoS \\
\texttt{sg-eq}              & 12  & snapped grid quantization~\cite{howard2019quantized} \\
\texttt{sg-eq-align16}      & 16  & \texttt{sg-eq} + 16B align \\
\\
\bottomrule
\end{tabular}
\end{minipage}
\hfill
\begin{minipage}[t]{0.48\textwidth}
\centering
\begin{tabular}{@{}lcl@{}}
\toprule
\textbf{8-ary Layout} & \textbf{Node Size (B)} & \textbf{Description} \\
\midrule
\texttt{bvh8}                  & 256 & unoptimized~\cite{woop2014hair} \\
\texttt{bvh8-align16}          & 256 & \texttt{bvh8} + 16B align \\
\texttt{bvh8-q8}               & 136 & 8-bit quantized~\cite{benthin2018compressed} \\
\texttt{bvh8-q8-align16}       & 144 & \texttt{bvh8-q8} + 16B align \\
\texttt{bvh8-q8-ci}            & 104 & \texttt{bvh8-q8} + compressed index \\
\texttt{bvh8-q8-ci-align16}    & 112 & \texttt{bvh8-q8-ci} + 16B align \\
\texttt{bvh8-q16}              & 184 & 16-bit quantized \\
\texttt{bvh8-q16-align16}      & 192 & \texttt{bvh8-q16} + 16B align \\
\texttt{bvh8-q16-ci}           & 152 & \texttt{bvh8-q16} + compressed index \\
\texttt{bvh8-q16-ci-align16}   & 160 & \texttt{bvh8-q16-ci} + 16B align \\
\bottomrule
\end{tabular}
\end{minipage}
\vspace{0.5em}
\caption{Binary (left) and octonary (right) BVH layouts with varying quantization schemes and memory alignment constraints. Node size refers to number of bytes to store a single node in the acceleration structure.}
\label{table:layouts}
\end{table}

Our evaluation explores four complementary optimization strategies: (1) bounding volume quantization\footnote{Bounding volume quantization guarantees functional equivalence; each quantized volume fully encloses its original.} ranging from 32-bit floating-point to $n$-bit parent-relative (\texttt{-q8,-q16}) and world-relative (\texttt{-eq}) encodings; (2) 16-byte alignment (\texttt{-align16}) to improve memory access alignment; (3) global memory transformation to improve spatial locality during traversal (\texttt{-soaos}); and (4) scene-specific optimizations: compressed indexing (\texttt{-ci}) exploits the known upper bound on primitives (28M triangles in our largest scene) to reduce memory footprint. We exclude the \texttt{ptr} layout from GPU evaluation as pointer-chasing produces a prohibitive slowdown (over two orders of magnitude). Reference implementations of \texttt{sg-eq} and \texttt{bvh8-q8-ci} are provided in Appendix~\ref{app:layout-implementations}.

\begin{wrapfigure}{r}{0.235\textwidth}
\vspace{-1.1em}
\centering
\tiny
\label{table:scenes}
\begin{tabular}{@{}lr@{}}
\toprule
\textbf{Scene} & \textbf{Triangles} \\
\midrule
lucy                   & 28,055,728 \\
power-plant            & 12,759,246 \\
san-miguel-x35-y22-z47 & 9,832,536 \\
sheep                  & 2,967,664 \\
hairball               & 2,880,000 \\
sponza                 & 262,267 \\
white-oak              & 36,760 \\
\bottomrule
\end{tabular}
\vspace{-0.5em}
\end{wrapfigure}

\textit{Scenes.}
Evaluated scenes retrieved from \citet{mcguire2017data} and \citet{stanford2025scan} contain triangle counts spanning three orders of magnitude and exhibit diverse geometry ranging from intricate organic models to expansive architectural environments. The only modified scene is \texttt{san-miguel-x35-y22-z47}, which we rotated about its axis by the prescribed degrees $(35^\circ, 22^\circ, 47^\circ)$ so that it is not axis-aligned.

\subsection{Design Space Exploration}
\label{sec:design-space}
We demonstrate that optimal data representations depend fundamentally on three factors: characteristics of the input data (data dependence), the target machine (machine dependence), and the traversal algorithm (algorithm dependence). Using \sys{} for systematic exploration, we also discover a novel layout that is Pareto-optimal in 35 of 42 evaluation contexts.

\subsubsection{Extended Methodology}
\label{sec:design-space-extended-methodology}

We evaluate across $2^{\{18,19,20,21,22,23\}}$ rays and compute weighted averages to determine latency (time per ray). We evaluate two ray distributions: (1) \emph{primary rays} originating from cameras and follow coherent paths with high locality, and (2) \emph{secondary rays} arising from light transport simulation, e.g., reflection and refraction, producing divergent, incoherent access patterns. On the CPUs, we parallelize the outer traversal loop using OpenMP dynamic scheduling and a block size of 64. On the GPU, we employ a one-thread-per-ray mapping with 256 threads per block. Presenting the results for the entire Cartesian product of scenes, machines, and ray distributions is impractical, so we present select plots to demonstrate our claims and provide a comprehensive dataset across all scenes, layouts, and platforms in Appendix~\ref{app:design-space-exploration-dataset}.

To isolate the impact of data representation, we standardize several design choices across all evaluation contexts. Every layout uses AABBs, applies explicit indexing into a contiguous primitive array for triangle retrieval, and employs Möller-Trumbore ray-triangle intersection algorithm~\cite{moller1997fast}. Tree construction follows a top-down recursive partitioning scheme with iterative binary refinement, guided by surface area heuristics (SAH) over 32 bins~\cite{wald2007fast}.

\begin{figure}[b]
\vspace{-1em}
\centering
\begin{subfigure}[t]{0.66\textwidth}
    \centering
    \includegraphics[width=\textwidth]{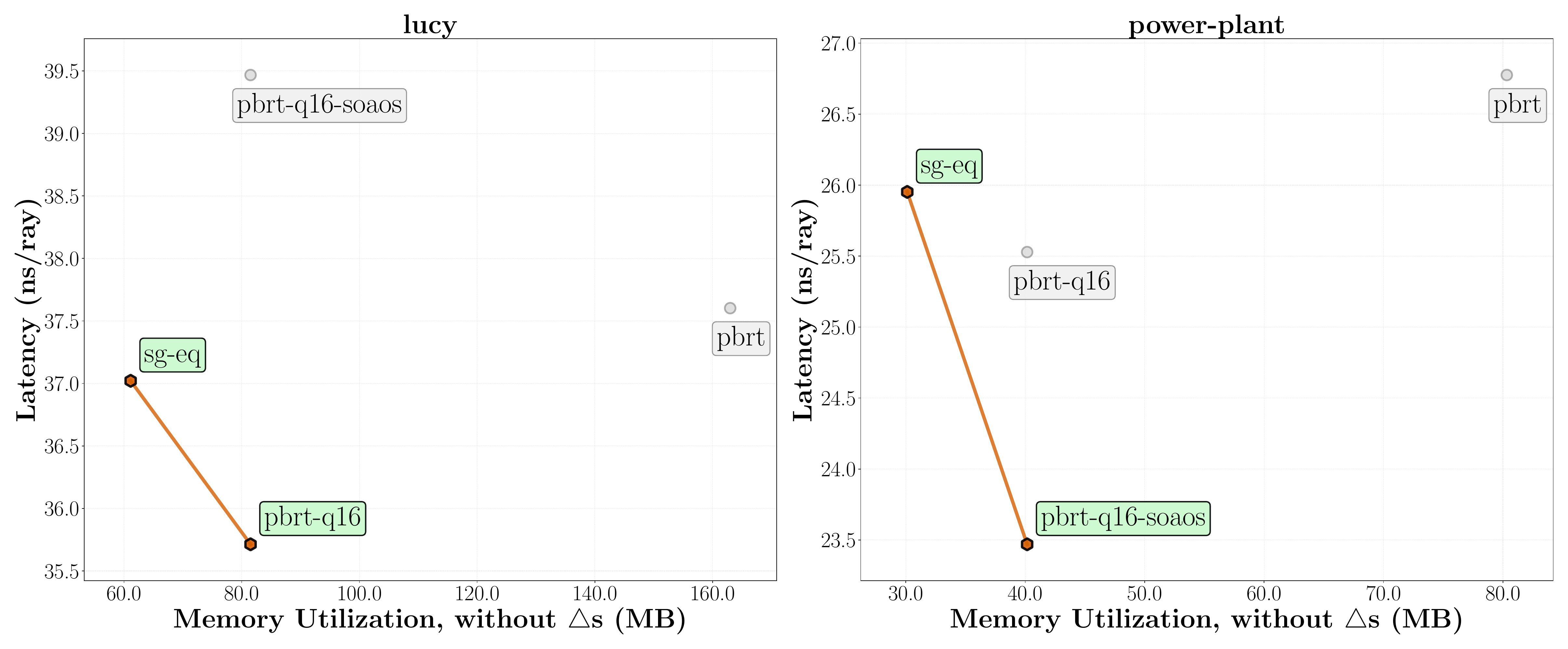}
    \caption{Scene-dependent}
    \label{figure:scene-dependent}
\end{subfigure}
\hfill
\begin{subfigure}[t]{0.33\textwidth}
    \centering
    \raisebox{0.5pt}{\includegraphics[width=\textwidth]{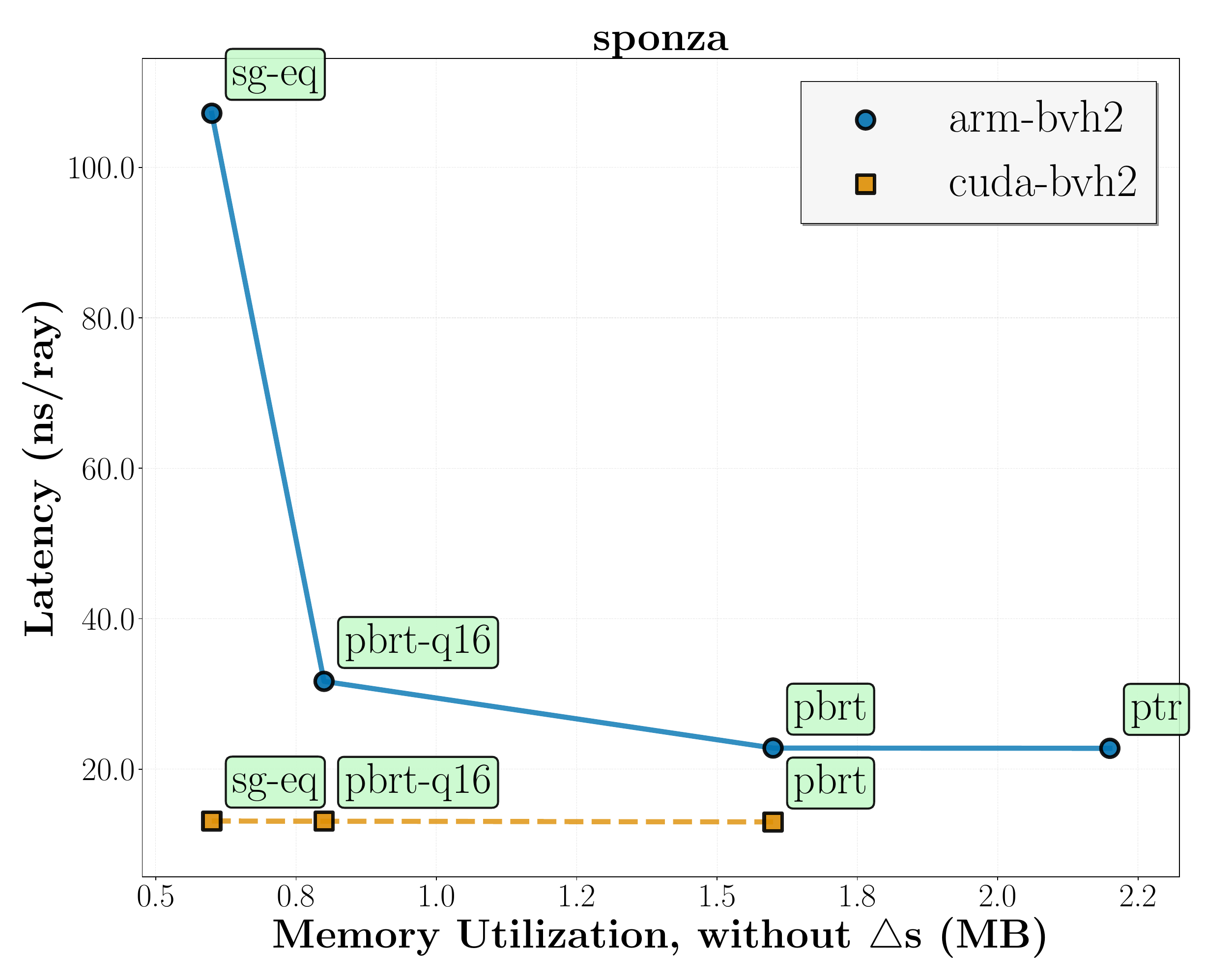}}
    \caption{Machine-dependent}
    \label{figure:machine-dependent}
\end{subfigure}
\caption{Performance variation across scenes and architectures for \texttt{bvh2} layouts. Lines denote the Pareto frontier, and gray points are non-Pareto optimal layouts.}
\label{figure:performance-variation}
\end{figure}

\subsubsection{Data-Dependent}
\label{sec:data-dependent}

Optimal data representations depend on input data characteristics, including both geometric properties of the indexed primitives and ray distribution patterns. \Cref{figure:performance-variation} compares Pareto frontiers of the \texttt{lucy} and \texttt{power-plant} scenes, revealing opposite latency orderings: \texttt{pbrt-q16} outperforms \texttt{pbrt-q16-soaos} by 11\% on \texttt{lucy} but underperforms by 9\% on \texttt{power-plant}, likely due to differing spatial distributions that affect cache behavior.

Ray distribution also affects layout performance independently of scene geometry. When tracing \texttt{lucy} on the GPU, the same layout yields divergent results for different ray types. Despite using 25\% less memory than \texttt{pbrt-q16} (the only other Pareto-optimal layout), \texttt{sg-eq} exhibits 3.7\% slowdown for coherent primary rays and 20.8\% slowdown for incoherent secondary rays. Thus, Pareto-optimal layout selection requires consideration of both scene characteristics and ray distribution.

\subsubsection{Machine-Dependent}
\label{sec:machine-dependent}

Optimal data layouts also depend on the architecture, reflecting differences in cache hierarchy organization, memory bandwidth characteristics, and instruction set capabilities. We demonstrate this dependence by comparing identical layouts and ray distributions (secondary) across the ARM and GPU architectures with \texttt{bvh2} layouts.

As illustrated in ~\Cref{figure:machine-dependent}, cache hierarchy differences produce divergent layout preferences across platforms. On ARM, eliminating index arithmetic operations in the traversal hot loop reduces memory-dependent operations, enabling \texttt{ptr} to marginally outperform \texttt{pbrt-align16} when the workload is memory latency bound. In contrast, the same \texttt{ptr} layout exhibits catastrophic performance degradation on the GPU (over 100$\times$ slowdown), as pointer chasing severely underutilizes the available memory bandwidth. Compact layouts like \texttt{sg-eq} and \texttt{pbrt-q16} achieve superior performance on the GPU by maximizing memory bandwidth utilization and enabling coalesced memory accesses. The performance gap between these two layouts on ARM is particularly notable: \texttt{sg-eq}'s dependence on directed rounding intrinsics during dequantization incurs substantial computational overhead.

\subsubsection{Algorithm-Dependent}
\label{sec:algorithm-dependent}

\begin{wrapfigure}{r}{0.4\textwidth}
\vspace{-2.5em}
\centering
\includegraphics[width=0.4\textwidth]{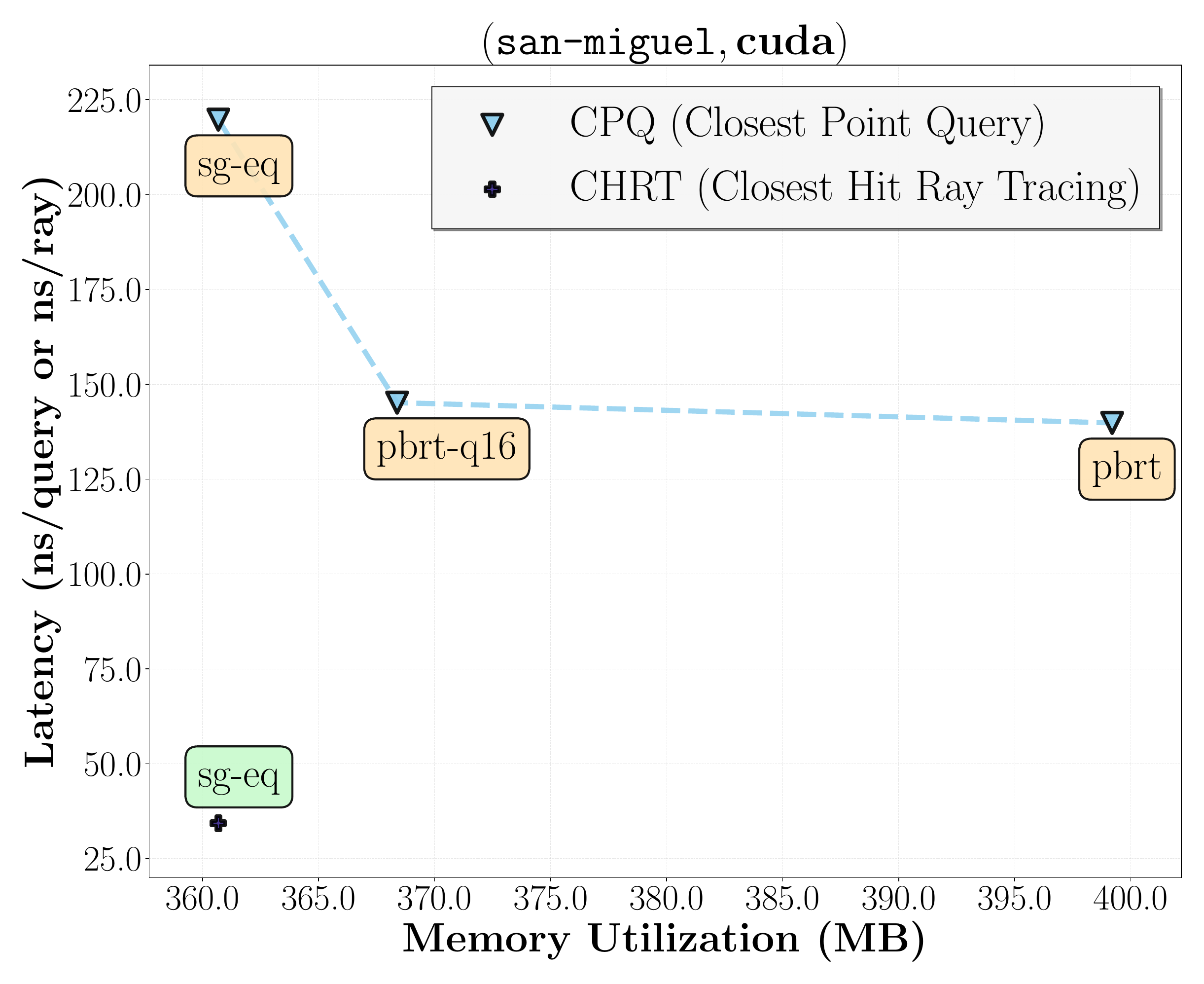}
\vspace{-1.75em}
\caption{Algorithm-dependent}
\label{figure:algorithm-dependent}
\vspace{-1.65em}
\end{wrapfigure}

\Cref{figure:algorithm-dependent} compares the Pareto frontiers of CHRT and CPQ of the San Miguel scene on the GPU. This comparison employs $2^{20}$ points and rays respectively with \texttt{bvh2} layouts. CHRT's Pareto-optimal layout is \texttt{sg-eq}, while CPQ sees better performance with \texttt{pbrt} variants. We attribute this to \texttt{sg-eq}'s quantization error: while the layout reduces memory footprint, the coarser bounds lead to additional node visits that degrade CPQ's pruning efficiency. These results demonstrate that the Pareto optimality of layouts can vary for different traversal algorithms.

\subsubsection{Exploring the Layout Design Space: \texttt{pbrt-q16}}
\label{sec:pbrt-q16}

By decoupling data specification from the logical specification, \sys{} enables rapid exploration of previously uninvestigated points in the representation design space. Our goal was to discover a layout that achieves Pareto-optimality across diverse evaluation contexts evaluated with our binary BVH layouts. We demonstrate this capability through \texttt{pbrt-q16}, a novel layout implemented by combining techniques from disparate sources: implicit indexing from PBRT~\cite{pharr2023pbrt}, global coordinate framing from molecular simulation neighbor search~\cite{howard2019quantized}, and quantization from ~\citet{benthin2018compressed}.

\textit{Rationale.} PBRTv4 employs an efficient bit-stealing scheme, but requires 32 bytes per node. To reduce the memory footprint, we can explore quantizing the bounding volume, which uses 24 of the 32 bytes. Most existing quantization schemes present a portability--performance tradeoff. For example, the snapped grid extent quantization (\texttt{sg-eq})~\cite{howard2019quantized} achieves aggressive compression at 12 bytes per node (a 62.5\% reduction from PBRTv4) through 10-bit fixed-point encoding relative to a global coordinate frame anchored at the root. However, \texttt{sg-eq} relies on directed rounding modes for their watertight dequantization algorithm, which are expensive on architectures that don't explicitly support such intrinsics, as discussed in ~\Cref{sec:machine-dependent}.

With a goal of portability, we adopt PBRTv4's implicit indexing and bit-stealing, then apply 16-bit quantization akin to the 8-bit quantization technique used in ~\citet{benthin2018compressed}. Unlike ~\citet{benthin2018compressed}, this quantization is applied relative to a global coordinate frame rather than the parent's frame. This design choice yields three advantages. First, this 16-bit quantization method requires only standard arithmetic operations, eliminating dependence on expensive rounding intrinsics and ensuring portability across CPU and GPU targets. Second, increasing the quantization grid to $2^{16}$ bins (a 64$\times$ increase over \texttt{sg-eq}'s $2^{10}$ bins) reduces representational errors in bounding volume extents. Consequently, fewer false negative intersections occur. Third, we reduce the node size to 16 bytes by using a global coordinate frame rather than storing the parent AABB and quantized offsets; this aligns naturally with cache line boundaries on contemporary architectures. We provide the full layout and build specification for \texttt{pbrt-q16} in Appendix~\ref{app:pbrt-q16-implementation}.

\begin{wrapfigure}{r}{0.5\textwidth}
\vspace{-1em}
\centering
\tiny
\begin{tabular}{@{}llll@{}}
\toprule
\textbf{Machine} & \textbf{Scene} & \textbf{Ray Dist.} & \textbf{Closest Pareto Point} \\
\midrule
x86 & lucy & primary & \texttt{pbrt-q16-soaos} \\
x86 & sponza & secondary & \texttt{pbrt-q16-soaos} \\
x86 & white-oak & secondary & \texttt{pbrt-q16-soaos} \\
cuda & power-plant & primary & \texttt{pbrt-q16-soaos} \\
cuda & san-miguel-x35-y22-z47 & primary & \texttt{sg-eq} \\
cuda & sponza & secondary & \texttt{sg-eq} \\
cuda & white-oak & primary & \texttt{sg-eq-align16} \\
\bottomrule
\end{tabular}
\vspace{-0.5em}
\caption{The set of (machine $\times$ scene $\times$ ray distribution) contexts where \texttt{pbrt-q16} is \textbf{not} Pareto-optimal.}
\label{figure:pbrt-q16-pareto-optimal}
\vspace{-1em}
\end{wrapfigure}

\textit{Results.} Among \texttt{bvh2} layouts, \texttt{pbrt-q16} achieves Pareto optimality in 35 of 42 evaluation contexts spanning three architectures, seven scenes, and two ray distribution patterns. Notably, 4 of the 7 non-dominating instances (shown in ~\Cref{figure:pbrt-q16-pareto-optimal}) are dominated by \texttt{pbrt-q16-soaos}, a variant of \texttt{pbrt-q16} that reorganizes the same representation into a struct-of-arrays-of-structs format. While we do not claim Pareto-optimality across all possible design spaces---indeed, the broader thesis of this paper is that no single layout can dominate universally---we demonstrate that systematic composition of orthogonal transformations can yield layouts that are Pareto-optimal within well-defined evaluation contexts.

\subsection{Comparison to State-of-the-Art Kernels}
\label{sec:sota-comparison}

Next, we evaluate \sys{}-generated code against hand-optimized baselines (1) to verify that our abstractions do not incur performance overhead, and (2) to contextualize the results of our design space exploration. While ~\Cref{sec:design-space} established data layout as a critical performance factor, the comparison below reveals that scheduling and tree topology yield significant performance variations, demonstrating that layout optimization is one of several orthogonal dimensions in the design space.

\subsubsection{Closest-Hit Ray Tracing: Comparison with Embree}
\label{sec:embree-comparison}

\begin{figure}[b]
\vspace{-1em}
\centering
\includegraphics[width=1.0\textwidth]{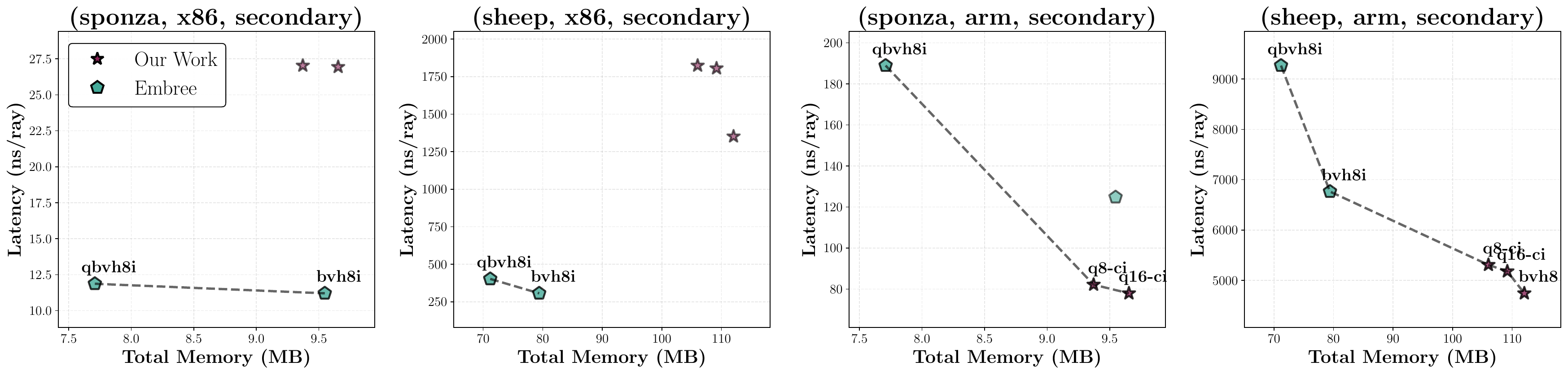}
\vspace{-1.5em}
\caption{Closest-hit ray tracing demonstrating (left) worst- and (right) best-case comparison with Embree.}
\label{figure:embree-comparison-rt}
\end{figure}

We compare (single ray) closest-hit ray tracing written in \sys{} to Intel Embree~\cite{wald2014embree}, the de facto standard for CPU-based ray tracing. Embree represents over a decade of production engineering effort by domain experts and employs sophisticated optimizations beyond data representation, e.g., SIMD-optimized traversal kernels.

\textit{Extended Methodology.} Our comparison employs Embree's 8-ary BVH layouts (\texttt{bvh8i} for unquantized, \texttt{qbvh8i} for quantized representations). We perform the same experiment as detailed in ~\Cref{sec:design-space-extended-methodology} for Embree and \texttt{bvh8} layouts. Embree defaults to Plücker coordinates~\cite{kay1986ray} for its unquantized layout, and uses Möller-Trumbore~\cite{moller1997fast} for its quantized layout. We employ the same intersection methods for our quantized and unquantized \texttt{bvh8} layouts (described in ~\Cref{table:layouts}). We note two system \emph{input} differences that prevent a direct comparison with Embree. First, while we attempt to approximate Embree's tree construction strategy, fully replicating all implementation details proved impractical, resulting in different tree topologies with corresponding effects on memory footprint and traversal performance. Second, Embree employs a two-level hierarchy separating geometric instances from primitives, while our system uses a single-level hierarchy directly over primitives. Despite these differences, the comparison demonstrates that decoupling layout from algorithm does not impose prohibitive abstraction overhead relative to hand-optimized kernels.

\textit{Results.} \Cref{figure:embree-comparison-rt} presents CHRT Pareto frontiers on the best- and worst-performing (scene, ray distribution, machine) contexts, selected by averaging performance over Pareto-optimal layouts. The worst-case comparison occurs on x86, an expected result as Embree is developed by Intel. In addition to hand-vectorized tree traversal and intersection, Embree also reduces memory utilization further through aggressive vertex compression. Despite no vectorization and the aforementioned differences in system input, \sys{} achieves Pareto-optimal performance in 14 of 28 evaluation contexts, three of which are on x86. Results for all contexts (7 scenes $\times$ 2 ray distributions $\times$ 2 architectures) are provided in Appendix~\ref{app:closest-hit-embree-comparison}.

\begin{figure}[b]
    \vspace{-1.25em}
    \centering
    \begin{subfigure}[t]{0.58\textwidth}
        \centering
        \includegraphics[width=\linewidth]{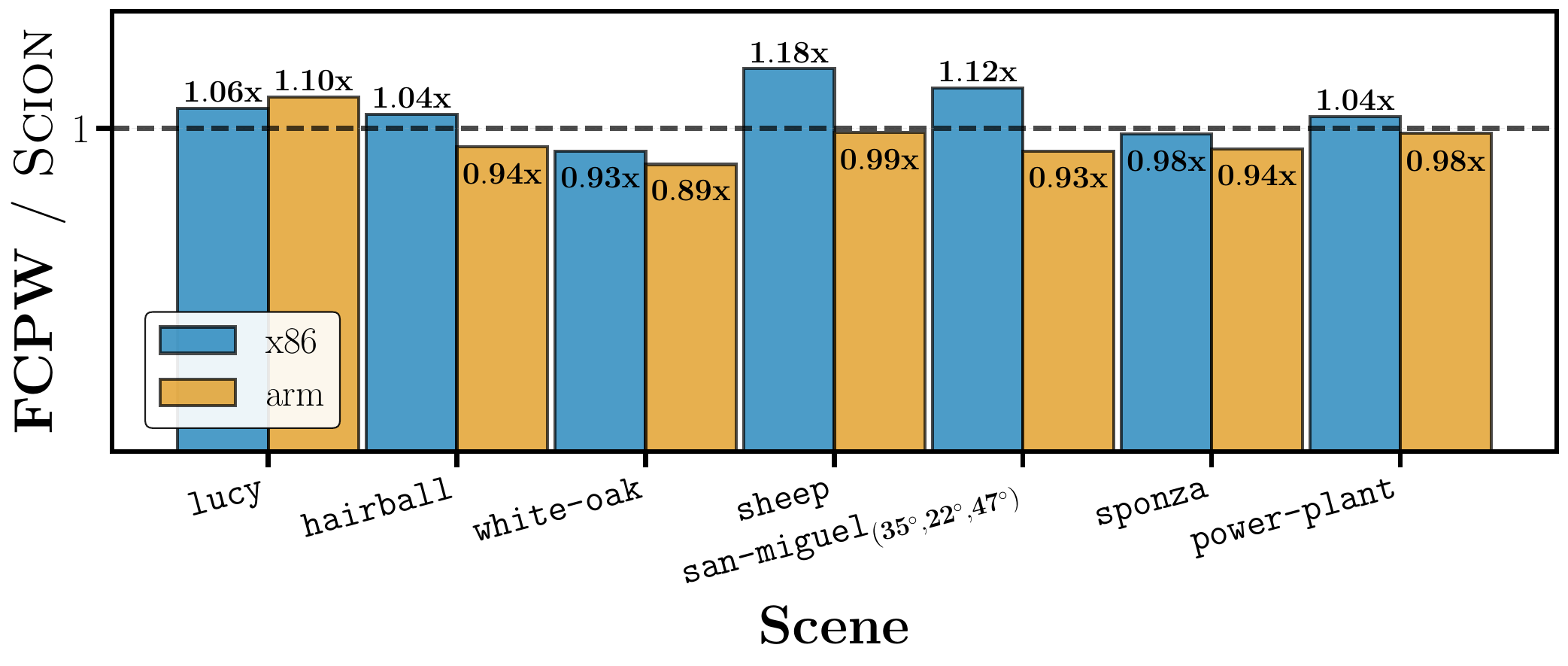}
        \caption{CPQ speedup vs FCPW (same layout).}
        \label{figure:fcpw-comparison-cpq}
    \end{subfigure}
    \hfill
    \begin{subfigure}[t]{0.34\textwidth}
        \centering
        \raisebox{12pt}{\includegraphics[width=\linewidth]{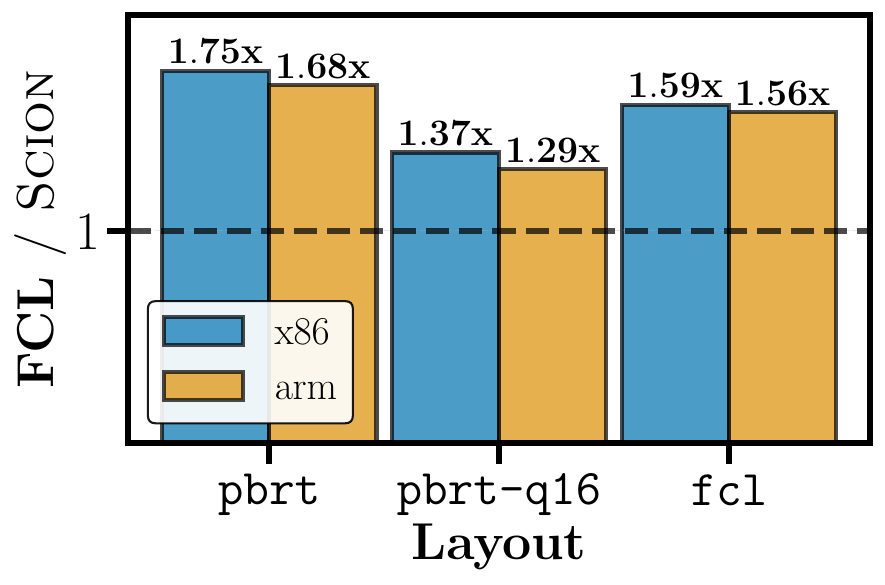}}
        \caption{CD speedup vs FCL (same scene).}
        \label{figure:fcl-comparison-cd}
    \end{subfigure}
    \vspace{-0.25em}
    \caption{Normalized performance comparison against state-of-the-art libraries (higher is better). }
    \label{figure:performance-comparison}
\end{figure}

\subsubsection{Closest Point Query: Comparison with FCPW}
\label{sec:fcpw-comparison-cpq}

The closest point query (CPQ) finds the nearest surface point to a given query position. We compare \sys{}-generated CPQ against Fastest Closest Points in the West (FCPW)~\cite{fcpw2021sawhney}, a hand-optimized library for geometric queries.

\textit{Extended Methodology.} For comparison, we use the same layout as FCPW (\texttt{pbrt}) and implement FCPW's tree construction algorithm: top-down recursive partitioning with binned surface area heuristic (SAH) splits: primitives are distributed into spatial buckets along each axis, and sweep operations compute partition costs to select optimal splits. Our implementation yields a nearly identical tree (less than 0.0005\% node count difference on our largest model, \texttt{lucy}). We issue 100,000 randomly sampled queries within each scene's bounding volume, executed single-threaded. Similar to FCPW, we additionally implement distance-based node sorting, which simply visits the children in an order determined by proximity to the point.

\textit{Results.} ~\Cref{figure:fcpw-comparison-cpq} illustrates that our performance is comparable to FCPW on both x86 and ARM architectures. Overall, \sys{} is 4\% slower on ARM and 5\% faster on x86 (geomean). The worst case ($0.89\times$ on \texttt{white-oak}) may be explained by two factors: (1) FCPW's use of highly optimized libraries, e.g., Eigen~\cite{eigen2010gael}, and (2) \sys{}'s default check for \texttt{SENTINEL} nodes since some layouts may have invalid children. FCPW assumes sentinels do not exist and elides this check.

\subsubsection{Collision Detection: Comparison with FCL}
\label{sec:fcpw-comparison-cpq}

Finally, we implement broad- and narrow-phase collision detection and compare to FCL~\cite{pan2012fcl}, a widely-deployed collision and proximity library for CPU-based applications.

\textit{Extended Methodology.} We evaluate collisions between two instances of the \texttt{hairball} scene (2.88M triangles each), where one instance is rotated by $(60^\circ, 70^\circ, 10^\circ)$ relative to the other. This configuration produces 5,118,441 colliding triangle pairs, providing substantial coverage of both tree traversal and primitive-level intersection tests. Notably, larger scene collisions resulted in stack overflow due to the recursive nature of FCL's tree traversal algorithm. To ensure fair comparison, we employ identical construction heuristics and intersection algorithms as FCL. Both systems construct bounding volume hierarchies using median split with one primitive per leaf. Both use the same algorithms for bounding volume overlap tests and employ the Separating Axis Theorem (SAT) for triangle-triangle intersection~\cite{tropp2006sat}. This is run single-threaded. Finally, unlike other evaluations, we use recursive function calls during culling to match FCL's traversal strategy.

\textit{Results.} ~\Cref{figure:fcl-comparison-cd} shows collision detection speedup relative to FCL; \sys{} is $1.68\times$ faster on ARM and $1.75\times$ faster on x86 (geomean). We observe two key results: first, when \sys{} employs the same memory layout as FCL, we achieve modest speedups across platforms. We attribute this improvement to reduced conditional branching overhead: FCL's implementation includes instrumentation for gathering statistics that adds additional branching to the traversal hot path. Second, when \sys{} employs optimized layouts originally designed for ray tracing, we observe further speedups ($1.29-1.75\times$), demonstrating that some layout optimizations can be effective across traversal algorithms.

\section{Related Work}
\label{sec:related-work}
Our work draws on several research threads in programming languages and compilers: fine-grain memory layout control, coarse-grain memory layout control, and separation of algorithm from representation. We position \sys{} relative to each area.

\textit{Fine-Grain Memory Layout Control.}
Recent work provides programmer control over memory layout at the granularity of individual fields and bits. \textsc{Ribbit}~\cite{baudon2023bitstealing} introduces tagged unions with bit-stealing, enabling bit-accurate, composable discriminated unions. QuanTaichi~\cite{hu2021quantaichi} supports per-field and shared exponent quantization in physical simulation software, trading precision for memory bandwidth. Virgil~\cite{teo2024virgil} enables customization of object layouts in an object-oriented, functional language. Dargent~\cite{chen2023dargent} applies verified data layout refinement to systems programming, providing formal guarantees about the correspondence between abstract specifications and concrete memory layouts. Many functional languages provide unboxed types to eliminate allocation overhead for small values~\cite{hall1994unboxing, jones1991unboxed, leroy1992unboxed}. Each of these systems operates at the level of individual elements, lacking control over coarse-grain transformations that we provide through reference types.

\textit{Coarse-Grain Memory Layout Control.}
Prior work also enables coarse-grain transformations such as serializing ADTs into flat layouts~\cite{vollmer2017compiling, vollmer2019local} and transforming AoS to SoA via a type declaration~\cite{pharr2012ispc}. Prior work in databases has explored improving cache utilization by tiling relations into \emph{minipages}~\cite{ailamaki2002data}. However, each of these systems only supports a predefined set of transformations and lacks fine-grain control, e.g., bit-stealing, necessary for optimizing BVH layouts.

\textit{Separation of Algorithm and Representation.}
More broadly, the principle of separating logical specification from physical implementation has deep roots in both databases and programming languages. Database systems achieve data independence through query optimization, where logical relational queries compile to diverse physical execution plans~\cite{codd1970relational}. Recent compiler frameworks extend this principle to specialized domains: the \textsc{Taco} compiler~\cite{chou2018format, kjolstad2017taco} decouples sparse tensors from data structure \emph{properties}; GraphIt~\cite{zhang2018graphit} and Taichi~\cite{hu2019taichi} perform a similar decoupling for graphs and spatially sparse grids, respectively. More recently, UniSparse~\cite{liu2024unisparse} proposed a series of composable transformations for rewriting sparse tensor representations.

\section{Conclusion}
\label{sec:conclusion}
We present \sys{}, a system that decouples data representation of bounding volume hierarchies from tree traversal algorithms through two domain-specific languages: a \emph{layout} language for specifying physical memory organization and a \emph{build} language for realizing the transformation from logical tree to physical representation. Through this bidirectional mapping, \sys{} automatically specializes build and traversal code to arbitrary layouts, enabling systematic exploration of the data representation design space across different algorithms, machines, and data characteristics.

\vspace{-0.2em} 
\section*{Data Availability Statement}
Performance results were generated using a publicly available artifact~\cite{this2026artifact} containing all benchmarking code, scripts, and instructions for reproducibility. Benchmarking results may vary across hardware platforms, which is precisely the thesis of this work.

\vspace{-0.2em} 
\begin{acks}
We thank our reviewers for their valuable feedback. We are especially grateful for Benjamin Driscoll's assistance in formalizing the language, and thank Andrew Adams for early
feedback on the initial development of the layout language. We also thank Scott Kovach, Shiv Sundram, Olivia Hsu, James Dong, Anderson Truong, and Katherine Mohr for feedback on early drafts of this paper. Finally, we extend special thanks to Rohan Yadav, who, by sheer misfortune, found himself seated in close proximity to the authors and graciously endured endless questions about scientific prose. 

We acknowledge Martin K\'{a}\v{c}erik for access to \texttt{sheep}; Morgan McGuire’s Computer Graphics Archive ~\cite{mcguire2017data} for access to \texttt{hairball}, \texttt{san-miguel}, \texttt{power-plant}, \texttt{white-oak}, and \texttt{sponza}; and Stanford University Computer Graphics Laboratory~\cite{stanford2025scan} for access to \texttt{lucy}. Christophe and Alexander were supported by the Qualcomm Innovation Fellowship and SystemX Alliance. This work was supported in part by the NSF under grant numbers 2216964 and 2143061, by the Stanford Portal Center, and by PRISM, one of seven centers in JUMP 2.0, a Semiconductor Research Corporation (SRC) program sponsored by DARPA.
\end{acks}
\vspace{-0.2em}

\bibliographystyle{ACM-Reference-Format}
\bibliography{references}

\vspace{-0.55em}

% \begingroup
%   \appendix
%   \setcounter{section}{0}
%   \setcounter{subsection}{0}
%   \refstepcounter{section}\label{app:path-uniqueness}%          A
%   \refstepcounter{section}\label{app:wf}%                       B
%   \refstepcounter{section}\label{app:design-space-exploration-dataset}% C
%   \refstepcounter{section}\label{app:layout-implementations}%   D
%     \refstepcounter{subsection}\label{app:sg-eq-implementation}%      D.1
%     \refstepcounter{subsection}\label{app:bvh-q8-ci-implementation}%  D.2
%     \refstepcounter{subsection}\label{app:pbrt-q16-implementation}%   D.3
%   \refstepcounter{section}\label{app:closest-hit-embree-comparison}%  E
%   \refstepcounter{section}\label{app:generated-code-pbrt}%      F
%   \refstepcounter{section}\label{app:tree-traversal}%           G
% \endgroup

%
% If your work has an appendix, this is the place to put it.
\appendix
\clearpage  % Start on a new page
% Reset counters for supplemental material.
% \setcounter{section}{0}
% \setcounter{figure}{0}
% \setcounter{table}{0}
% \setcounter{equation}{0}
% \setcounter{algorithm}{0}

\begin{center}
    {\LARGE \textbf{Appendix}} \\[1em]
\end{center}

\vspace{-1.5em} \section{Algorithm for Layout Path Uniqueness}
\label{app:path-uniqueness}

\vspace{-1em} \begin{algorithm}
\vspace{-0.25em}
\caption{Checking uniqueness of paths for field $F$ of variant type $T_v$.}
\tiny
\label{alg:uniqueness}
\begin{algorithmic}[1]
\State $M$, layout member to traverse; $F$, target field to resolve;
\State $T_{v}$, current variant type for \textsc{Split} disambiguation
\State $I$, map of indirect group identifier to layouts
\Function{CountPaths}{$M, F, T_{v}$}
    \State \textbf{match} $M$ \textbf{with}
    \State \hspace{1em} $|$ \textsc{Field}$(id)$ $\Rightarrow$
    \State \hspace{1em} $|$ \textsc{Derive}$(id, E)$ $\Rightarrow$
    \State \hspace{2em} \textbf{if} $id = F$ \textbf{then return} 1
    \State \hspace{2em} \textbf{else return} 0
    \State \hspace{1em} $|$ \textsc{Group}$(T_{G}, body, index)$ $\Rightarrow$ 
    \State \hspace{2em} \textbf{match} $T_{G}$ \textbf{with}
    \State \hspace{3em} $|$ \textsc{Direct} $\Rightarrow$ \textbf{return} \Call{CountPaths}{$body$, $F$, $T_{v}$, $I$}
    \State \hspace{3em} $|$ \textsc{Indirect}($name$) $\Rightarrow$ \textbf{insert} ($name$, $body$) into $I$; \textbf{return} $0$
    \State \hspace{1em} $|$ \textsc{From}$(name, key)$ $\Rightarrow$
    \State \hspace{2em} \textbf{let} $body \leftarrow I[name]$ \textbf{in} \Comment{Fail if $name \notin I$}
    \State \hspace{2em} \textbf{return} \textsc{CountPaths}($body$, $F$, $T_{v}$, $I$)
    \State \hspace{1em} $|$ \textsc{Split}$(discriminant, arms)$ $\Rightarrow$
    \State \hspace{2em} \textbf{for each} $Arm(C, T_{a}, body) \in arms$ \textbf{do}
    \State \hspace{3em} \textbf{if} $T_{a} = T_{v}$ \textbf{then}
    \State \hspace{4em} \textbf{then return} \textsc{CountPaths}($body$, $F$, $T_{v}$, $I$)
    \State \hspace{1em} $|$ \textsc{Sequence}$(layouts)$ $\Rightarrow$
    \State \hspace{2em} \textbf{let} $c \leftarrow 0$ \textbf{in}
    \State \hspace{2em} \textbf{for each} $L \in layouts$ \textbf{do}
    \State \hspace{3em} $c \leftarrow$ $c$ $+$ \textsc{CountPaths}($L$, $F$, $T_{v}$, $I$)
    \State \hspace{2em} \textbf{if} $c > 1$ \textbf{then fail} \Comment{Multiple paths to field $F$}
    \State \hspace{2em} \textbf{else return} $c$
\EndFunction
\end{algorithmic}
\vspace{-0.25em}
\end{algorithm}

\vspace{-1.75em}
\section{Well-Formedness of the Derivation-Free Fragment}
\label{app:wf}
The \emph{derivation-free fragment} is a subset of the \sys{} languages where all fields are stored; this disallows the use of relational primitives and derived fields. We assume the reference type $\mathcal{T}_\ell$ is injective, so distinct logical elements produce distinct references and therefore disjoint store regions. We state the well-formedness conditions summarized in Section~\ref{sec:well-formedness} for this subset of \sys{}, proving layout completeness (Theorem~\ref{thm:layout-complete}) and providing a proof sketch for round-trip soundness (Theorem~\ref{thm:round-trip}).

\subsection{Logical Representation}
\label{app:logical-representation}

A \emph{named field} is a pair $(x, T)$ of an identifier~$x$ and a type~$T$, written $x : T$. A \emph{named variant} is a pair $(V, \overline{x : T})$ of a constructor
name~$V$ and a finite sequence of distinctly named fields, written
$V(\overline{x : T})$. We formalize over recursive algebraic data types of the form:
\begin{align*}
  \tau \;&::=\;
    \mu\alpha.\;
    V_1(\overline{x_1 : T_1})
    + \cdots +
    V_m(\overline{x_m : T_m})
  \\[2pt]
  T \;&::=\; t \;\mid\; \alpha \;\mid\; T_1 \times \cdots \times T_n
        \;\mid\; [T]
\end{align*}
where $t$ ranges over primitive types and $\alpha$ is the recursive
variable bound by $\mu$. The surface language permits arbitrarily nested sum types. For the formalization, we elaborate each source type $\tau^\star$ to a sum-of-products normal form $\mathsf{NF}(\tau^\star)$ by hoisting nested sums to fresh top-level variants. The formal development below ranges over $\tau = \mathsf{NF}(\tau^\star)$. We write $\mathcal{V}(\tau)$ for the set of variants, and
\[
  \mathcal{N}(V)
  \;=\;
  \{(x, T) \mid x : T \text{ is a named field of } V\}
\]
for the named fields of variant~$V$. Let $\lfloor \cdot \rfloor_\ell$ denote the substitution $[\alpha \mapsto \mathcal{T}_\ell]$ applied structurally, where $\mathcal{T}_\ell$ is the
reference type declared by layout~$\ell$:
\[
  \lfloor t \rfloor_\ell \;=\; t
  \qquad
  \lfloor \alpha \rfloor_\ell \;=\; \mathcal{T}_\ell
  \qquad
  \lfloor T_1 \times \cdots \times T_n \rfloor_\ell
    \;=\; \lfloor T_1 \rfloor_\ell \times \cdots \times \lfloor T_n \rfloor_\ell
  \qquad
  \lfloor [T] \rfloor_\ell \;=\; [\lfloor T \rfloor_\ell]
\]
Finally, we define the \emph{resolved field set} of variant~$V$ for layout~$\ell$ as the following:
\[
  \mathcal{F}_\ell(V)
  \;=\;
  \{(x, \lfloor T \rfloor_\ell) \mid (x, T) \in \mathcal{N}(V)\}.
\]

\subsection{Physical Paths}
\label{app:physical-paths}

In the derivation-free fragment, layout members have the grammar:
\[
  M \;::=\; (x : T) \;\mid\; \mathbf{group}\; g\; \mathbf{by}\; I\; \{M\} \;\mid\; \mathbf{split}\; e\; \{C_i \to V_i\{M_i\}\}_{i=1}^n \;\mid\; M_0;\; M_1
\]
A \emph{physical path} records how a terminal (stored field) is reached from the root reference:
\[
  \pi \;::=\; \varepsilon \;\mid\; \mathcal{G}(g, I) \cdot \pi \;\mid\; \mathcal{S}(e, V) \cdot \pi
\]
where $\mathcal{G}(g, I)$ addresses group~$g$ via~$I$, and $\mathcal{S}(e, V)$ selects the arm labeled~$V$ based on discriminant~$e$. The judgment $V \vdash M : (\pi, x : T)$ states that terminal $x : T$
is reachable in member~$M$ by path~$\pi$ when decoding variant~$V$:
\[
\frac{
}{
 V \vdash (x : T) : (\varepsilon,\, x : T)
}
\tag{\textsc{P-Fld}}
\]

\[
\frac{
  V \vdash M : (\pi,\, x : T)
}{
  V \vdash \mathbf{group}\; g\; \mathbf{by}\; I\; \{M\} : (\mathcal{G}(g, I) \cdot \pi,\, x : T)
}
\tag{\textsc{P-Grp}}
\]

\[
\frac{
  V_j = V
  \qquad
  V \vdash M_j : (\pi,\, x : T)
}{
  V \vdash \mathbf{split}\; e\; \{C_i \to V_i\{M_i\}\}_{i=1}^n
        : (\mathcal{S}(e, V) \cdot \pi,\, x : T)
}
\tag{\textsc{P-Arm}}
\]

\[
\frac{
  V \vdash M_k : (\pi,\, x : T)
  \qquad
  k \in \{0, 1\}
}{
  V \vdash M_0;\; M_1 : (\pi,\, x : T)
}
\tag{\textsc{P-Seq}}
\]

\textsc{P-Fld} terminates at a named field. \textsc{P-Grp} prepends a group step. \textsc{P-Arm} selects the arm $V_j = V$ and prepends a split step. \textsc{P-Seq} propagates into either member without modifying the path.

\subsection{Layout Well-Formedness}
\label{app:layout-well-formed}

\begin{definition}[Layout completeness]

\label{def:layout-complete}
Let $\mathcal{P}(V, \ell) = \{(\pi, x : T) \mid V \vdash M_\ell : (\pi, x : T)\}$, where $M_\ell$ is the body of layout~$\ell$. A layout~$\ell$ is \emph{complete} if, for every $V \in \mathcal{V}(\tau)$ and every $(x, T) \in \mathcal{F}_\ell(V)$, there exists exactly one $(\pi, x : T) \in \mathcal{P}(V, \ell)$.
\end{definition}

Completeness says every logical named field has exactly one unambiguous physical path with the correct resolved type. A scope environment $\Gamma$ is a finite map from names to scope classes:
\[
  \Gamma \;::=\; x : T \;\mid\; g : \mathsf{grp}.
\]
Entries of the form $x : T$ encompass both reference parameters and field names, while $g : \mathsf{grp}$ are group names. A \emph{terminal binding set} $\Delta$ is a finite set of typed bindings
$\{x : T\}$. The judgment $\Gamma;\,V \vdash M\ \mathsf{wf} : \Delta$ states that member~$M$ is well-formed in scope~$\Gamma$ for variant~$V$ and exposes terminal bindings~$\Delta$.

% names must be fresh, and if the field is in the logical specification, it must conform to this floor typing relation.
\[
\frac{
  x \notin \mathsf{dom}(\Gamma)
  \qquad
  (x, U) \in \mathcal{N}(V) \implies T = \lfloor U \rfloor_\ell
}{
  \Gamma;\, V \vdash (x : T)\ \mathsf{wf} : \{x : T\}
}
\tag{\textsc{WF-Fld}}
\]

% I must be in scope, names must be fresh, the body must be well-formed
\[
\frac{
  (I : T) \in \Gamma
  \qquad
  g \notin \mathsf{dom}(\Gamma)
  \qquad
  (\Gamma,\, g : \mathsf{grp});\, V \vdash M\ \mathsf{wf} : \Delta
}{
  \Gamma;\, V \vdash \mathbf{group}\; g\; \mathbf{by}\; I\; \{M\}\ \mathsf{wf} : \Delta
}
\tag{\textsc{WF-Grp}}
\]

% 0. The split is exhaustive
%
% 1. There must exist only one choice for the variant V_j. The reason for this
% is that we want to prove statically that only one path exists, so something like
% this is ill-formed, for an ADT like T = A(x: i32) | ...;
%
% split e { 
%   0 -> A { x } // path 1
%   1 -> A { x } // path 2
% }
%
\[
\frac{
  (e : T) \in \Gamma
  \qquad
   \bigsqcup_{i=1}^{n} C_i = \llbracket T \rrbracket
  \qquad
  \exists!\, j.\; V_j = V
  \qquad
  \Gamma;\, V \vdash M_j\ \mathsf{wf} : \Delta
}{
  \Gamma;\, V \vdash
  \mathbf{split}\; e\; \{C_i \to V_i\{M_i\}\}_{i=1}^{n}\ \mathsf{wf} : \Delta
}
\tag{\textsc{WF-Split}}
\]

% A sequence of fields are well-formed if they are individually well-formed 
% and their bindings are disjoint.
\[
\frac{
  \Gamma;\, V \vdash M_0\ \mathsf{wf} : \Delta_0
  \qquad
  (\Gamma,\, \Delta_0);\, V \vdash M_1\ \mathsf{wf} : \Delta_1
}{
  \Gamma;\, V \vdash M_0;\; M_1\ \mathsf{wf} : \Delta_0 \cup \Delta_1
}
\tag{\textsc{WF-Seq}}
\]

% A layout is well-formed if the bindings cover the resolved field set of variant V_i
\[
\frac{
  \forall V \in \mathcal{V}(\tau).\quad
  \Gamma_0;\, V \vdash M_\ell\ \mathsf{wf} : \Delta_V
  \qquad
  \mathcal{F}_\ell(V) \subseteq \Delta_V
}{
  \vdash \ell\ \mathsf{wf}
}
\tag{\textsc{WF-Layout}}
\]

\textsc{WF-Fld} registers a fresh terminal. When $x$ names a logical field of~$V$, the type must match. 
\textsc{WF-Grp} requires the address $I : T$ is in scope, and that group names are fresh.
\textsc{WF-Split} requires that $e : T$ is in scope, that the conditions $\{C_i\}$ form a disjoint partition of the domain of $T$, and that there exists exactly one arm labeled by the variant $V$, extending the path by this branch.
\textsc{WF-Seq} extends $\Gamma$ with $\Delta_0$ before checking $M_1$, implicitly enforcing $\mathsf{dom}(\Delta_0) \cap \mathsf{dom}(\Delta_1) = \varnothing$ via \textsc{WF-Fld} freshness.
Finally, \textsc{WF-Layout} checks each variant independently, beginning with an initial environment~$\Gamma_0$ containing the layout's reference parameters, and requires the terminals to cover all resolved logical fields.

\subsection{Build Well-Formedness}
\label{app:build-well-formed}

In the derivation-free setting, the only possible outcome is copying the fields of the na\"ive layout, so well-formedness checking verifies there is a surjective mapping onto the physical layout. A \emph{field build} $b ::= \mathbf{build}\;x$ copies the value of named field~$x$ directly to its physical terminal. A \emph{variant build} is $\mathcal{B}_V = \langle b_0, \ldots, b_{n-1};\; \mathbf{return}\; e_V \rangle$. A \emph{build specification} $\beta = \{\mathcal{B}_V\}_{V \in \mathcal{V}(\tau)}$ is a set of variant builds, one per variant. The judgment $V ; \Delta_V \vdash b\ \mathsf{bwf} : \Psi$ checks that a field build is well-formed with respect to terminal bindings~$\Delta_V$ provided by \textsc{WF-Layout} for $V$, and records the field set~$\Psi$:

% A field build is well-formed if the typing is correct.
\[
\frac{
  (x,\, U) \in \mathcal{N}(V)
  \qquad
  (x,\, \lfloor U \rfloor_\ell) \in \Delta_V
}{
  \Delta_V \vdash \mathbf{build}\;x\ \mathsf{bwf} : \{x\}
}
\tag{\textsc{WF-BFld}}
\]

% similar to WF-Seq
\[
\frac{
  \Delta_V \vdash b_0\ \mathsf{bwf} : \Psi_0
  \qquad
  \Delta_V \vdash b_1\ \mathsf{bwf} : \Psi_1
  \qquad
  \Psi_0 \cap \Psi_1 = \varnothing
}{
  \Delta_V \vdash b_0;\; b_1\ \mathsf{bwf} : \Psi_0 \cup \Psi_1
}
\tag{\textsc{WF-BSeq}}
\]

% A build is well-formed if all variants cover the logical fields and the 
% returned value e_V has the reference type (by construction in this fragment).
\[
\frac{
  \forall V \in \mathcal{V}(\tau).\quad
  \Delta_V \vdash \mathcal{B}_V \; \mathsf{bwf} : \Psi_V
  \qquad
  \{x \mid (x, U) \in \mathcal{F}_\ell(V)\} \subseteq \Psi_V
  \qquad
  e_V : \mathcal{T}_\ell
}{
  \vdash \beta\ \mathsf{bwf}
}
\tag{\textsc{WF-Build}}
\]

\textsc{WF-BFld} requires that $x$ is a logical field of~$V$ and that the terminal exists in $\Delta_V$.
\textsc{WF-BSeq} prevents double initialization. 
Finally, \textsc{WF-Build} requires coverage of all fields in $\mathcal{F}_\ell(V)$ and a $\mathsf{return}$ expression $e_V$ of type $\mathcal{T}_\ell$.

\subsection{Layout Completeness}
\label{app:layout-complete}

\begin{lemma}[Field Coverage]
\label{lem:delta-phys} \mbox{}\\
If\; $\Gamma;\,V \vdash M\ \mathsf{wf} : \Delta$, then
\[
  \{\,x : T \mid \exists\,\pi.\;V \vdash M : (\pi, x : T)\,\} = \Delta.
\]
\end{lemma}

\begin{proof}
By induction on~$M$.
\begin{description}
\item[$M = (x : T)$.]
  Immediate from \textsc{P-Fld} and \textsc{WF-Fld}; both sides equal
  $\{x : T\}$.

\item[$M = \mathbf{group}\;g\;\mathbf{by}\;I\;\{M'\}$.]
  \textsc{P-Grp} prepends $\mathcal{G}(g, I)$ without changing the
  terminal, and \textsc{WF-Grp} passes $\Delta$ unchanged. Apply IH to $M'$.

\item[$M = \mathbf{split}\;e\;\{C_i \to V_i\{M_i\}\}$.]
  By \textsc{WF-Split}, there is exactly one arm $j$ with $V_j = V$; only $M_j$ is checked and $\Delta$ is unchanged. By \textsc{P-Arm}, the same arm contributes terminals. Apply IH to $M_j$.

\item[$M = M_0;\,M_1$.]  \mbox{}
  \begin{itemize}
  \item[$(\subseteq)$:] Let $x : T$ be reachable in $M$. By
    \textsc{P-Seq}, the derivation comes from $M_0$ or $M_1$. By IH
    on that member, $x : T \in \Delta_0$ or $x : T \in \Delta_1$,
    hence $x : T \in \Delta_0 \cup \Delta_1$.
  \item[$(\supseteq)$:] Any $x : T \in \Delta_k$ is reachable in $M_k$
    by IH, hence reachable in $M$ by \textsc{P-Seq}.\qedhere
  \end{itemize}
\end{description}
\end{proof}

\begin{lemma}[Path Determinism]
\label{lem:unique-path} \mbox{}\\
If\; $\Gamma;\,V \vdash M\ \mathsf{wf} : \Delta$, and
$V \vdash M : (\pi_0, x : T)$ and $V \vdash M : (\pi_1, x : T)$,
then $\pi_0 = \pi_1$.
\end{lemma}

\begin{proof}
By induction on~$M$.
\begin{description}
\item[$M = (x : T)$.]
  Both paths are $\varepsilon$ by \textsc{P-Fld}.

\item[$M = \mathbf{group}\;g\;\mathbf{by}\;I\;\{M'\}$.]
  Both paths have prefix $\mathcal{G}(g, I)$ by \textsc{P-Grp}. Apply
  IH to $M'$ for the suffix.

\item[$M = \mathbf{split}\;e\;\{C_i \to V_i\{M_i\}\}$.]
  By \textsc{WF-Split}, there is exactly one arm $j$ with $V_j = V$.
  By \textsc{P-Arm}, both derivations must use arm~$j$, so both paths
  share prefix $\mathcal{S}(e, V)$. Apply IH to $M_j$.

\item[$M = M_0;\,M_1$.]
  By \textsc{P-Seq}, each path passes through $M_0$ or $M_1$. If both
  pass through the same $M_k$, apply IH to $M_k$. Otherwise one passes
  through $M_0$ and the other through $M_1$. By
  Lemma~\ref{lem:delta-phys}, $x : T \in \Delta_0$ and $x : T \in
  \Delta_1$, so $x \in \mathsf{dom}(\Delta_0) \cap \mathsf{dom}(\Delta_1)$.
  This contradicts $\mathsf{dom}(\Delta_0) \cap \mathsf{dom}(\Delta_1) =
  \varnothing$, which follows from the freshness condition of
  \textsc{WF-Fld} propagated through \textsc{WF-Seq}.\qedhere
\end{description}
\end{proof}

\begin{theorem}[Layout Completeness]
\label{thm:layout-complete}
If\; $\vdash \ell\ \mathsf{wf}$, then $\ell$ is complete
(Definition~\ref{def:layout-complete}).
\end{theorem}

\begin{proof}
Let $V \in \mathcal{V}(\tau)$ and $(x, T) \in \mathcal{F}_\ell(V)$. By
\textsc{WF-Layout}, $\mathcal{F}_\ell(V) \subseteq \Delta_V$, so $(x : T)
\in \Delta_V$. By Lemma~\ref{lem:delta-phys}, there exists a path $(\pi,
x : T)$. Uniqueness of~$\pi$ follows from Lemma~\ref{lem:unique-path}.
\end{proof}

\subsection{Round-Trip Soundness}
\label{app:round-trip}

For a physical layout~$\ell$, let
\[
  \Pi_\ell
  \;=\;
  \{\pi \mid \exists V, x, T.\; V \vdash M_\ell : (\pi, x : T)\}
\]
be the set of physical paths admitted by the layout body~$M_\ell$. An
\emph{abstract store} is a finite partial map
$
\sigma : \mathcal{T}_\ell \times \Pi_\ell \rightharpoonup \mathsf{Val}
$. 
The store is indexed by paths rather than variants, so two variants may share the same location $(r, \pi)$ for a common field; injectivity of $\mathcal{T}_\ell$ ensures that cells $(r, \_)$ and $(r', \_)$ are disjoint whenever $r \neq r'$. We define 
\[
\mathsf{enc}_\beta : \tau \to (\Sigma \times  \mathcal{T}_\ell) 
\qquad
\mathsf{dec}_\ell : \Sigma \times \mathcal{T}_\ell \to \tau
\]
informally, where $\Sigma$ denotes the set of all abstract stores. A full operational semantics for both is left for future work. Instead, we sketch why the structural argument below is sound under the well-formedness conditions above.

\begin{theorem}[Round-Trip Soundness]
\label{thm:round-trip}
If\; $\vdash \ell\ \mathsf{wf}$ and $\vdash \beta\ \mathsf{bwf}$, then
$\mathsf{dec}_\ell(\mathsf{enc}_\beta(t)) = t$.
\end{theorem}

\begin{proof}[Proof Sketch]
By induction on $t = V(\overline{v_x})$ with $\mathcal{N}(V) =
\{(x_i, T_i)\}_{i=1}^k$. The sketch assumes that $\mathsf{enc}_\beta$
allocates a fresh reference $r$, writes each field to its unique terminal
path (as guaranteed by Theorem~\ref{thm:layout-complete}), and that
$\mathsf{dec}_\ell$ is a syntax-directed traversal reading values back
from those same paths. By \textsc{WF-Build}, variant~$V$ has a unique build procedure $\mathcal{B}_V$ covering all names in $\mathcal{F}_\ell(V)$; by \textsc{WF-BSeq}, each name is initialized exactly once. By Theorem~\ref{thm:layout-complete}, every field in $\mathcal{N}(V)$ has exactly one terminal path. For each $(x_i, T_i) \in \mathcal{N}(V)$:

\begin{description}
\item[$T_i \in T \setminus \{\alpha\}$.]
  By \textsc{WF-BFld}, the encoder writes $v_{x_i}$ to the unique
  terminal $x_i : T_i \in \Delta_V$. The decoder follows path~$\pi_i$
  and reads $v_{x_i}$ directly.
\item[$T_i = \alpha$.]
  The encoder recursively computes $(\sigma_i, r_i) =
  \mathsf{enc}_\beta(v_{x_i})$, writes $r_i : \mathcal{T}_\ell$ to
  terminal $x_i : \mathcal{T}_\ell \in \Delta_V$, and accumulates
  $\sigma_i$ by disjoint union. The decoder follows path~$\pi_i$, reads
  $r_i$, and applies IH to obtain $\mathsf{dec}_\ell(\sigma_i, r_i) = v_{x_i}$.
\end{description}
Since every field is recovered, $\mathsf{dec}_\ell(\mathsf{enc}_\beta(t))
= V(\overline{v_x}) = t$.
\end{proof}

\vspace{-1em} \section{Design Space Exploration: Dataset \& Analysis Script}
\label{app:design-space-exploration-dataset}

For the layout design space exploration in ~\Cref{sec:design-space}, we provide the complete dataset and analysis script in the Supplemental Material. The script requires Python 3 with \texttt{numpy}, \texttt{matplotlib}, and \texttt{pandas} installed. CSV column descriptions and example usage are provided in the script.

\newpage

\section{\sys{} Layout \& Build Specifications}
\label{app:layout-implementations}

\subsection{Layout \& Build Specifications for \texttt{sg-eq}}
\label{app:sg-eq-implementation}

\begin{lstlisting}[style=layout-style]
type BVH(low : f32x3, high : f32x3) = Interior(left : BVH, right : BVH) | Leaf(nprims: u4, data : Triangle[nprims]);

func dequantize(v: u30, bins: f32x3) -> f32x3 {
  let x_: u32 = (v >> 20) & 1023; let y_: u32 = (v >> 10) & 1023; let z_: u32 = (v >> 0) & 1023;
  return f32x3 { fmul_rd(x_ as f32, bins.x), fmul_rd(y_ as f32, bins.y), fmul_rd(z_ as f32, bins.z) };
}
func construct_bins_inverse(low: f32x3, high: f32x3) -> f32x3 {
  let L1: f32x3 = f32x3{fsub_ru(high.x, low.x), fsub_ru(high.y, low.y), fsub_ru(high.z, low.z)}; 
  let L2: mut f32x3 = {1.0, 1.0, 1.0};
  if (L1.x > 0.0) { L2.x = L1.x; } if (L1.y > 0.0) { L2.y = L1.y; } if (L1.z > 0.0) { L2.z = L1.z; }
  return f32x3{fdiv_rd(1023.0, L2.x), fdiv_rd(1023.0, L2.y), fdiv_rd(1023.0, L2.z)};
}
func construct_bins(low: f32x3, high: f32x3) -> f32x3 {
  let bins_inverse: f32x3 = construct_bins_inverse(low, high);
  return f32x3 {frcp_rd(bins_inverse.x), frcp_rd(bins_inverse.y), frcp_rd(bins_inverse.z)};
}
func quantize_lo(current: f32x3, world: f32x3, bin_inverse: f32x3) -> u32 {
  let x: u32 = floorf(fmul_rd(fsub_rd(current.x, world.x), bin_inverse.x)) as u32;
  let y: u32 = floorf(fmul_rd(fsub_rd(current.y, world.y), bin_inverse.y)) as u32;
  let z: u32 = floorf(fmul_rd(fsub_rd(current.z, world.z), bin_inverse.z)) as u32;
  return ((x & 1023u) << 20u) | ((y & 1023u) << 10u) | (z & 1023u);
}
func quantize_hi(current: f32x3, world: f32x3, bin_inverse: f32x3) -> u32 {
  let x: u32 = floorf(fmul_rd(fsub_rd(world.x, current.x), bin_inverse.x)) as u32;
  let y: u32 = floorf(fmul_rd(fsub_rd(world.y, current.y), bin_inverse.y)) as u32;
  let z: u32 = floorf(fmul_rd(fsub_rd(world.z, current.z), bin_inverse.z)) as u32;
  return ((x & 1023u) << 20u) | ((y & 1023u) << 10u) | (z & 1023u);
}
// Snapped-grid extent quantization using 2^10-1 bins.
layout BVH(index: u32) {
  primitive_count : u32;
  primitives : Triangle[primitive_count];
  wlow: f32x3; whigh: f32x3;
  bins: f32x3; bins_inv: f32x3;
  node_count : u32;
  group nodes[size=node_count] by index {
    q_min: u30; q_max: u30; nprims: u4;
    low  = fadd_rd(wlow,  dequantize(q_min, bins));
    high = fsub_ru(whigh, dequantize(q_max, bins));
    split nprims {
      0 -> Interior {
        offset : u32; left = index + 1; right = index + offset;
      };
      > 0 -> Leaf {
        poffset : u32; data = primitives[poffset : poffset + nprims];
      };
    };
  };
};
build BVH[order=pre] {
  build Interior(low: f32x3, high: f32x3, left: BVH, right: BVH) {
    build root {
      build wlow = low;
      build whigh = high;
      build bins_inv = construct_bins_inverse(low, high);
      build bins = construct_bins(low, high);
    };
    build q_min = quantize_lo(low, wlow, bins_inv);
    build q_max = quantize_hi(high, whigh, bins_inv);
    build nprims = 0;
    build left;
    let R: u32 = build right;
    build offset = R - this;
    return this;
  };
  
  build Leaf(low: f32x3, high: f32x3, nprims: u4, data: Triangle[nprims]) {
    build q_min = quantize_lo(low, wlow, bins_inv);
    build q_max = quantize_hi(high, whigh, bins_inv);
    build nprims;
    build poffset = append(data, nprims);
    return this;
  };
};
\end{lstlisting}

\newpage
\subsection{Layout \& Build Specifications for \texttt{bvh8-q8-ci}}
\label{app:bvh-q8-ci-implementation}

\begin{lstlisting}[style=layout-style]
type BVH = Interior(children: BVH[8], lo: f32x3x8, hi: f32x3x8) | Leaf(nprims: u8, data: Triangle[nprims]);
type qbox3(lo: u8x3, hi: u8x3);

func dequantize_bounds_lo(mlo: f32x3, mex: f32x3, bound: qbox3x8) -> f32x3x8 {
  let rcp: f32 = 1 / 255.0;
  return f32x3x8 { 
    mlo + (bound[0].lo as f32x3 * rcp) * mex, mlo + (bound[1].lo as f32x3 * rcp) * mex,
    mlo + (bound[2].lo as f32x3 * rcp) * mex, mlo + (bound[3].lo as f32x3 * rcp) * mex,
    mlo + (bound[4].lo as f32x3 * rcp) * mex, mlo + (bound[5].lo as f32x3 * rcp) * mex,
    mlo + (bound[6].lo as f32x3 * rcp) * mex, mlo + (bound[7].lo as f32x3 * rcp) * mex
  };
}
func dequantize_bounds_hi(mlo: f32x3, mex: f32x3, bound: qbox3x8) -> f32x3x8 {
  let rcp: f32 = 1 / 255.0;
  return f32x3x8 { 
    mlo + (bound[0].hi as f32x3 * rcp) * mex, mlo + (bound[1].hi as f32x3 * rcp) * mex,
    mlo + (bound[2].hi as f32x3 * rcp) * mex, mlo + (bound[3].hi as f32x3 * rcp) * mex,
    mlo + (bound[4].hi as f32x3 * rcp) * mex, mlo + (bound[5].hi as f32x3 * rcp) * mex,
    mlo + (bound[6].hi as f32x3 * rcp) * mex, mlo + (bound[7].hi as f32x3 * rcp) * mex
  };
}
func tfloor(f: f32x3) -> u8x3 { let f1: f32x3 = floorf(f); let f2: f32x3 = max(0.0, min(f1, 255.0)); return f2 as u8x3; }
func tceil(f: f32x3) -> u8x3 { let f1: f32x3 = ceilf(f); let f2: f32x3 = max(0.0, min(f1, 255.0)); return f2 as u8x3; }
// Use 32-bit reference type; this still allows for 2^25 primitives. 
// The first Interior node will have value 1, thus we provide this as the default.
layout BVH(I: u32 = 1u) {
  primitive_count: u32; primitives : Triangle[primitive_count]; interior_count: u32;
  indirect group Interiors[size=interior_count] {
    mlo: f32x3; mex: f32x3; child_bounds: qbox3x8; children: u32x8;
    lo = dequantize_bounds_lo(mlo, mex, child_bounds); hi = dequantize_bounds_hi(mlo, mex, child_bounds);
  };
  group by I {
    split I[0:1] {
      1 -> Interior from Interiors[I[2:31]];
      0 -> Leaf { let O: u32 = I[7:31]; nprims = (I[2:6] + 1) as u8;  data = primitives[O : O + nprims];  };
    };
  };
};

func quantize_bounds(low: f32x3x8, high: f32x3x8) -> qbox3x8 {
  let mlo: f32x3 = compute_merged_low(low); let mex: f32x3 = compute_merged_extent(low, high); 
  let rcp: f32x3 = (1.0 / mex) * 255.0;
  return qbox3x8 {
    qbox3 {tfloor((low[0] - mlo) * rcp), tceil((high[0] - mlo) * rcp) },
    qbox3 {tfloor((low[1] - mlo) * rcp), tceil((high[1] - mlo) * rcp) },
    qbox3 {tfloor((low[2] - mlo) * rcp), tceil((high[2] - mlo) * rcp) },
    qbox3 {tfloor((low[3] - mlo) * rcp), tceil((high[3] - mlo) * rcp) },
    qbox3 {tfloor((low[4] - mlo) * rcp), tceil((high[4] - mlo) * rcp) },
    qbox3 {tfloor((low[5] - mlo) * rcp), tceil((high[5] - mlo) * rcp) },
    qbox3 {tfloor((low[6] - mlo) * rcp), tceil((high[6] - mlo) * rcp) },
    qbox3 {tfloor((low[7] - mlo) * rcp), tceil((high[7] - mlo) * rcp) }
  };
}
func compute_merged_low(low: f32x3x8) -> f32x3 {
  return min(low[0], min(low[1], min(low[2], min(low[3], 
         min(low[4], min(low[5], min(low[6], low[7])))))));
}
func compute_merged_extent(lo: f32x3x8, hi: f32x3x8) -> f32x3 {
  let mlo: f32x3 = min(lo[0], min(lo[1], min(lo[2], min(lo[3], 
                   min(lo[4], min(lo[5], min(lo[6], lo[7])))))));
  let mhi: f32x3 = max(hi[0], max(hi[1], max(hi[2], max(hi[3], 
                   max(hi[4], max(hi[5], max(hi[6], hi[7])))))));
  return mhi - mlo;
}
build BVH[order=pre] {
  build Interior(children: BVH[8], lo: f32x3x8, hi: f32x3x8) {
    build mlo = compute_merged_low(lo); build mex = compute_merged_extent(lo, hi);
    build child_bounds = quantize_bounds(lo, hi);
    build children;
    return ((this << 2u) | 1u) as u32; // `1` tag for Interior.
  };
  build Leaf(nprims: u8, data: Triangle[nprims]) {
    let poffset: u32 = append(data, nprims);
    return ((poffset << 7u) | ((nprims - 1u) << 2u) | 0u) as u32; // `0` tag for Leaf.
  };
};
\end{lstlisting}

\newpage
\subsection{Layout \& Build Specifications for \texttt{pbrt-q16}}
\label{app:pbrt-q16-implementation}

\begin{lstlisting}[style=layout-style]
type BVH(low : f32x3, high : f32x3) = Interior(left : BVH, right : BVH) | Leaf(nprims: u4, data : Triangle[nprims]);
type q16x3(lo: u16x3, hi: u16x3);

func dequantize_lo(mlo: f32x3, mex: f32x3, bound: q16x3) -> f32x3 {
  let rcp: f32 = 1.0 / 65535.0;
  return mlo + ((bound.lo as f32x3) * rcp) * mex;
}
func dequantize_hi(mlo: f32x3, mex: f32x3, bound: q16x3) -> f32x3 {
  let rcp: f32 = 1.0 / 65535.0;
  return mlo + ((bound.hi as f32x3) * rcp) * mex;
}
func vu_floor(f: f32x3) -> u16x3 {
    let f1: f32x3 = floorf(f);
    let f2: f32x3 = max(0.0, min(f1, 65535.0));
    return f2 as u16x3;
}
func vu_ceil(f: f32x3) -> u16x3 {
    let f1: f32x3 = ceilf(f);
    let f2: f32x3 = max(0.0, min(f1, 65535.0));
    return f2 as u16x3;
}
// PBRT implicit indexing and bit stealing with snapped-grid extent quantization using 2^16-1 bins. 
// Additionally, this uses the AABB quantization scheme described in the Compressed-Leaf BVH work.
layout BVH(index: u32) {
  primitive_count : u32;
  primitives : Triangle[primitive_count];
  world_low : f32x3; world_extent : f32x3;
  node_count : u32;
  group nodes[size=node_count, align=16] by index {
    bounds_q : q16x3;
    low = dequantize_lo(world_low, world_extent, bounds_q);
    high = dequantize_hi(world_low, world_extent, bounds_q);
    nprims: u4;
    split nprims {
      0 -> Interior { 
        c_offset: u28;
        left = index + 1; 
        right = index + c_offset; 
      };
      > 0 -> Leaf { 
        p_offset: u28;
        data = primitives[p_offset : p_offset + nprims]; 
      };
    };
  };
};
func quantize_bounds(low: f32x3, high: f32x3, mlo: f32x3, mex: f32x3) -> q16x3 {
  let rcp: f32x3 = (1.0 / mex) * 65535.0;
  return q16x3 { vu_floor((low - mlo) * rcp), vu_ceil((high - mlo) * rcp) };
}
build BVH[order=pre] {  
  build Interior(low: f32x3, high: f32x3, left: BVH, right: BVH) {
    build root {
      build world_low = low;
      build world_extent = high - low;
    };
    build bounds_q = quantize_bounds(low, high, world_low, world_extent);
    build left;
    let R: u32 = build right;
    build c_offset = R - this;
    return this;
  };
  
  build Leaf(low: f32x3, high: f32x3, nprims: u4, data: Triangle[nprims]) {
    build bounds_q = quantize_bounds(low, high, world_low, world_extent);
    build p_offset = append(data, nprims);
    build nprims;
    return this;
  };
};
\end{lstlisting}

\newpage
\section{Closest Hit Ray Tracing Pareto Frontier (Embree vs \sys{})}
\label{app:closest-hit-embree-comparison}

\includegraphics[width=0.945\linewidth]{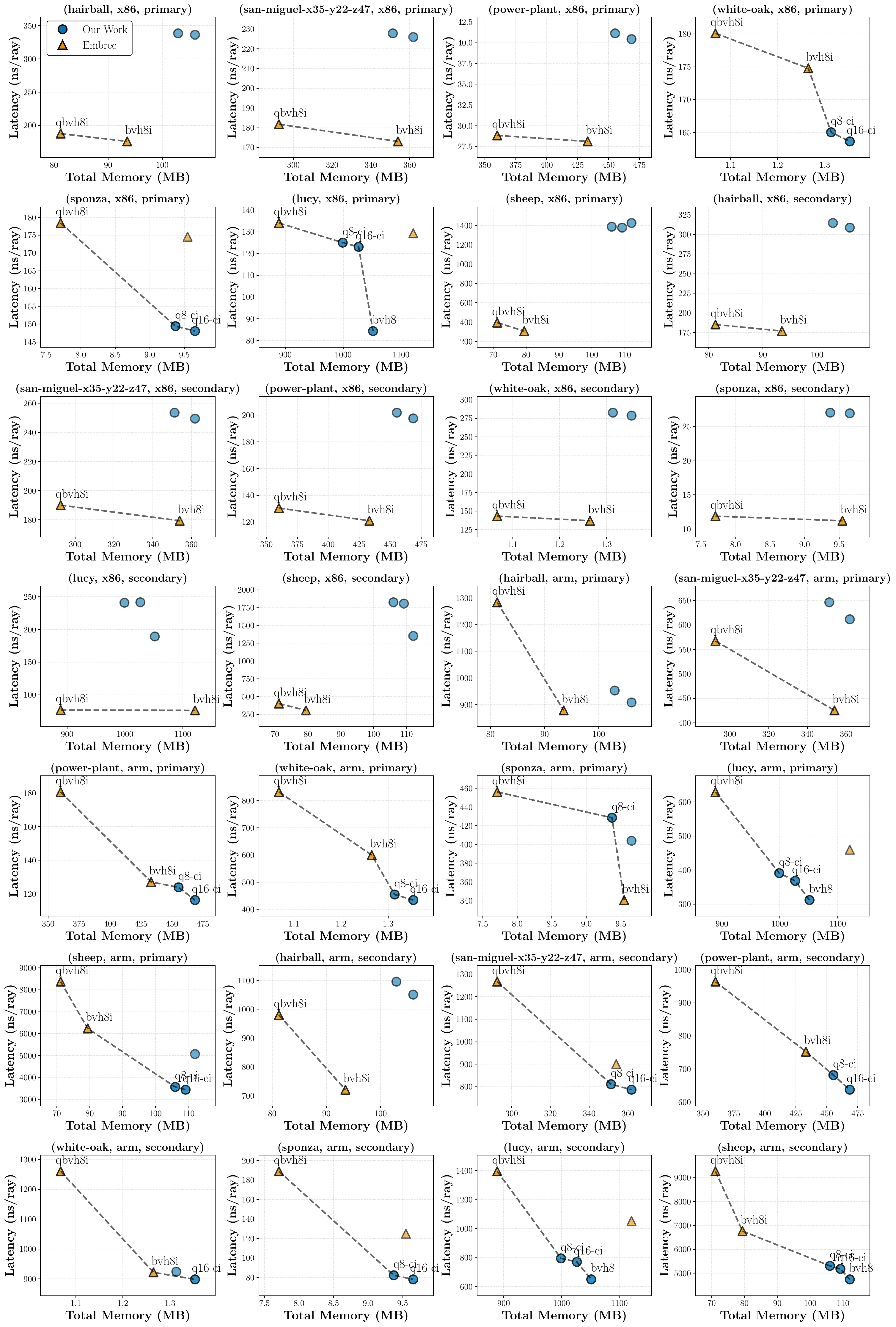}

\section{Generated C++ Code for Closest Hit Ray Tracing with PBRTv4}
\label{app:generated-code-pbrt}

\begin{figure}[H]
\centering
\begin{minipage}{0.49\textwidth}
\begin{lstlisting}[style=cpp-style, basicstyle=\ttfamily\fontsize{5}{6}\selectfont]
// Logical Tree
struct Interior;
struct Leaf;
using BVH = std::variant<Interior, Leaf>;
struct Interior { float3 low; float3 high; BVH* left; BVH* right; };
struct AABB { float3 low; float3 high; };
struct Leaf { float3 low; float3 high; uint16_t nprims; Triangle* data; };

struct Ray { float3 origin; float3 direction; float tmax = INF; };
struct Triangle { float3 p0; float3 p1; float3 p2; };

// Physical Tree
struct InteriorArm {
    uint32_t c_o;
} __attribute__((packed));
struct LeafArm {
    uint32_t p_o;
} __attribute__((packed));
struct alignas(32) Nodes {
    float3 low; float3 high; uint16_t nprims;
    uint8_t S[4]; // p_o or c_o
} __attribute__((packed));
struct LinearBVH {
    uint32_t P; uint32_t N; Triangle* primitives; Nodes* nodes;
} __attribute__((packed));

// Concretization, generated from the Tree Traversal and Layout Language.
void closest_hit(
  const uint32_t I,
  const LinearBVH* __restrict__ PT,
  const Ray* __restrict__ ray,
  std::tuple<float, Triangle>* __restrict__ _best0
) {
  if ((*PT).nodes[I].nprims == 0u) {
    if (intersects_Ray_AABB(
          ray, &AABB{
            .low  = (*PT).nodes[I].low, 
            .high = (*PT).nodes[I].high
        }) &&
        distmin_Ray_AABB(
          ray, &AABB{
            .low  = (*PT).nodes[I].low, 
            .high = (*PT).nodes[I].high
        }) < std::get<0>(* _best0)) {
      closest_hit(I + 1u, PT, ray, _best0);
      closest_hit(
        I + reinterpret<InteriorArm>((*PT).nodes[index].S).c_o,
        PT, ray, _best0
      );
    }
  } else {
    for (uint32_t j =
           reinterpret<LeafArm>((*PT).nodes[I].S).p_o;
         j <
           (reinterpret<LeafArm>((*PT).nodes[I].S).p_o +
             (*PT).nodes[I].nprims); ++j) {
      if (intersects_Ray_AABB(
            ray, &AABB{
              .low  = (*PT).nodes[I].low, 
              .high = (*PT).nodes[I].high
          }) &&
          distmin_Ray_AABB(
            ray, &AABB{
              .low  = (*PT).nodes[I].low, 
              .high = (*PT).nodes[I].high
          }) < std::get<0>(* _best0)) {
        if (intersects_Ray_Triangle(ray, &(*PT).primitives[j]) 
            && distmin_Ray_Triangle(ray, &(*PT).primitives[j]) 
            < std::get<0>(* _best0)) {
          (*_best0) = {
            distmin_Ray_Triangle(ray, &(*PT).primitives[j]), 
            (*PT).primitives[j]
          };
        }
      }
    }
  }
  return;
}


\end{lstlisting}
\end{minipage}
\hfill
\begin{minipage}{0.48\textwidth}
\tiny
\begin{lstlisting}[style=cpp-style, basicstyle=\ttfamily\fontsize{5}{6}\selectfont]
// Recursive build construction. Every variant has its own procedure.
uint32_t rec_build(
  const BVH* __restrict__ node, 
  LinearBVH* __restrict__ PT, 
  size_t* __restrict__ nodes_index, 
  size_t* __restrict__ primitives_index
) {
  return std::visit(overloaded{
    [&](const Interior& node) {
      const size_t this_index = (*nodes_index);
      (*nodes_index) += 1u;
      (*PT).nodes[this_index].low = node.low;
      (*PT).nodes[this_index].high = node.high;
      (*PT).nodes[this_index].nprims = 0;
      rec_build(node.left, PT, nodes_index, primitives_index);
      const uint32_t right_index = 
        rec_build(node.right, PT, nodes_index, primitives_index);
      reinterpret_cast<InteriorArm *>(
        &(*PT).nodes[this_index].S
      )->c_o = right_index - this_index;
      return this_index;
    },
    [&](const Leaf& node) {
      const size_t this_index = (*nodes_index);
      (*nodes_index) += 1u;
      (*PT).nodes[this_index].low = node.low;
      (*PT).nodes[this_index].high = node.high;
      (*PT).nodes[this_index].nprims = node.nprims;
      reinterpret_cast<LeafArm *>(
        &(*PT).nodes[this_index].S
      )->p_o = (*primitives_index);
      for (uint16_t __p = 0u; __p < node.nprims; ++__p) {
        (*PT).primitives[(__p + (*primitives_index))] = node.data[__p];
      }
      (*primitives_index) += node.nprims;
      return this_index;
    }
  }, *node);
}
// Recursive count procedure for allocation. The compiler can infer 
// which groups should be incremented by referring to the layout.
void rec_count(
  const BVH* __restrict__ node, 
  LinearBVH* __restrict__ PT
) {
  return std::visit(overloaded{
    [&](const Interior& node) {
      rec_count(node.left, PT);
      rec_count(node.right, PT);
      (*PT).N += 1u;
    },
    [&](const Leaf& node) {
      (*PT).P += node.nprims;
      (*PT).N += 1u;
    }
  }, *node);
}
// Constructor, generated from the Build Language.
LinearBVH build(const BVH* __restrict__ LT) {
  LinearBVH PT;
  size_t primitives_index = 0u;
  size_t nodes_index = 0u;
  PT.P = 0u;
  PT.N = 0u;
  rec_count(LT, (&PT));
  Triangle* primitives = reinterpret_cast<Triangle*>(
    malloc(sizeof(Triangle) * PT.P)
  );
  PT.primitives = primitives;
  Nodes* nodes = reinterpret_cast<Nodes*>(
    std::aligned_alloc(32, ((sizeof(Nodes) * PT.N) + 31) / 32) * 32
  );
  PT.nodes = nodes;
  rec_build(LT, (&PT), (&nodes_index), (&primitives_index));
  return PT;
}
\end{lstlisting}
\end{minipage}
\caption{C++ code generated by \sys{} for Closest Hit Ray Tracing with the PBRTv4 layout. We provide the unoptimized version because renaming occurs during the optimization passes, resulting in inscrutable code.}
\label{figure:pbrt-cpp-code-generation}
\end{figure}

\newpage
\section{\sys{}-Implemented Tree Traversal Algorithms}
\label{app:tree-traversal}

\subsection*{Closest Hit Ray Tracing}
\begin{lstlisting}[style=layout-style, mathescape=true]
type Ray(origin: f32x3, direction: f32x3, tmax: f32 = $\infty$);
type FInterval(low: f32, high: f32);
type TriangleIntersection(b0: f32, b1: f32, b2: f32, t: f32);
type AABB(low: f32x3, high: f32x3);
type Triangle(p0: f32x3, p1: f32x3, p1: f32x3);
type BVH(low: f32x3, high: f32x3) 
  = Interior(left: BVH, right: BVH) 
  | Leaf(nprims: u8, data: Triangle[nprims]);
// Ray-AABB intersection following Embree's approach.
func intersectsp_ray_aabb(r: Ray, b: AABB) -> option[FInterval] {
    let rdir: f32x3 = 1.0 / r.direction; let is_n: boolx3 = r.direction < 0.0;
    let nx: f32 = select(is_n.x, b.high.x, b.low.x); let fx: f32 = select(is_n.x, b.low.x, b.high.x); 
    let ny: f32 = select(is_n.y, b.high.y, b.low.y); let fy: f32 = select(is_n.y, b.low.y, b.high.y); 
    let nz: f32 = select(is_n.z, b.high.z, b.low.z); let fz: f32 = select(is_n.z, b.low.z, b.high.z);
    let t_nx: f32 = (nx - r.origin.x) * rdir.x; let t_fx: f32 = (fx - r.origin.x) * rdir.x;
    let t_ny: f32 = (ny - r.origin.y) * rdir.y; let t_fy: f32 = (fy - r.origin.y) * rdir.y;
    let t_nz: f32 = (nz - r.origin.z) * rdir.z; let t_fz: f32 = (fz - r.origin.z) * rdir.z;
    let t_near: f32 = max(0.0, max(t_nx, max(t_ny, t_nz)));  let t_far: f32 = min(r.tmax, min(t_fx, min(t_fy, t_fz)));
    if t_near <= t_far { return FInterval{t_near, t_far}; } return {};
}
// Moeller-Trumbore ray-triangle intersection following Embree's approach.
func intersectsp_ray_tri_mt(ray: Ray, tri: Triangle) -> option[TriangleIntersection] {
    let e1: f32x3 = tri.p0 - tri.p1; let e2: f32x3 = tri.p2 - tri.p0; let ng: f32x3 = cross_(e2, e1);
    let c: f32x3 = tri.p0 - ray.origin; let r: f32x3 = cross_(c, ray.direction); let D: f32 = dot(ng, ray.direction);
    if D == 0.0 { return {}; }
    let abs_D: f32 = abs(D); let sgn_D: u32 = (D to u32) & 2147483648u;
    let u_raw: f32 = ((dot(r, e2) to u32) ^ sgn_D) to f32; let v_raw: f32 = ((dot(r, e1) to u32) ^ sgn_D) to f32;
    if !(u_raw >= 0.0 && v_raw >= 0.0 && u_raw + v_raw <= abs_D) { return {}; }
    let t_raw: f32 = (((dot(ng, c)) ^ sgn_D) to u32) to f32;
    if !(abs_D * 0.0 < t_raw && t_raw <= abs_D * ray.tmax) { return {}; }
    let inv_abs_D: f32 = 1.0 / abs_D;
    let t: f32 = t_raw * inv_abs_D; let u: f32 = u_raw * inv_abs_D; let v: f32 = v_raw * inv_abs_D;
    let b0: f32 = 1.0 - u - v; let b1: f32 = u; let b2: f32 = v;
    return TriangleIntersection{b0, b1, b2, t};
}
// Pluecker coordinates ray-triangle intersection following Embree's approach.
func intersectsp_ray_tri_pc(ray: Ray, tri: Triangle) -> option[TriangleIntersection] {
    let v0: f32x3 = tri.p0 - ray.origin; let v1: f32x3 = tri.p1 - ray.origin; let v2: f32x3 = tri.p2 - ray.origin;
    let e0: f32x3 = v2 - v0; let e1: f32x3 = v0 - v1; let e2: f32x3 = v1 - v2;
    let u_raw: f32 = dot(cross_(e0, v2 + v0), ray.direction); 
    let v_raw: f32 = dot(cross_(e1, v0 + v1), ray.direction); 
    let w_raw: f32 = dot(cross_(e2, v1 + v2), ray.direction);
    let uvw: f32 = u_raw + v_raw + w_raw; let e: f32 = $\epsilon$ * abs(uvw);
    let min_uvw: f32 = min({u_raw, v_raw, w_raw}); let max_uvw: f32 = max({u_raw, v_raw, w_raw});
    if !(min_uvw >= -e || max_uvw <= e) { return {}; }
    let ng: f32x3 = cross_(e0, e1); let den: f32 = 2.0 * dot(ng, ray.direction); let t_raw: f32 = 2.0 * dot(v0, ng); 
    let t: f32 = t_raw / den; if !(t >= 0.0 && t <= ray.tmax) { return {}; } if den == 0.0 { return {}; }
    let inv_uvw: f32 = 1.0 / uvw; let b0: f32 = w_raw * inv_uvw; let b1: f32 = u_raw * inv_uvw; let b2: f32 = v_raw * inv_uvw;
    if b0 < 0.0 || b1 < 0.0 || b2 < 0.0 { return {}; } return TriangleIntersection{b0, b1, b2, t};
}
func intersects(r : Ray, b : AABB) -> bool {
    let I: option[FInterval] = intersectsp_ray_aabb(r, b); if I { return I.low < r.tmax && I.high > 0; } return false;
}
func distmin(r : Ray, b : AABB) -> f32 {
    let interval: option[FInterval] = intersectsp_ray_aabb(r, b); if interval { return interval.low; } return $\infty$;
}
func distmin(ray : Ray, tri : Triangle) -> f32 {
    // *NOTE*: chosen ray-triangle intersection method is specified in the Evaluation.
    let I: option[TriangleIntersection] = intersectsp_ray_tri(ray, tri); if I { return I.t; } return $\infty$;
}
func closest_hit(ray: Ray, bvh: BVH, best: mut (f32, Triangle)) =
  match bvh {
  | Interior(low, high, left, right) ->
    if intersects(ray, AABB(low, high)) && (distmin(ray, AABB(low, high)) < best[0]) { 
      closest_hit(ray, left,  best); 
      closest_hit(ray, right, best);
    }
  | Leaf(low, high, nprims, data) ->
    if intersects(ray, AABB(low, high)) {
      foreach t in data {
        if intersects(ray, t) && distmin(ray, t) < best[0] { 
          best = (distmin(ray, t), t); 
        } 
      } 
    } 
  }
\end{lstlisting}

\newpage
\subsection*{Closest Point Query}
\begin{lstlisting}[style=layout-style, mathescape=true]
type AABB(low: f32x3, high: f32x3);
type Point(v: f32x3);
type Triangle(p0: f32x3, p1: f32x3, p2: f32x3);
type BVH(low: f32x3, high: f32x3) 
  = Interior(left: BVH, right: BVH) 
  | Leaf(nprims: u16, data: Triangle[nprims]);

// Returns the closest point and barycentric coordinates.
// Ref: Real-Time Collision Detection [Ericson et al.] Ch. 5.1.5
func distmin_point_triangle(pt: Point, tri: Triangle) -> (Point, Point) {
    let p: f32x3 = pt.v;
    let a: f32x3 = tri.p0; let b: f32x3 = tri.p1; let c: f32x3 = tri.p2;
    let ab: f32x3 = b - a; let ac: f32x3 = c - a; let ap: f32x3 = p - a;
    let d1: f32 = dot(ab, ap); let d2: f32 = dot(ac, ap);
    if d1 <= 0.0 && d2 <= 0.0 { return (a, {{1.0, 0.0, 0.0}}); }
    let bp: f32x3 = p - b; let d3: f32 = dot(ab, bp); let d4: f32 = dot(ac, bp);
    if d3 >= 0.0 && d4 <= d3 {  return (b, {{0.0, 1.0, 0.0}}); }
    let vc: f32 = d1*d4 - d3*d2;
    if vc <= 0.0 && d1 >= 0.0 && d3 <= 0.0 {
        let v0: f32 = d1 / (d1 - d3); return (a + v0 * ab, {{1.0 - v0, v0, 0.0}});
    }
    let cp: f32x3 = p - c; let d5: f32 = dot(ab, cp); let d6: f32 = dot(ac, cp);
    if d6 >= 0.0 && d5 <= d6 { return (c, {{0.0, 0.0, 1.0}}); }
    let vb: f32 = d5*d2 - d1*d6;
    if vb <= 0.0 && d2 >= 0.0 && d6 <= 0.0 {
        let w0: f32 = d2 / (d2 - d6); return (a + w0 * ac, {{1.0-w0, 0.0, w0}});
    }
    let va: f32 = d3*d6 - d5*d4;
    if va <= 0.0 && (d4 - d3) >= 0.0 && (d5 - d6) >= 0.0 {
        let w1: f32 = (d4 - d3) / ((d4 - d3) + (d5 - d6)); return (b + w1 * (c - b), {{0.0, 1.0-w1, w1}});
    }
    let D: f32 = 1.0 / (va + vb + vc); let v: f32 = vb * D; let w: f32 = vc * D; let u: f32 = va * D;
    return (a + ab * v + ac * w, {{u, v, w}}); 
}
func distmin(p: Point, tri: Triangle) -> f32 {
    let ps: (Point, Point) = distmin_point_triangle(p, tri); let x: f32x3 = p.v - ps[0].v; return dot(x, x);
}
// Ref: Real-Time Collision Detection [Ericson et al.] Ch. 5.1.3.1
func square_distance_point_aabb(pt: Point, a: AABB) -> f32 {
    let v: f32x3 = pt.v;
    let sq_low: f32x3 = (a.low - v) * (a.low - v); let low: f32x3 = select(v < a.low, sq_low, 0.0);
    let sq_high: f32x3 = (v - a.high) * (v - a.high); let high: f32x3 = select(v > a.high, sq_high, 0.0);
    return sum(low + high);
}
func distmin(pt: Point, a: AABB) -> f32 {
    return square_distance_point_aabb(pt, a);
}
func distmax(pt: Point, a: AABB) -> f32 {
    let u: f32x3 = a.low - pt.v; let v: f32x3 = pt.v - a.high; let d: f32x3 = min(u, v); return dot(d, d);
}
func closest_point(p: Point, bvh: BVH, best: mut (f32, Point)) =
  match bvh {
  | Interior(low, high, left, right) ->
    if distmin(p, AABB(low, high)) < best[0] {
      let upper_bound: f32 = distmax(p, AABB(low, high));
      if upper_bound < best[0] { best = (upper_bound, best[1]); }
      // Children sorting optimization.
      let La: AABB = match left {
        | Interior(ll, hl, _, _) -> AABB(ll, hl)
        | Leaf(ll, hl, _, _) -> AABB(ll, hl)
      }
      let Ra: AABB = match right {
        | Interior(lr, hr, _, _) -> AABB(lr, hr)
        | Leaf(lr, hr, _, _) -> AABB(lr, hr)
      }
      let L: f32 = distmin(p, La); let R: f32 = distmin(p, Ra);
      if L < R {
        closest_triangle(p, left,  best); closest_triangle(p, right, best);
      } else {
        closest_triangle(p, right, best); closest_triangle(p, left,  best);
      }
    }
  | Leaf(low, high, nprims, data) ->
    if distmin(p, AABB(low, high)) < best[0] {
      foreach t in data {
        if distmin(p, t) < best[0] { best = (distmin(p, t), t); } 
      } 
    } 
  }
\end{lstlisting}

\newpage
\subsection*{Collision Detection}
\begin{lstlisting}[style=layout-style, mathescape=true]
type AABB(low: f32x3, high: f32x3);
type Triangle(p0: f32x3, p1: f32x3, p2: f32x3);
type BVH(low: f32x3, high: f32x3) 
  = Interior(left: BVH, right: BVH) 
  | Leaf(nprims: u16, data: Triangle[nprims]);

func project6(ax: f32x3, p1: f32x3, p2: f32x3, p3: f32x3, 
              q1: f32x3, q2: f32x3, q3: f32x3) -> i32 {
    let P1: f32 = dot(ax, p1); let P2: f32 = dot(ax, p2); let P3: f32 = dot(ax, p3);
    let Q1: f32 = dot(ax, q1); let Q2: f32 = dot(ax, q2); let Q3: f32 = dot(ax, q3);
    let mn1: f32 = min(min(P1, P2), P3); let mx2: f32 = max(max(Q1, Q2), Q3); if mn1 > mx2 { return 0; }
    let mx1: f32 = max(max(P1, P2), P3); let mn2: f32 = min(min(Q1, Q2), Q3); if mn2 > mx1 { return 0; }
    return 1;
}
// SAT triangle-triangle intersection following FCL's approach.
func SAT_triangle_intersection(
  P1: f32x3, P2: f32x3, P3: f32x3, Q1: f32x3, Q2: f32x3, Q3: f32x3
) -> bool {
    let p1: f32x3 = {0.0, 0.0, 0.0}; let p2: f32x3 = P2 - P1; let p3: f32x3 = P3 - P1;
    let q1: f32x3 = Q1 - P1; let q2: f32x3 = Q2 - P1; let q3: f32x3 = Q3 - P1;
    let e1: f32x3 = p2 - p1; let e2: f32x3 = p3 - p2; let n1: f32x3 = cross(e1, e2);
    if project6(n1, p1, p2, p3, q1, q2, q3) == 0 { return false; }
    let f1: f32x3 = q2 - q1; let f2: f32x3 = q3 - q2; let m1: f32x3 = cross(f1, f2);
    if project6(m1, p1, p2, p3, q1, q2, q3) == 0 { return false; }
    let ef11: f32x3 = cross(e1, f1); if project6(ef11, p1, p2, p3, q1, q2, q3) == 0 { return false; }
    let ef12: f32x3 = cross(e1, f2); if project6(ef12, p1, p2, p3, q1, q2, q3) == 0 { return false; }
    let f3: f32x3 = q1 - q3; let ef13: f32x3 = cross(e1, f3); if project6(ef13, p1, p2, p3, q1, q2, q3) == 0 { return false; }
    let ef21: f32x3 = cross(e2, f1); if project6(ef21, p1, p2, p3, q1, q2, q3) == 0 { return false; }
    let ef22: f32x3 = cross(e2, f2); if project6(ef22, p1, p2, p3, q1, q2, q3) == 0 { return false; }
    let ef23: f32x3 = cross(e2, f3); if project6(ef23, p1, p2, p3, q1, q2, q3) == 0 { return false; }
    let e3: f32x3 = p1 - p3; let ef31: f32x3 = cross(e3, f1); if project6(ef31, p1, p2, p3, q1, q2, q3) == 0 { return false; }
    let ef32: f32x3 = cross(e3, f2); if project6(ef32, p1, p2, p3, q1, q2, q3) == 0 { return false; }
    let ef33: f32x3 = cross(e3, f3); if project6(ef33, p1, p2, p3, q1, q2, q3) == 0 { return false; }
    let g1: f32x3 = cross(e1, n1); if project6(g1, p1, p2, p3, q1, q2, q3) == 0 { return false; }
    let g2: f32x3 = cross(e2, n1); if project6(g2, p1, p2, p3, q1, q2, q3) == 0 { return false; }
    let g3: f32x3 = cross(e3, n1); if project6(g3, p1, p2, p3, q1, q2, q3) == 0 { return false; }
    let h1: f32x3 = cross(f1, m1); if project6(h1, p1, p2, p3, q1, q2, q3) == 0 { return false; }
    let h2: f32x3 = cross(f2, m1); if project6(h2, p1, p2, p3, q1, q2, q3) == 0 { return false; }
    let h3: f32x3 = cross(f3, m1); if project6(h3, p1, p2, p3, q1, q2, q3) == 0 { return false; }
    return true;
}
func intersects(a: Triangle, b: Triangle) -> bool {
    return SAT_triangle_intersection(a.p0, a.p1, a.p2, b.p0, b.p1, b.p2);
}
func intersects(a: AABB, b: AABB) -> bool {
    let low: f32x3 = max(a.low, b.low); let high: f32x3 = min(a.high, b.high); return all(low <= high);
}

func collision_detection(bvh1: BVH, bvh2: BVH, r: mut set[(Triangle, Triangle)]) =
  match bvh1 {
  | Interior(low1, high1, left1, right1) ->
    match bvh2 {
    | Interior(low2, high2, left2, right2) ->
      if intersects(AABB(low1, high1), AABB(low2, high2)) {
        collision_detection(left1,  left2,  r); collision_detection(left1,  right2, r);
        collision_detection(right1, left2,  r); collision_detection(right1, right2, r);
      }
    | Leaf(low2, high2, _, _) ->
      if intersects(AABB(low1, high1), AABB(low2, high2)) {
        collision_detection(left1,  bvh2, r); collision_detection(right1, bvh2, r);
      }
    }
  | Leaf(low1, high1, _, data1) ->
    match bvh2 {
    | Interior(_, _, left2, right2) ->
      if intersects(AABB(low1, high1), AABB(low2, high2)) {
        collision_detection(bvh1, left2,  r); collision_detection(bvh1, right2, r);
      }
    | Leaf(low2, high2, _, data2) ->
      if intersects(AABB(low1, high1), AABB(low2, high2)) {
        foreach t1 in data1 {
          foreach t2 in data2 {
            if intersects(t1, t2) { r.insert((t1, t2)); } 
          } 
        } 
      }
    }
  }
\end{lstlisting}

\end{document}